\documentclass[journal]{IEEEtran}

\IEEEoverridecommandlockouts
\usepackage{multirow}%
\usepackage{bbm}%
\usepackage{bm}%
\usepackage{cite}%
\usepackage{amsmath,amssymb,amsfonts,tabularx}%
\usepackage{algorithm, algorithmic}%
\usepackage{graphicx}%
\usepackage{graphics}
\usepackage{textcomp}%
\usepackage{dsfont}%
\usepackage{cleveref}
\usepackage{comment}
\usepackage{gensymb}
\usepackage[long, c3, nocomma]{optidef}
\usepackage{epstopdf}
\usepackage{refcount}
\usepackage{threeparttable}
\usepackage{amsthm}
\usepackage{rotating}
\usepackage{tabularx}
\usepackage{pifont}
\usepackage{multirow}
\usepackage[activate={true,nocompatibility},final,tracking=true,kerning=true,spacing=true,factor=1100,stretch=10,shrink=10]{microtype}
\newtheorem{theorem}{Theorem}[section]

\allowdisplaybreaks[1]

\def\BibTeX{{\rm B\kern-.05em{\sc i\kern-.025em b}\kern-.08em
    T\kern-.1667em\lower.7ex\hbox{E}\kern-.125emX}}
\ifCLASSOPTIONcompsoc
    \usepackage[caption=false, font=normalsize, labelfont=sf, textfont=sf]{subfig}
\else
\usepackage[caption=false, font=footnotesize]{subfig}
\fi

\makeatletter
\begin{document}

	\def\correspondingauthor{\footnote{Corresponding author.}}
	
	\title{SA-HMTS: A Secure and Adaptive Hierarchical Multi-timescale Framework for Resilient Load Restoration Using A Community Microgrid}
	
	\author{Ashwin~Shirsat,~\IEEEmembership{Student Member,~IEEE},             Valliappan~Muthukaruppan,~\IEEEmembership{Student Member,~IEEE},
	Rongxing~Hu,~\IEEEmembership{Student Member,~IEEE},
	Victor~Paduani,~\IEEEmembership{Student Member,~IEEE},
	Bei~Xu,~\IEEEmembership{Student Member,~IEEE},
	Lidong~Song,~\IEEEmembership{Student Member,~IEEE},
	Yiyan~Li,~\IEEEmembership{Member,~IEEE},
	Ning~Lu,~\IEEEmembership{Fellow,~IEEE},
	Mesut~Baran,~\IEEEmembership{Fellow,~IEEE},
	David~Lubkeman,~\IEEEmembership{Fellow,~IEEE},
	and Wenyuan~Tang,~\IEEEmembership{Member,~IEEE}
		\thanks{\textit{(Corresponding author: Wenyuan Tang.)}
			\par The authors are with the Department of Electrical and Computer Engineering, North Carolina State University, Raleigh, NC 27695 USA (e-mail:
			\{ashirsa, vmuthuk2, rhu5, vdaldeg, bxu8, lsong4, yli257, nlu2, baran, dllubkem, wtang8\}@ncsu.edu). This material is based upon work supported by U.S. Department of Energy's Office of Energy Efficiency and Renewable Energy (EERE) under Solar Energy Technologies Office Award Number DE-EE0008770.}}

\maketitle
%%%%%%%%%%%%%%%%%%%Abstract%%%%%%%%%%%%%%%%%%
\begin{abstract}
Distribution system integrated community microgrids (CMGs) can partake in restoring loads during extended duration outages. At such times, the CMG is challenged with limited resource availability, absence of robust grid support, and heightened demand-supply uncertainty. This paper proposes a secure and adaptive three-stage hierarchical multi-timescale framework for scheduling and real-time (RT) dispatch of CMGs with hybrid PV systems to address these challenges. The framework enables the CMG to dynamically expand its boundary to support the neighboring grid sections and is adaptive to the changing forecast error impacts. The first stage solves a stochastic extended duration scheduling (EDS) problem to obtain referral plans for optimal resource rationing. The intermediate near-real-time (NRT) scheduling stage updates the EDS schedule closer to the dispatch time using newly obtained forecasts, followed by the RT dispatch stage. To make the dispatch decisions more secure and robust against forecast errors, a novel concept called delayed recourse is proposed. The methodology is evaluated via numerical simulations on a modified IEEE 123-bus system and validated using OpenDSS/hardware-in-loop simulations. The results show superior performance in maximizing load supply and continuous secure CMG operation under numerous operating scenarios.

\iffalse Lastly, the proposed approach is modified to make contingency aware decisions using a hybrid stochastic+robust optimization formulation. \fi 
\end{abstract}


	\section*{Nomenclature}
	\addcontentsline{toc}{section}{Nomenclature}
	
	\subsection{Abbreviations}
	\begin{IEEEdescription}[\IEEEusemathlabelsep\IEEEsetlabelwidth{$\underline{SOC}/ \overline{SOC}$}]
		%\small
		\item[$\text{CL}$] Critical load.
		\item[$\text{CMG}$] Community microgrid.
		\item[$\text{D}$] Demand.
		\item[$\text{EDS}$] Extended duration scheduling.
		\item[$\text{ES}$] Energy storage.
		\item[$\text{DG}$] Diesel generator.
		\item[$\text{DR}$] Demand response.
		\item[$\text{G}$] Generation.
		\item[$\text{HIL}$] Hardware-in-loop.
		\item[$\text{MG}$] Microgrid.
		\item[$\text{MSD}$] Minimum service duration.
		\item[$\text{NCL}$] Non-critical load.
		\item[$\text{NG}$] Node group.
		\item[$\text{NRT}$] Near-real-time.
		\item[$\text{PV}$] Photovoltaic system.
		\item[$\text{RR}$] DG Ramp rate.
		\item[$\text{RT}$] Real-time.
		\item[$\text{SA-HMTS}$] Secure and adaptive hierarchical multi-timescale.
	\end{IEEEdescription}
	\vspace{-0.0cm}
	
	\subsection{Sets, Indices, and Functions}
	\begin{IEEEdescription}[\IEEEusemathlabelsep\IEEEsetlabelwidth{$\underline{SOC}/ \overline{SOC}$}]
		%\small
        \item[$\mathcal{N}^{\text{NG}}$] NG index set.
        \item[$\mathcal{N}_n$] Node set of NG $n \in \mathcal{N}^{\text{NG}}$.
        \item[$\mathcal{N}^{\text{PV/ES/DG}}_n$] Node set of PV/ES/DG units belonging to NG $n$.
        \item[$\mathcal{N}^{\text{PV-UC}}_n$] Node set of uncontrollable PV generators within NG $n$, where $\mathcal{N}^{\text{PV-UC}}_n \in \mathcal{N}^{\text{PV}}_n$.
        \item[$\mathcal{N}^{\text{PV-C}}_n$] Node set of controllable PV generators within NG $n$, where $\mathcal{N}^{\text{PV-C}}_n \in  \mathcal{N}^{\text{PV}}_n$.
        \item[$\mathcal{N}^{\text{CL/NCL}}_n$] Node set of CL/NCL belonging to NG $n$.
        \item[$\mathcal{E}$] Network edge set.
        \item[$\mathcal{N}$] Network node set, where $\cup_{n \in \mathcal{N}^{\text{NG}}}{\mathcal{N}_n} = \mathcal{N}$.
        \item[$\mathcal{P}$] Network phase set.
        \item[$\mathcal{T}/\mathcal{H}_t/\mathcal{K}_{t,h}$] Time slot set for EDS/NRT/RT stages.
        \item[$\Omega$] EDS scenario set.
        
		\item[$i,j$] Network node index, where $i,j \in \mathcal{N}$.
		\item[$ij$] Network edge index, where $ij \in \mathcal{E}$.
		\item[$p$] Network node phase, where $p \in \mathcal{P}$.
		\item[$n$] NG index, where $n \in \mathcal{N}^{NG}$.
		\item[$s$] Scenario index, where $s \in \Omega$.
		\item[$t/h/k$] Time slot index for EDS/NRT/RT stages, where $t \in \mathcal{T}$, $h \in \mathcal{H}_t$, and $k \in \mathcal{K}_{t,h}$.
		
		\item[$\langle \boldsymbol{\cdot}, \boldsymbol{\cdot} \rangle$] Inner product of two vectors.
		\item[$(\tilde{\boldsymbol{\cdot}})$] Circularly shifted array by one position.For example, if $\textbf{P} = [a,b,c]$, then $\tilde{\textbf{P}} = [b,c,a]$.
		\item[$(\overline{\overline{\boldsymbol{\cdot}}})$] Average of all vector elements. 
		\item[$(\hat{\boldsymbol{\cdot}})$] Reference values obtained from the immediate previous stage results.
		\item[${[\boldsymbol{\cdot}]}_{+}$] Maximum number when compared with $0$.
	\end{IEEEdescription}
	\vspace{-0.0cm}

	\subsection{Common Parameters}
	\begin{IEEEdescription}[\IEEEusemathlabelsep\IEEEsetlabelwidth{$\underline{SOC}/ SOC \overline{SOC}$}]
		\item[$\overline{S}_{i}^{\text{PV/ES/DG}}$] kVA rating of the PV/ES/DG unit.	
		\item[$\overline{E}_{i}^{\text{ES}}$] kWh rating of ES unit.	
		\item[$\{\overline{SOC}, \underline{SOC}\}_{i}^{\text{ES}}$] Max./Min. SOC limits of ES unit.
		\item[$\{\overline{F},\underline{F}\}_{i}^{\text{DG}}$] Max./Min. fuel limits of DG unit.
		\item[$\alpha_i, \beta_i$] Fuel consumption coefficients of DG (L/kW-hr).
		\item[$\omega_i^{\text{1}}$] Priority weight based on load type.
		\item[$\omega_{i,t}^{\text{2}}$] Priority weight based on load service duration.
		\item[$\gamma$] Reserve factor for ES and DG units $\in \{1,2\}$. 		
		
	\end{IEEEdescription}
	
	\subsection{EDS Stage Parameters}
	\begin{IEEEdescription}[\IEEEusemathlabelsep\IEEEsetlabelwidth{$\underline{SOC}/ SOC \overline{SOC}$}]
		%\small
		\item[$\pi_{s}$] Scenario probability.
		\item[$\{\overline{P}, \overline{Q}\}_{i,t,s}^{\text{PV}}$] Maximum real and reactive power output of PV unit.
		\item[$\{\overline{P}, \overline{Q}\}_{i}^{\text{ES}}$] Maximum real and reactive power input/output of ES unit.
		\item[$\{\overline{P}, \overline{Q}\}_{i}^{\text{DG}}$] Maximum real and reactive power output of DG unit.
		\item[$\{\underline{P}, \underline{Q}\}_{i}^{\text{DG}}$] Minimum real and reactive power output of DG unit.
		\item[$\{\overline{P}, \overline{Q}\}_{i,t,s}^{\text{D}}$] Real and reactive power forecast.
		\item[$\{\underline{P}, \underline{Q}\}_{i,t,s}^{\text{D}}$] Minimum must supply real and reactive power.
		\item[$\overline{{P}}_{i}^{\text{DG,RR}}$] Maximum ramp rate of DG unit.
		\item[$\nu$] Minimum service duration of a NG by CMG.
		\item[$\eta_n$] Minimum load supply requirement for connectivity of NG $n\in \mathcal{N}^{NG}$.
		\item[$\epsilon_n$] Chance constraint violation threshold for NG $n\in \mathcal{N}^{NG}$.
		\item[$\Delta t$] EDS simulation time slot duration.

	\end{IEEEdescription}
	\vspace{-0cm}

	\subsection{NRT and RT Stage Parameters}
	\begin{IEEEdescription}[\IEEEusemathlabelsep\IEEEsetlabelwidth{$\underline{SOC}/ SOC \overline{SOC}$}]

	    \item[$\{\overline{\textbf{P}}, \overline{\textbf{Q}}\}$] Network line flow limits.	
	    \item[$\{r, x\}_{ij}^{pp'}$] Line resistance and reactance between phase $p$ and $p'$.
		\item[$\{\overline{\textbf{P}}, \overline{\textbf{Q}}\}_{i,h/k}^{\text{PV}}$] Maximum real and reactive power output of PV unit.
		\item[$\{\overline{\textbf{P}}, \overline{\textbf{Q}}\}_{i}^{\text{ES}}$] Maximum real and reactive power input/output of ES unit.
		\item[$\{\overline{\textbf{P}}, \overline{\textbf{Q}}\}_{i}^{\text{DG}}$] Maximum real and reactive power output of DG unit.
		\item[$\{\underline{\textbf{P}}, \underline{\textbf{Q}}\}_{i}^{\text{DG}}$] Minimum real and reactive power output of DG unit.
		\item[${\textbf{P}}^{\text{G,FE}}_t$] Forecast error induced generation over/under-consumption observed in RT during hour $t$.
		\item[$\{\overline{\textbf{V}}, \underline{\textbf{V}}\}$] ANSI voltage limits.
	    \item[$\{\overline{\textbf{P}}, \overline{\textbf{Q}}\}_{i,h/k}^\text{D}$] Real and reactive demand forecast.
	    \item[$\{\overline{\textbf{P}}, \overline{\textbf{Q}}\}_{i,h/k}^\text{CLPU}$] Real and reactive cold load demand estimate.
	    \item[$\overline{\boldsymbol{\delta}}_{i}^{\text{DG}}$] DG unit phase imbalance limit. 
		\item[$\Delta h/\Delta k$] NRT/RT simulation time slot duration.
		
	\end{IEEEdescription}

	\subsection{EDS Stage Decision Variables}
	\begin{IEEEdescription}[\IEEEusemathlabelsep\IEEEsetlabelwidth{$\underline{SOC}/ \overline{SOC}$}]
		%\small
		\item[$\{P,Q\}_{i,t,s}^{\text{PV}}$] PV output power.
		\item[$\{P,Q\}_{i,t,s}^{\text{ES}}$] ES charge/discharge power.
		\item[$\{P,Q\}_{i,t}^{\text{DG}}$] DG output power.		
		\item[$\{P,Q\}_{i,t,s}^{\text{D}}$] Load allocated for supply.
		\item[$SOC_{i,t,s}^{\text{ES}}$] ES SOC value.
		\item[$F_{i,t}^{\text{DG}}$] DG fuel consumption.
		\iffalse
		\item[$X^{\text{DG/ES/PV}}_{i,s}$] Connectivity status indicator of DG/ES/PV (1=connected, 0=disconnected).
		\fi
		\item[$\phi_{n,t}$] Probability of supplying $\eta_n\%$ demand of NG $n$.
		\item[$\theta_{n,t}$] Connectivity status indicator of NG $n$ (1=connected, 0=disconnected).	
		\iffalse
		\item[{$\Gamma_s$}] Budget of uncertainty for robust contingency analysis.
		\fi
	\end{IEEEdescription}

	\subsection{NRT and RT Stage Decision Variables}
	\begin{IEEEdescription}[\IEEEusemathlabelsep\IEEEsetlabelwidth{$\underline{SOC}/ \overline{SOC}$}]
		%\small
		\item[$\{\textbf{P},\textbf{Q}\}_{ij,h/k}$] Power flowing between nodes $i$ and $j$.
		\item[$\{\textbf{P},\textbf{Q}\}_{i,h/k}^{\text{PV}}$] PV output power.
		\item[$\{\textbf{P},\textbf{Q}\}_{i,h/k}^{\text{ES}}$] ES charge/discharge power.
		\item[$\{\textbf{P},\textbf{Q}\}_{i,h/k}^{\text{DG}}$] DG output power.		\item[$\{\textbf{P},\textbf{Q}\}_{i,h/k}^{\text{D}}$] Load scheduled to be supplied.
		\item[${SOC}_{i,h/k}^{\text{ES}}$] ES SOC value.
		\item[$F_{i,h/k}^{\text{DG}}$] DG fuel consumption.
		\item[$\textbf{V}_{i,h/k}$] Squared voltage of node $i$.
		\item[$\boldsymbol{\rho}_{ij,h/k}$] Direction indicator of power flow between nodes $i$ and $j$.
		\item[$\zeta_{ij,h/k}$] Slack variable to capture voltage difference between disconnected nodes $i$ and $j$.		\item[$\textbf{x}_{i,t}$] Load connectivity status of node $i$.
		\item[$\boldsymbol{\delta}_{h}^{\text{D}}$] Network load phase imbalance.
		\item[$\boldsymbol{\delta}_{i,h}^{\text{D}}$] DG phase imbalance.
	\end{IEEEdescription}	

%	\subsection{RT Stage Decision Variables}
%	\begin{IEEEdescription}[\IEEEusemathlabelsep\IEEEsetlabelwidth{$\underline{SOC}/ \overline{SOC}$}]
		%\small
%		\item[$\{\textbf{P},\textbf{Q}\}_{ij,k}$] Power flowing between nodes $i$ and $j$.
%		\item[$\{\textbf{P},\textbf{Q}\}_{i,k}^{\text{G,RT}}$] Total RT scheduled generation.
%		\item[$\{\textbf{P},\textbf{Q}\}_{i,k}^{\text{PV,RT}}$] Total RT PV output power.
%		\item[$\{\textbf{P},\textbf{Q}\}_{i,k}^{\text{ES,RT}}$] Total RT ES charge/discharge power.
%		\item[$\{\textbf{P},\textbf{Q}\}_{i,k}^{\text{DG,RT}}$] Total RT DG output power.		\item[$\{\textbf{P},\textbf{Q}\}_{i,k}^{\text{D,RT}}$] Total RT load scheduled to be supplied.
%		\item[${SOC}_{i,k}^{\text{ES,RT}}$] EDS ES SOC value.
%		\item[$F_{i,k}^{\text{DG,RT}}$] RT DG fuel consumption.
%		\item[$\textbf{V}_{i,k}$] Voltage of node $i$.
%		\item[$\rho_{ij,k}$] Direction indicator of power flow between nodes $i$ and $j$.
%		\item[$\zeta_{ij,k}$] Slack variable to capture voltage difference between disconnected nodes $i$ and $j$.		\item[$x_{i}$] Load connectivity status for node $i$.
%		\item[$\boldsymbol{\delta}_{i,k}^{\text{DG,RT}}$] DG phase imbalance. 
%	\end{IEEEdescription}
	
\section{Introduction}
%%%%%%%%%%%%%%%%%%%Introduction%%%%%%%%%%%%%%%%%%
The resiliency of the distribution grid needs to be enhanced for withstanding, operating, and recovering from the disruptions caused by extreme natural events such as wildfires, storms, hurricanes, and manufactured threats such as cyber-security attacks. Such extreme events pose severe short-term and long-term consequences to power systems in terms of financial losses, infrastructure damages, customer inconvenience due to loss of electricity supply for a long duration, and harm to human life \cite{outage_impact1}. Events like these tend to fall on the high-impact low-frequency (HILF) risk spectrum \cite{lfhi_riskspectrum}. However, the intensity and frequency of such extreme weather events are on the rise and are expected to soar due to climate change. Further, scientific advancements in cyber warfare technologies also pose an alarming threat to the electricity grid. The year 2020 witnessed 22 extreme weather events costing 95 billion USD in damages alone and shattering the record of 16 events for the years 2011 and 2017 \cite{lfhi_stats1}. Loss of electricity during such adverse events does take a significant toll on human life. For example, the 2021 Winter Storm Uri drove the electric grid in Texas toward its breakpoint, which resulted in the death of 210 people, most of which could be attributed to hypothermia due to the unavailability of electricity supply to keep themselves warm in freezing weather conditions. 

Enhancement of the power system resiliency and making its operation robust to extreme events is the key to bolstering the system's operational performance. Various studies performed for analyzing and improving the network resiliency propose two measures, i.e., long-term hardening measures and short-term operational measures \cite{resilience_review1}. Long-term hardening measures incorporate strengthening of and modifications to the physical components of the grid before the occurrence of the extreme event. In contrast, short-term operational measures focus on developing energy management and topology reconfiguration algorithms for the safe operation of the system. 

For hardening the electric grid and improving its reliability and resiliency, MGs are considered as a practical and viable solution \cite{microgridbenefit3}. This is due to the ability of the MGs to operate in an islanded manner and smoothly integrate distributed generation, especially renewables \cite{microgrid_benefit1, microgrid_benefit2}. Incorporating a significantly high portion of intermittent renewable generation poses a significant challenge to grid stability and security. In the traditional centralized grids, conventional generation, renewable generation, and energy consumption are distant from each other. Hence, coordinating the conventional generation and the consumption based on the uncertainty and fast variations in the renewable generation becomes challenging and can lead to grid stability issues. On the contrary, by colocating generation and consumption within a small geographic area, such as that of a MG, the renewable energy uncertainty is absorbed locally \cite{microgridbenefit4}. This minimizes the negative impact on the stability of the macrogrid. These two critical abilities provide numerous benefits to ensure increased macrogrid resiliency by using MGs. Although cyber-security attacks can affect any system, MGs provide certain benefits over large-scale interconnected grids due to their inherent decentralized structure. The MG paradigm also offers resiliency against natural disasters due to the colocation of active components such as distributed generators and flexible loads, minimized distance between generation and demand, and reduced susceptibility of conventional centralized grid and control architecture \cite{microgrid_benefit2}.

When it comes to the MG paradigm, the existing literature can be categorized into two, i.e., MGs as a resiliency resource and MG operation strategy for network resiliency improvement. The key focus of this paper is to propose an approach that combines the above two categories. Numerous studies are available in the existing literature that portray the use of MGs as a resilience resource during extended outages caused by extreme events \cite{mg_as_resilienceresource1, mg_as_resilienceresource2, mg_as_resilienceresource3, mg_as_resilienceresource4, mg_as_resilienceresource5,mg_as_resilienceresource6}. \cite{mg_as_resilienceresource1} and \cite{ mg_as_resilienceresource2} propose resiliency oriented load restoration using a MG by prioritizing CL supply. \cite{mg_as_resilienceresource1} uses strategy table approach wherein the ideal feasible restoration path is chosen from a dictionary of paths stored in the strategy table. \cite{mg_as_resilienceresource2} uses a coverage maximization approach for load restoration aimed at maximizing the load restored by prioritizing CLs. Apart from using the existing constructed MGs for improving the resiliency of the macrogrid, a new trend is dynamically partitioning the existing macrogrid into multiple MGs by isolating the faults in the grid \cite{mg_as_resilienceresource3, mg_as_resilienceresource4}. \cite{mg_as_resilienceresource5} considers resiliency improvement from a planning perspective wherein the optimal placement of distributed generation is considered, which will assist in the dynamic formation of self-adequate MGs. Lastly, to quantify the impact of MGs on the grid resiliency, \cite{mg_as_resilienceresource6} proposes four indices that serve the purpose of key performance indicators. There is growing literature on the topic of MGs as a resiliency resource. However, to achieve the merits highlighted in this literature, it is necessary to also focus on the operational strategies adopted by the MGs to enhance their resiliency. 

A lot of recent studies have focused on developing new algorithms and strategies for the secure and resilient operation of the MGs \cite{mg_resilient_operation1, mg_resilient_operation2, mg_resilient_operation3, mg_resilient_operation4, mg_resilient_operation5, mg_resilient_operation6, da_mg_sch_op1, da_mg_sch_op2}. In \cite{mg_resilient_operation1}, the authors have proposed a MG resiliency enhancement during floods by modifying the MG operation by identifying the vulnerable network components and then proactively tripping them. In \cite{mg_resilient_operation2}, Gholami \textit{et al.} have proposed a robust optimization-based day-ahead (DA) scheduling for MG resiliency enhancement via pre-event preparedness. \cite{mg_resilient_operation3} proposes a survivability approach for MG resiliency enhancement for long-duration outages. For ensuring survivability of the MG CLs, \cite{mg_resilient_operation4} proposes a resiliency-constrained MG operation using ES units as resiliency resources. \cite{mg_resilient_operation5} proposes a resiliency-oriented approach for MG scheduling for extended outages by splitting the scheduling problem into a normal scheduling problem and a resilient scheduling problem. Further, the use of resiliency cuts is proposed to ensure secure MG scheduling. \cite{mg_resilient_operation6} proposes a two-stage stochastic approach for the resilient MG scheduling capable of mitigating the damaging impacts caused by electricity interruptions. In \cite{da_mg_sch_op1}, Yang \textit{et al.} have proposed a two-stage approach for MG scheduling and RT dispatch. The scheduling problem is solved for the projected outage duration, and the optimal power flow (OPF) based RT dispatch problem is solved using the scheduling results. In \cite{da_mg_sch_op2}, Qiu \textit{et al.} propose a three-stage formulation for optimal dispatch of MGs for islanded operation under normal conditions. The existing literature covers various aspects of MG energy management. However, a holistic approach for proactive scheduling and dispatch of MGs during emergencies emphasizing uncertainty mitigation, CL priority, optimal resource allocation for self-sustained operation, and MG support expansion to the neighboring grid has not been addressed.

From the above literature, the use of MGs for resiliency enhancement is either studied from how MGs can be formed and used to enhance resiliency or how MGs can be proactively scheduled and operated. For the former approach, dynamically partitioning the electric grid into self-sufficient MGs looks like a very viable option for resiliency improvement theoretically. The dynamic MG formation approach in distribution grids with increased behind-the-meter (BTM) PV systems will significantly increase resiliency under certain extreme operating conditions. This is due to the grid support provided by the MG necessary for integrating BTM systems with the grid. However, to realize this concept in the real world, significant modifications and reinforcements to the existing grid will be required, and it does not appear to be a viable solution to reinforce the existing infrastructure. To achieve a smooth realization of this approach, the existing distribution grid would be required to be reinforced with additional switches, a new protection mechanism to handle bidirectional flows, and an advanced control system capable of controlling the MGs with non-stationary boundaries. To avoid this issue and yet have flexibility over the MG formation, we propose the dynamic boundary expansion of existing CMGs. This will ensure that there will be limited combinations of the dynamic boundary, and the power flow direction from a network protection perspective will be known since the CMG will expand its boundary in one particular direction due to the necessity of maintaining network radiality. Further, allowing pre-existing CMGs to expand their boundaries will ensure that the network islands formed are secure due to the availability of robust grid-support from the CMG and will be able to maximize the use of BTM PV systems that exist outside the CMG, which otherwise would not be able to participate in load restoration due to lack of grid-support.   

For the latter, various stochastic or robust optimization-based approaches are used for uncertainty-aware scheduling and operation of the MGs. With more renewable generation being plugged into the grid, the MG scheduling and operation uncertainty rises. This uncertainty issue is further plagued due to the extremely low probability of high-impact events. Since such events do not occur regularly, and each event is unique from the other, it is not easy to train the forecasting algorithms to obtain accurate forecasts for such events. Further, a forecaster trained using large historical data of normal operating conditions would produce biased forecasts during such extreme events. \cite{forecast_error_general} estimates that the DA forecasting error for residential load and PV generation is approximately 20\%, even when real-life data is used for developing the forecasting model. A similar study performed in \cite{forecast_new_2} for reviewing state-of-the-art methods for PV forecasting states the average MAPE is 21.76\%. The energy management scheduling decisions using the DA forecast will not be executable for this degree of error. This places much stress on developing uncertainty-aware algorithms capable of reliable MG operation decision-making.

Various approaches using stochastic optimization and robust optimization with or without receding horizon control have been proposed in the literature for MG scheduling and operation under uncertainty. By incorporating uncertainties in renewable generation and demand, \cite{mgmanagement1_aro} has proposed an adaptive robust optimization (ARO) based two-stage optimization model for a grid-connected MG. The first stage handles the unit-commitment problem and the second stage solves a RT economic dispatch problem. \cite{mg_resilient_operation2} has proposed an ARO-based approach for MG resiliency enhancement by incorporating uncertainty in net demand, RT electricity price, and islanding events. In \cite{mgmanagement2}, a two-stage approach is proposed that alters the DA scheduling plans before the RT dispatch using the data obtained closer to the actual time of dispatch. In \cite{mgmanagement3}, the authors have proposed a rolling horizon-based optimization model for RT energy management of a grid-connected MG. Their staggered approach consists of a high-level scheduling horizon that obtains the preliminary schedule of the MG for the considered duration, followed by a prediction horizon that is initiated in a rolling horizon manner. Finally, the RT dispatch follows the prediction horizon implementing the decisions of the first time-interval in the prediction horizon. The limitation of this approach is that the impact of uncertainty in the scheduling horizon is not considered. This approach will work under a grid-connected setting, in which any forecast error impact can be easily addressed using grid interaction. However, the limited availability of resources in an islanded mode would eliminate this flexibility. Thus, the deterministic scheduling horizon would limit the robustness of the proposed approach against uncertainty \cite{mgmanagement4}. A similar multi-timescale approach is proposed in \cite{mgmanagement4_1,mgmanagement4_2} for multi-energy MGs with combined cooling, heating, and power functionalities. The uncertainties in heating and electric demand are addressed by using a rolling horizon approach on hourly and intra-hourly timescales. 

Summarizing the existing literature on using MGs as a resiliency resource and developing energy management algorithms for MGs, we notice the following limitations. Firstly, from the algorithm perspective, the most commonly used approach is having an DA stage followed by RT dispatch. If the uncertainties are accounted for in the DA stage using stochastic/robust/ARO optimization, the RT dispatch decisions are likely to be feasible given the realization of uncertainty owing to their conservative nature. However, the algorithm will likely fail with deterministic DA stage formulation due to forecast errors. Second, most approaches consider grid-connected MGs wherein the grid is actively providing the necessary support required for MG sustainability. In such cases, ample flexibility is available to correct the forecast error by altering grid interactions. However, this flexibility disappears during extreme events due to the loss of robust grid support, which the algorithms must proactively factor in. Third, most approaches compute the DA decisions with a longer look-ahead horizon only once, and the subsequent lower stages are computed in a rolling horizon manner. However, during extreme conditions, constrained resource availability and the occurrence of unforeseen contingencies may require the extended horizon schedule to update frequently. Fourth, the stochastic/robust/ARO approach heavily relies on a priori probability distributions of the uncertain variables or worst-case uncertainty set. However, HILF events tend to impact the probability distributions, which cannot be estimated correctly due to limited data available on such events. This would result in biased scenarios for stochastic optimization and a substantially large uncertainty set for robust/ARO optimization. The former results may not provide the necessary robustness against uncertainty, and those of the latter would be difficult to compute due to increased computation required due to a larger size of the uncertainty set. 

Further, the MGs may have access to limited computing resources due to the conditions created by the extreme conditions, thus requiring the energy management system to be highly computationally efficient. In addition, the existing literature focuses on develping new solution algorithms to solve the proposed complex optimization probloem but does not emphasize demonstrating the implementability of the scheduling decisions on real-world systems using OpenDSS/HIL simulations. The literature proposing MGs with dynamic boundaries has demonstrated a feasible results on paper but has not validated the concept using HIL network simulators. In short, the combined approach of using a MG with a dynamic boundary for resilient extended-duration system operation under extreme conditions, which is validated in a dynamic operating environment using HIL simulation, has not been explored yet. Table \ref{tab:lit_review} summarizes the literature review and compares the existing literature with our proposed approach.

\begin{table*}[]
\caption{Summary of literature review.}
\centering
\label{tab:lit_review}
\begin{tabular}{c|c|c|c|c|c|c||c|c|c}
\hline
\multirow{2}{*}{Reference}       & \multicolumn{6}{c||}{Framework specifics}                                                                                                                                                                                                                                                                                                                                                                                                        & \multicolumn{3}{c}{Numerical simulations}                                                                                                                                                                                                          \\ \cline{2-10} 
                                 & \multicolumn{1}{c|}{\begin{tabular}[c]{@{}c@{}}Load \\ maximization\end{tabular}} & \multicolumn{1}{c|}{Stages} & \multicolumn{1}{c|}{\begin{tabular}[c]{@{}c@{}}Rolling \\ horizon\end{tabular}} & \multicolumn{1}{c|}{\begin{tabular}[c]{@{}c@{}}Dynamic MG \\ formation\end{tabular}} & \multicolumn{1}{c|}{\begin{tabular}[c]{@{}c@{}}Uncertainty \\ aware\end{tabular}} & \begin{tabular}[c]{@{}c@{}}RT dispatch \\ frequency\end{tabular} & \multicolumn{1}{c|}{\begin{tabular}[c]{@{}c@{}}Digital twin \\ validation$^{*}$\end{tabular}} & \multicolumn{1}{c|}{\begin{tabular}[c]{@{}c@{}}Forecast \\ error analysis\end{tabular}} & \begin{tabular}[c]{@{}c@{}}Temporal \\ generalizability\end{tabular} \\ \hline \hline
\cite{mg_as_resilienceresource1} & \checkmark  &  2   & x           & \checkmark  & \checkmark  & NA      & \checkmark  & x  & x \\ \hline

\cite{mg_as_resilienceresource2} & \checkmark  &  1   & x           & \checkmark  & x           & NA      & \checkmark  & x  & x \\ \hline

\cite{mg_as_resilienceresource3} & \checkmark  &  1   & \checkmark  & \checkmark  & \checkmark  & 1 hour  & x           & x  & x \\ \hline

\cite{mg_as_resilienceresource4} & \checkmark  &  1   & x           & \checkmark  & x           & NA      & x           & x  & x \\ \hline

\cite{mg_as_resilienceresource5} & \checkmark  &  1   & x           & \checkmark  & \checkmark  & NA      & x           & x  & x \\ \hline

\cite{mg_resilient_operation1} & \checkmark  &  1   & x           & x  & x            & NA          & \checkmark  & x  & x \\ \hline

\cite{mg_resilient_operation2} & x           &  1   & x           & x  & \checkmark   & NA          & \checkmark  & x  & x \\ \hline

\cite{mg_resilient_operation3} & \checkmark  &  1   & x           & x  & \checkmark   & 1 hour      & x           & x  & \checkmark \\ \hline

\cite{mg_resilient_operation4} & \checkmark  &  1   & x           & x  & \checkmark   & 1 hour      & x           & x  & \checkmark \\ \hline

\cite{mg_resilient_operation5} & \checkmark  &  1   & x           & x  & \checkmark  & 1 hour      & x           & \checkmark  & x \\ \hline

\cite{mg_resilient_operation6} & x           &  1   & x           & x  & \checkmark  & 1 hour      & x           & x  & x \\ \hline

\cite{da_mg_sch_op1} & x           &  2   & \checkmark           & x  & x            & 5 minutes          & \checkmark  & x  & x \\ \hline

\cite{da_mg_sch_op2} & x           &  3   & \checkmark           & x  & \checkmark   & 5 minutes          & x           & x  & x \\ \hline

\cite{mgmanagement2} & x           &  2   & \checkmark           & x  & x           & 15 minutes          & x           & x  & x \\ \hline

\cite{mgmanagement3} & x           &  1   & \checkmark           & x  & x           & 1 hour              & x           & x  & x \\ \hline

\cite{mgmanagement4} & x           &  3   & \checkmark           & x  & \checkmark  & 15 minutes          & x           & x  & x \\ \hline

\cite{mgmanagement4_1, mgmanagement4_2} & x           &  2   & \checkmark           & x  & \checkmark  & 5 minutes      & x           & x  & x \\ \hline

SA-HMTS & \checkmark           &  3   & \checkmark           & \checkmark  & \checkmark  & 5 minutes      & \checkmark           & \checkmark  & \checkmark \\ \hline
\end{tabular}
\begin{tablenotes}
\centering
\item[*] ${}^{*}$ Using OpenDSS or HIL simulations.    
\end{tablenotes}
\end{table*}

%elaborate then on forecast errors, need for proactive management under extreme conditions, 

To address the above limitations, the contributions of our paper are summarized as follows: 
\begin{enumerate}
    \item A secure and adaptive three-stage hierarchical multi-timescale model for proactive scheduling and RT operation of a CMG with high penetration of residential BTM PV generators.
    \item A CMG with dynamic boundary is considered, i.e., the CMG can dynamically expand its boundary to support neighboring distribution system nodes.
    \item The proposed energy management system is designed to maintain a sustained and secure operation during extended duration outages, prioritizing a continuous supply to the CLs.
    \item A new approach called delayed recourse is developed for adapting the decisions to the time-varying forecast error and mitigating its impact.
    \item The proposed approach is designed for load restoration in PV-dense networks during extended duration outages and is validated using OpenDSS/HIL simulations. 
\end{enumerate}
The paper is organized as follows. Section II introduces the SA-HMTS optimization framework. Section III describes the novel delayed recourse concept for proactive forecast error impact mitigation. Section IV describes the conducted case studies, and section V concludes the paper. The Appendix section describes the polygonal relaxation methodology, chance constraint approximation, and the proofs for the theorems listed in Section III. 

%%%%%%%%%%%%%%%%%%%Methodology%%%%%%%%%%%%%%%%%%
\vspace{-0cm}
\section{SA-HMTS Problem Formulation}
\begin{figure}[tb]
  \centering
  \includegraphics[width = \linewidth ,keepaspectratio, trim={1.95cm 3.25cm 3.0cm 4cm},clip]{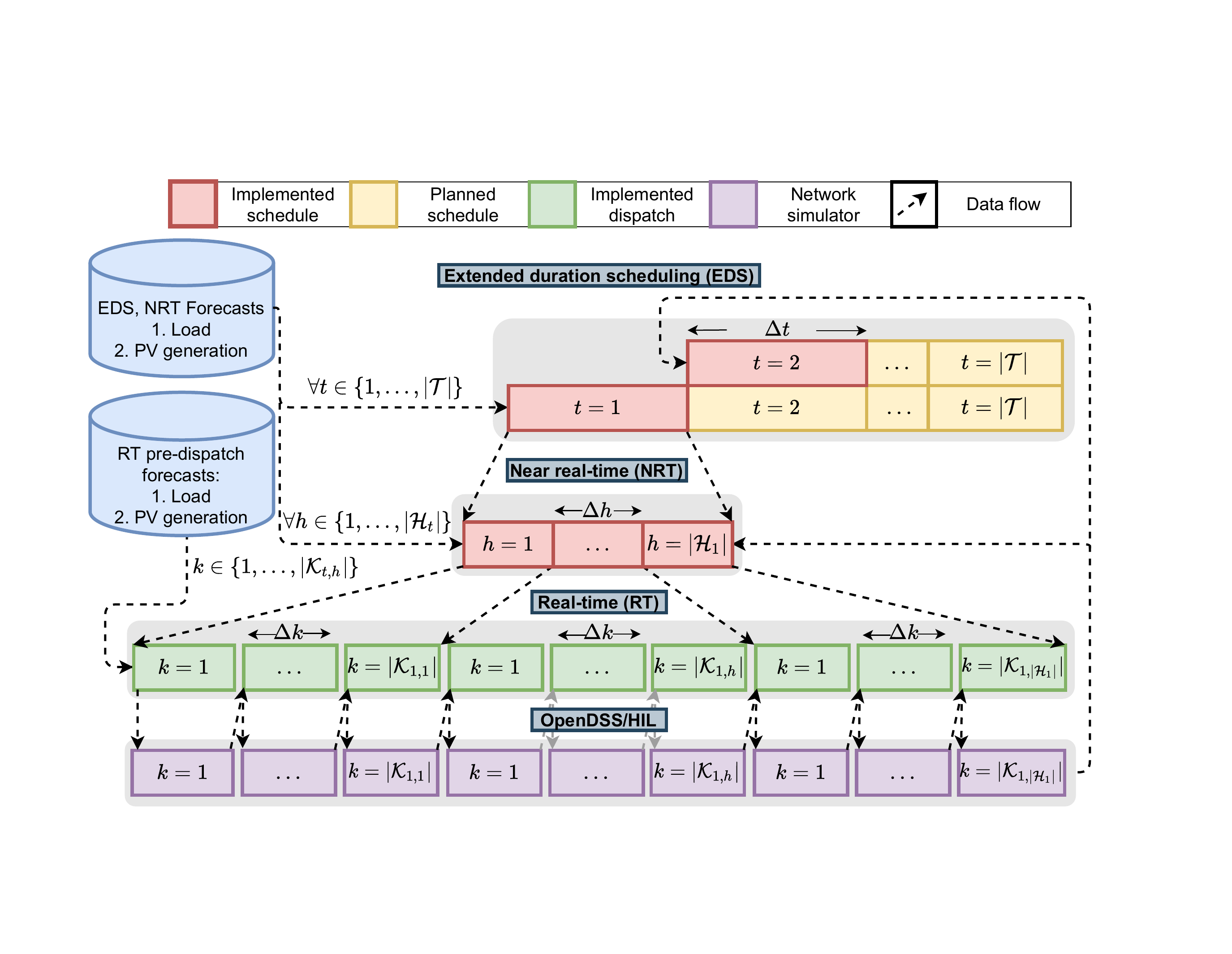}
  \caption{Schematic layout of the proposed HMTS framework.}
  \label{fig:layout}
  \vspace{-0.5cm}
\end{figure}

\begin{figure}[tb]
  \centering
  \includegraphics[width = \linewidth ,keepaspectratio, trim={2cm 6.25cm 3.5cm 3.75cm},clip]{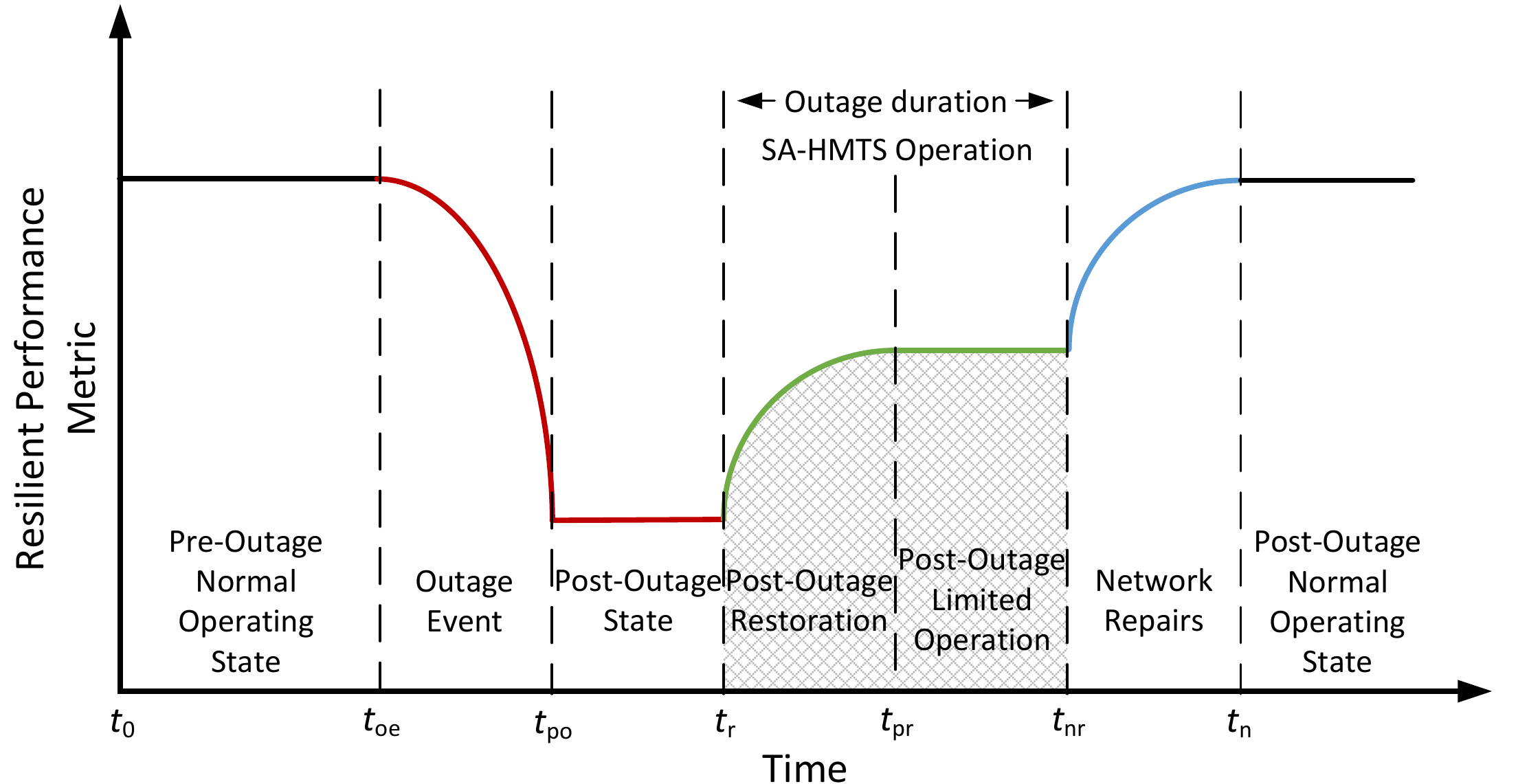}
  \caption{Distribution network resiliency curve during extreme events.}
  \label{fig:resilience_curve}
  \vspace{-0.5cm}
\end{figure}

Minimal load and renewable generation data are available for HILF events for making accurate forecasts. Hence, a novel three-stage hierarchical multi-timescale approach is proposed for the resilient energy management of CMGs under extended outage duration cases. A new intermediate stage is added between the EDS and RT stage to mitigate the impact of forecast errors, which uses updated forecasts made closer to the RT dispatch time. The underlying ideology is that increased forecast accuracy is observed as the forecast interval gets closer to the actual dispatch time \cite{da_mg_sch_op2}. Hence, improvising decisions before final dispatch with new obtained forecasts will ensure more proactive and secure decision making. A pictorial representation of the proposed HMTS framework is shown in Fig. \ref{fig:layout}. Fig. \ref{fig:resilience_curve} shows the resilience curve that is typically observed when the normal system operations are impacted by extreme events. After an extreme event occurs, the system usually transitions through five stages before restoring the normal operating state. The proposed SA-HMTS approach is designed for resilient load restoration. Hence, it will be used during the restorative operation states between the time intervals $t_{\text{r}}$ and $t_{\text{nr}}$. In the remainder of the paper, the duration between these two time intervals is referred to as the total outage duration $\mathcal{T}$.

\subsection{Stage 1: Extended Duration Scheduling}
The emphasis on proactive CMG scheduling begins with the EDS stage. While operating in an islanded mode for extended duration outages during adverse operating conditions, the CMG is challenged with the following: minimal generation resources at hand, heightened impact of uncertainty, and high likelihood of component failure. Owing to these issues, the EDS stage of the HMTS model emphasizes optimal resource allocation for the entire outage duration to avoid premature generation depletion and integration of uncertainty in load and PV generation. 

%The EDS stage problem is solved in a receding horizon fashion for the entire forecasted outage duration. The EDS forecasts are know to demonstrate low accuracy, with a further dip in the accuracy as the forecasted outage duration increases. Hence, we adopt a stochastic formulation to model the uncertainty in load and PV generation. This reduces the stress on the forecasting model to generate highly accurate EDS forecasts. %Further, the uncertainty of occurrence of a contingency in the CMG is integrated via a robust optimization formulation. The combined uncertainty integration in the EDS stage decision making results in a hybrid stochastic-robust optimization (HSRO) formulation. The two-fold reason behind the choice of using the HSRO approach is as follows: 1) It is difficult to estimate the probability distributions of the CMG component failure rate; 2) Using the robust formulation, the system operator can exercise flexibility over the choice of different risk levels.  

The EDS forecasts demonstrate low accuracy, with a further dip in the accuracy as the forecasted outage duration increases. Hence, we adopt a stochastic formulation to model load and PV generation uncertainty. This reduces the stress on the forecasting model to generate highly accurate EDS forecasts. To further alleviate this issue, the EDS stage problem can be solved in a receding horizon fashion after every hour or when there is an unforeseen event. %such as a change in the outage duration, network topology, network generation portfolio, equipment failure, or when a significantly high forecasting error is observed. 
The event-driven EDS scheduling stage can thus be considered as a flexible time frame scheduling stage, wherein the scheduling horizon, network topology, and generation portfolio can accommodate the changing operating conditions. The decision on the format of the receding horizon approach can be made based on the availability of the computational resources. This proposed approach differs from most of the existing approaches which use a single DA stage in place of the EDS stage and only obtain its decisions once before the start of the outage duration. 

Furthermore, we introduce the concept of dynamic CMG, wherein the CMG can expand its support to the neighboring distribution system nodes. This concept is demonstrated in Fig. \ref{fig:dynamic_cmg}. The dynamic CMG functionality provides a two-fold benefit to the system. First, the CMG can use additional resources from the external distribution grid by providing grid-support functionality to the neighboring nodes. This symbiotic relationship will benefit the CMG customers and the customers connected to the external grid. Second, the amount of load (including the cold load) that needs to be picked up during the final restoration phase would reduce. This would mitigate the burden on the upstream transmission network during the final restoration. The dynamic CMG boundary decisions are denoted by the binary variable $\theta_{n,t}$, where $\theta_{1,t}$ denotes the CMG NG, which is always connected. 

The objective function (\ref{eq:da_opt_obj}) aims at maximizing the total expected served load by prioritizing the CLs:  
\begin{figure}[tb]
  \centering
  \includegraphics[width = \linewidth ,keepaspectratio, trim={0.5cm 9.5cm 3.0cm 0.75cm},clip]{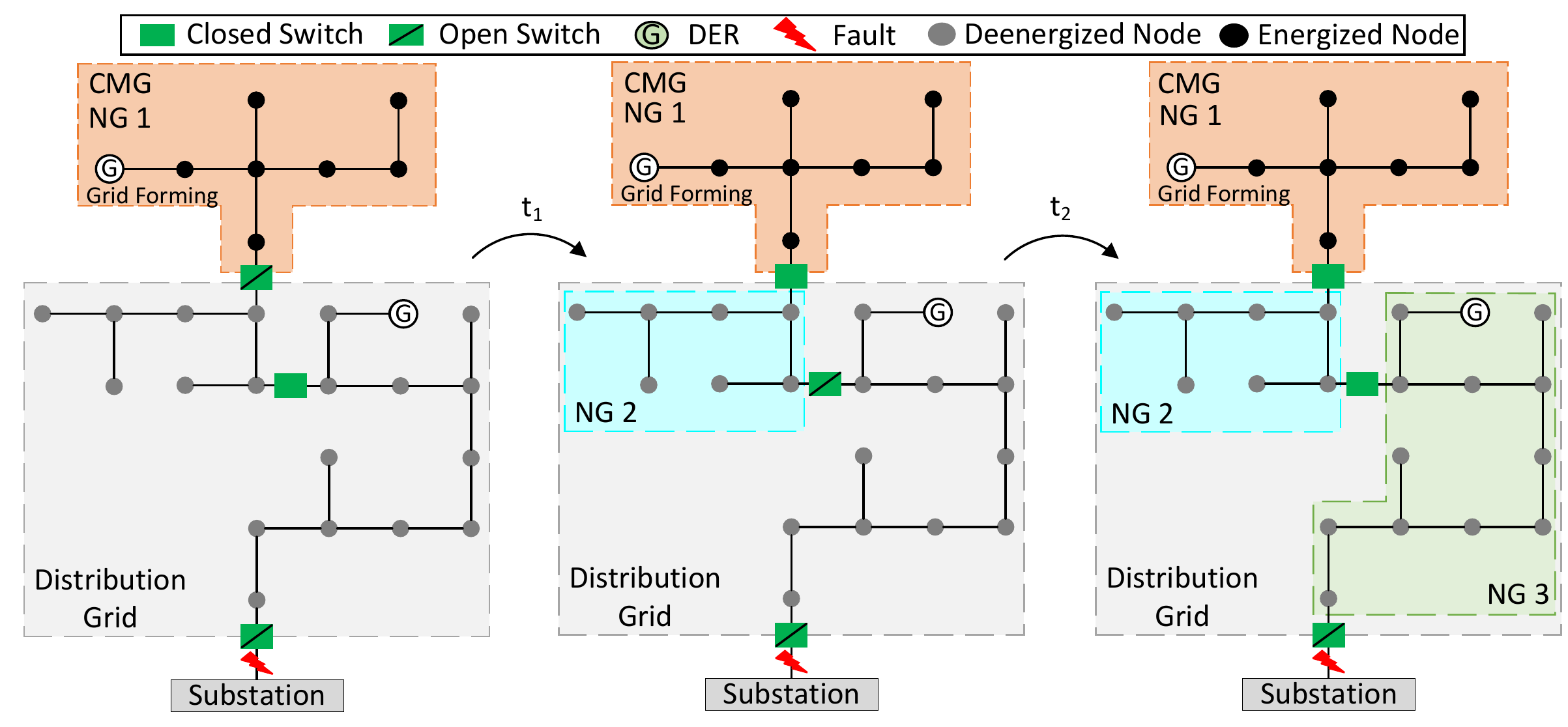}
  \caption{CMG boundary concept (the CMG extends its reach and provides grid support to NG 2 and subsequently to NG 3).}
  \label{fig:dynamic_cmg}
  \vspace{-0.5cm}
\end{figure}
\begin{equation}
\begin{aligned}
\vspace{-0.2cm}
%\min_{X^{\text{DG}}, X^{\text{ES}}, X^{\text{PV}}}
\max_{P_{}^{\text {D}}} &\quad\sum\limits_{t \in \mathcal{T}} \sum\limits_{s \in \Omega} {{\pi _s}} {{\sum\limits_{n \in \mathcal{N}^{\text NG}, i \in \mathcal{N}_n} {{\omega _i^{\text{1}}}P_{i,t,s}^\text{D}}}}  \label{eq:da_opt_obj}.
\vspace{-0.5cm}
\end{aligned}
\end{equation}
The load priority weight, $\omega_i^{\text{1}}$, is assigned the highest value for the CLs in the entire network, followed by the NCLs of the CMG. The NCLs external to the CMG take the lowest priority. The constraints are incorporated $\forall$ $n \in \mathcal{N}^{\text NG}$, $t \in \mathcal{T}$, and $s \in \Omega$, unless explicitly stated. Equations (\ref{eq:da_opt_p})--(\ref{eq:da_opt_q}) list the power balance constraints. This stage does not account for the detailed OPF constraints as a measure for reducing the computational complexity, thus allowing enough computational room for handling the stochastic decision making. Adding power flow constraints in the EDS stage, which uses a receding-horizon approach for the entire projected outage duration, will overcomplicate this stage without adding significant benefits \cite{eds_ignorepowerflow}. The decisions concerning power flow for future time intervals, excluding the current time interval, will be discarded due to the model predictive control approach followed by the EDS stage, resulting in redundant decisions. Hence, a single-phase equivalent model of the system is considered by aggregating the values of variables and parameters for different phases at every node: %Owing to this reason, the EDS stage only focuses on contingencies pertaining to generator failure: 
\begin{subequations}
\begin{gather}
 \sum\limits_{n \in \mathcal{N}^{\text {NG}}}( \sum \limits_{ i \in \mathcal{N}^{\text {PV}}_n} {P_{i,t,s}^{\text {PV}}} + \sum \limits_{ i \in \mathcal{N}^{\text {ES}}_n} {P_{i,t,s}^{\text {ES}}} + \sum \limits_{ i \in \mathcal{N}^{\text {DG}}_n} {P_{i,t}^{\text {DG}}})   = \nonumber \\ \sum\limits_{n \in \mathcal{N}^{\text {NG}}, i \in \mathcal{N}_n} {P_{i,t,s}^{\text {D}}},  \label{eq:da_opt_p}\\
 \sum\limits_{n \in \mathcal{N}^{\text {NG}}}( \sum \limits_{ i \in \mathcal{N}^{\text {PV}}_n} {Q_{i,t,s}^{\text {PV}}} + \sum \limits_{ i \in \mathcal{N}^{\text {ES}}_n} {Q_{i,t,s}^{\text {ES}}} + \sum \limits_{ i \in \mathcal{N}^{\text {DG}}_n} {Q_{i,t,s}^{\text {DG}}})   = \nonumber \\ \sum\limits_{n \in \mathcal{N}^{\text {NG}}, i \in \mathcal{N}_n} {Q_{i,t,s}^{\text {D}}}.  \label{eq:da_opt_q}
%
%P_{i,t,s}^{\text G} = P_{i,t,s}^{\text {PV}} + P_{i,t,s}^{\text {DG}} + P_{i,t}^{\text {DG}} \quad \forall i \in \mathcal{N}^{\text {PV/ES/DG}}_n  \label{eq:da_opt_pg}\\
%
%Q_{i,t,s}^{\text G} = Q_{i,t,s}^{\text {PV}} + Q_{i,t,s}^{\text {DG}} + Q_{i,t}^{\text {DG}} \quad \forall i \in \mathcal{N}^{\text {PV/ES/DG}}_n \label{eq:da_opt_qg}
%
\end{gather}
\end{subequations}
Constraints (\ref{eq:da_opt_pvds1})--(\ref{eq:da_opt_pvds3}) represent the real, reactive, and apparent power limits of the controllable PV generators $\forall$ $i \in \mathcal{N}^{\text {PV-C}}_n$, and (\ref{eq:da_opt_pvds4})--(\ref{eq:da_opt_pvds5}) represent the real and reactive power limits of the uncontrollable residential BTM PV generators $\forall$ $i \in \mathcal{N}^{\text {PV-UC}}_n$. $\overline P_{i,t,s}^\text{PV}$ and $\overline Q_{i,t,s}^\text{PV}$ take the EDS forecast value. $\theta_{n,t}$ is a binary variable indicating the connectivity status of NG $n$: 
\begin{subequations}
\begin{gather}
0 \le P_{i,t,s}^{\text {PV}} \le \theta_{n,t}\overline P _{i,t,s}^{\text {PV}},  \label{eq:da_opt_pvds1}\\
%0 \le P_{i,t,s}^{\text {PV}}, Q_{i,t,s}^{\text {PV}} \le \theta_{n,t}\overline P _{i,t,s}^{\text {PV,EDS}}, \theta_{n,t}\overline Q _{i,t,s}^{\text {PV,EDS}}  \label{eq:da_opt_pvds1}\\
% 
0 \le Q_{i,t,s}^{\text {PV}} \le \theta_{n,t}\overline Q _{i,t,s}^{\text {PV}},  \label{eq:da_opt_pvds2}\\
[{(P_{i,t,s}^{\text {PV}})^2} + {(Q_{i,t,s}^{\text {PV}})^2}] \le {(\overline S _i^{\text {PV}})^2}. \label{eq:da_opt_pvds3} \\
P_{i,t,s}^{\text {PV}} = \theta_{n,t}\overline P _{i,t,s}^{\text {PV}},  \label{eq:da_opt_pvds4}\\
Q_{i,t,s}^{\text {PV}} = 0.  \label{eq:da_opt_pvds5}
\end{gather}
\end{subequations}
Equations (\ref{eq:da_opt_esds1})--(\ref{eq:da_opt_esds5}) represent the ES real power, reactive power, apparent power, and inter-temporal SOC change constraints for all $i \in \mathcal{N}^{\text {ES}}_n$. $\gamma$ is the reserve factor that ensures the ES is not operating at its limits to allow reserve availability. $P_{i,t,s}^{\text {ES}} > 0$ indicates battery discharge, and vice versa:
\begin{subequations}
\begin{gather}
 - \theta_{n,t}\overline P _i^{\text { ES}} \le \gamma P_{i,t,s}^{\text {ES}} \le \theta_{n,t} \overline P _i^{\text {ES}},  \label{eq:da_opt_esds1}\\
 0 \le \gamma Q_{i,t,s}^{\text {ES}} \le \theta_{n,t}\overline Q _{i}^{\text {ES}} , \label{eq:da_opt_esds2}\\
  [{(P_{i,t,s}^{\text {ES}})^2} + {(Q_{i,t,s}^{\text {ES}})^2}] \le {(\overline S _i^{\text {ES}})^2} , \label{eq:da_opt_esds3}\\
 SO{C_{i,t,s}^{\text {ES}}} = SO{C_{i,t - 1,s}^{\text {ES}}} - \frac{{P_{i,t,s}^{\text {ES}}}}{{\overline{E}_i^{\text {ES}}}}\Delta t\ , \label{eq:da_opt_esds4}\\
 \underline {SOC} _i^{\text {ES}} \le SOC_{i,t,s}^{\text {ES}} \le \overline {SOC} _i^{\text {ES}}.   \label{eq:da_opt_esds5}
\end{gather}
\end{subequations}
Equations (\ref{eq:da_opt_dgds1})--(\ref{eq:da_opt_dgds6}) represent the DG real power limit, reactive power limit, apparent power limit, ramp up/down limits, inter-temporal fuel consumption, and fuel limit constraints for all $i \in \mathcal{N}^{\text {DG}}_n$. The DGs corresponding to the NGs that are connected to the CMG are required to stay on and operate above a predetermined minimum value at all times. $\gamma$ ensures that the DG is not operating at its limits to allow space for reserves. $\alpha_i$ and $\beta_i$ are the the fuel consumption coefficients \cite{dg_fuel_eq}:
\begin{subequations}
\begin{gather}
\gamma\theta_{n,t}\underline P _i^{\text {DG}} \le P_{i,t}^{\text {DG}} \le \gamma^{-1} \theta_{n,t} \overline P _i^{\text {DG}}, \label{eq:da_opt_dgds1}\\ 
\gamma\theta_{n,t} \underline Q _i^{\text {DG}} \le  Q_{i,t}^{\text {DG}} \le \gamma^{-1} \theta_{n,t}\overline Q _i^{\text {DG}}, \label{eq:da_opt_dgds2} \\
[{(P_{i,t}^{\text {DG}})^2} + {(Q_{i,t}^{\text {DG}})^2}] \le {(\overline S _i^{\text {DG}})^2},  \label{eq:da_opt_dgds3}\\
-\overline{P}_i^{\text {DG,RR}} \le P_{i,t}^{\text {DG}} - P_{i,t - 1}^{\text {DG}} \le \overline{P}_i^{\text {DG,RR}} , \label{eq:da_opt_dgds4}\\
F_{i,t}^{\text {DG}} = F_{i,t - 1}^{\text {DG}} - ({\alpha_i}P_{i,t}^{\text {DG}}+ {\beta_i}\theta_{n,t}\overline P _i^{\text {DG}})\Delta t .  \label{eq:da_opt_dgds5} \\
\underline{F}_{i}^{\text {DG}} \le F_{i,t}^{\text {DG}} \le \overline{F}_{i}^{\text {DG}} .  \label{eq:da_opt_dgds6}
%
%\underline F _i^{\text { DG,EDS}} \le F_{i,t}^{\text {DG}} \le \overline F _i^{\text { DG,EDS}}  \label{eq:da_opt_dgds6}
\end{gather}
\end{subequations}
Refer to Appendix \ref{appendix:Hex} for more details on the polygonal relaxation approach for constraints (\ref{eq:da_opt_pvds3}), (\ref{eq:da_opt_esds3}), (\ref{eq:da_opt_dgds3}). Equations (\ref{eq:da_opt_ld6})--(\ref{eq:da_opt_ld8}) place limits on CL and NCL. $\overline P_{i,t,s}^{\text {D}}$, $\overline Q_{i,t,s}^{\text {D}}$ take the EDS forecast value. $\underline P_{i,t,s}^{\text {D}}$, $\underline Q_{i,t,s}^{\text {D}}$ are set to $0$ $\forall \ i \in \mathcal{N}^{\text {NCL}}_n$ and to a minimum must-supply value $\forall \ i \in \mathcal{N}^{\text {CL}}_n$, which is a fixed percentage of the load forecast value:
\begin{subequations}
\begin{gather}
%%
%0 \le P_{i,t,s}^{\text {D}} \le \theta_{n,t}\overline P _{i,t,s}^{\text {D}}  \label{eq:da_opt_ld1}\\
%
%0 \le Q_{i,t,s}^{\text {D}} \le \theta_{n,t}\overline Q _{i,t,s}^{\text {D}}  \label{eq:da_opt_ld2}\\
%
%P_{i,t,s}^{\text {D}} = P_{i,t,s}^{\text {D,C}} + P_{i,t,s}^{\text {D,UC}}  \label{eq:da_opt_ld3}\\
%
%Q_{i,t,s}^{\text {D}} = Q_{i,t,s}^{\text {D,C}} + Q_{i,t,s}^{\text {D,UC}}  \label{eq:da_opt_ld4}\\
%
%0 \le P_{i,t,s}^{\text {D,C}} \le \theta_{n,t}\overline P _{i,t,s}^{\text {D,C}}  \label{eq:da_opt_ld5}\\
%
\theta_{n,t}\underline P_{i,t,s}^{\text {D}} \le P_{i,t,s}^{\text {D}} \le \theta_{n,t} \overline P _{i,t,s}^{\text {D}},   \label{eq:da_opt_ld6}\\
%
%0 \le Q_{i,t,s}^{\text {D,C}} \le \theta_{n,t}{\mu}\overline Q _{i,t,s}^{\text {D,C}}  \label{eq:da_opt_ld7}\\
%
\theta_{n,t}\underline Q _{i,t,s}^{\text {D}} \le Q_{i,t,s}^{\text {D}} \le \theta_{n,t}\overline Q _{i,t,s}^{\text {D}}.   \label{eq:da_opt_ld8}
\end{gather}
\end{subequations}
Constraint (\ref{eq:da_opt_ng1}) establishes a MSD of $\upsilon$ consecutive hours for which an NG $n \in \mathcal{N}^{NG} \backslash \{1\}$ must remain connected and (\ref{eq:da_opt_ng2}) ensures that the NG $1$, belonging to the CMG, always stays connected. Equation (\ref{eq:da_opt_ng3}) adds a constraint to maintain the NG connectivity sequence with the CMG $\forall \ n, n' \in \mathcal{N}^{NG} \backslash \{1\}$. This ensures that a NG $n'$ having a direct connectivity to CMG via NG $n$ can only be connected to the CMG if the NG $n$ is connected:   
\begin{subequations}
\begin{gather}
\sum\limits_{t' = t}^{t + (\upsilon  - 1)} {{\theta _{n,t'}}}  \ge \upsilon ({\theta _{n,t}} - {\theta _{n,t - 1}}), \label{eq:da_opt_ng1} \\
\theta _{1,t} = 1, \label{eq:da_opt_ng2} \\
\theta _{n,t} - \theta _{n',t} \ge 0. \label{eq:da_opt_ng3}
\end{gather}
\end{subequations}
Lastly, to determine the connectivity status of the external NGs, we propose a chance-constrained (CC) formulation. The underlying idea of these constraints is that a CMG may expand its support to an NG only if the probability of satisfying a fixed value of NG load exceeds a predetermined probability threshold:    
\begin{subequations}
\begin{gather}
\text{Pr}(\sum\limits_{n \in \mathcal{N}^{\text{NG}}\backslash \{1\}, i \in \mathcal{N}_n} P_{i,t}^{\text {D}} \ge \eta_n \sum\limits_{n \in \mathcal{N}^{\text{NG}}\backslash \{1\}, i \in \mathcal{N}_n} \overline{P}_{i,t}^{\text {D}}) = \phi_{n,t},   \label{eq:chanceconstrained_ngload1} \\
-\phi_{n,t} \le -(1-\epsilon_n) + 1 - \theta_{n,t}. \label{eq:probabilityswitchrelation}
\end{gather}
\end{subequations}
Constraint (\ref{eq:chanceconstrained_ngload1}) computes the probability, $\phi_{n,t} \in [0,1]$, of supplying at least $\eta_n\%$ of the total forecasted load for all time intervals. Constraint (\ref{eq:probabilityswitchrelation}) ensures that the NG $n$ will be disconnected from the CMG if $\phi_{n,t}$ is less than the threshold of $1-\epsilon_n$ where $\epsilon_n \in [0,1]$. Evaluating the CC (\ref{eq:chanceconstrained_ngload1}) is difficult due to the lack of knowledge on the probability density function of the random variable. Hence, approximations are needed to ensure such constraints can be smoothly incorporated. A scenario-based approximation, as shown in Appendix \ref{appendix:cc_approx}, is used to approximate constraint (\ref{eq:chanceconstrained_ngload1}). Further, it is to be noted that each NG is unique and has different characteristics. Hence, in order to avoid scenarios wherein the NG with CLs has to be disconnected over an NG without CLs due to the constraint (\ref{eq:probabilityswitchrelation}), we ensure that the values of $\eta_n$ and $\epsilon_n$ are much stricter for NGs without CLs and slightly relaxed for NGs with CLs.     

\subsection{Stage 2: Near-Real-Time Schedule Update}
 This OPF-based problem is solved on an hourly basis by taking decisions for sub-hourly intervals after an outage. The output is the updated plan for generation resources and decisions for DR. This stage is solved closer to the start of the hourly RT dispatch to ensure that network reconfiguration and DR decisions can be successfully communicated to the field devices on time. Unlike \cite{da_mg_sch_op2}, this stage does not consider a receding-horizon based decision making for the following reasons: 
 \begin{enumerate}
    \item \textbf{Computational complexity}: Owing to a mixed-integer formulation with numerous binary variables, a receding-horizon-based approach will add significantly to the computational complexity. 
     \item \textbf{Computation time}: The NRT stage requires computing and deploying the decisions closer to RT, making it necessary to obtain the problem solution within small time duration. 
     \item \textbf{Scalability}: Mixed-integer optimization problems do not scale well for systems with a large number of decision variables. 
 \end{enumerate}
 In order to overcome the limitations of not having a receding-horizon-based control approach, some key NRT stage decisions are coupled with the EDS decisions. This coupling with the EDS stage helps add the impact of stochastic and receding-horizon decisions to the deterministic NRT stage. The coupling with the EDS stage is achieved by adding constraints that ensure that the dynamic CMG boundary decision is directly inherited from the EDS stage. In addition, the objective function is modified to minimize the deviations between the EDS decisions and the updated NRT decisions. This form of the objective function allows room for modifications to the decisions without encountering any infeasibilities, instead of hard coupling constraints. The objective function (\ref{eq:obj_nrt}) is modeled to maximize the squared weighted load supplied, minimize the network load phase imbalance, and minimize the squared deviation between the following: DG power output and the reference EDS value; ES SOC and the EDS reference SOC value:
 
\begin{subequations}
\small
\begin{gather}
\label{eq:obj_nrt}
{\max_{{\textbf{\tiny P}}, {\text{\tiny SOC}}}}\sum\limits_{h \in \mathcal{H}_t}[ \sum\limits_{i \in \mathcal{N}} [{\omega_i^{\text{1}}\omega_{i,t}^{\text{2}}} \langle {\bf{1}},  {\bf{P}}_{i,h}^{\text {D}} \rangle]^2 - {{\boldsymbol{\delta }}_{h}^{\text {D}}}^2 - \sum\limits_{i \in \mathcal{N}^{\text{DG}}}[\hat{P}_{i,t}^{\text {DG}} - \langle {\bf{1}},{\bf{P}}_{i,h}^{\text {DG}}\rangle]^2]  \nonumber\\
- \sum\limits_{i \in \mathcal{N}^{\text{ES}}}[( \mathbb{E} [\hat {{SOC}}_{i,t,s}^{\text {ES}}]- {SOC}_{i,\vert \mathcal{H}_t\vert}^{\text {ES}}) E_i^{\text {ES}}/\Delta h ]^2. \tag{9} 
\end{gather}
\end{subequations}
Note that $h \in \mathcal{H}_t$ is a sub-hourly interval within the $t^{\text{th}}$ hour. For every hour, the updated NG connectivity status ($\theta_{n,t}$) is taken from the EDS stage solution and the CMG node set is appended with the nodes of the connected NGs, thus forming the larger node set $\mathcal{N}$. For any given hour $t$, the constraints are incorporated for all $i \in \mathcal{N}$, $(i,j) \in \mathcal{E}$, and $h \in \mathcal{H}_t$, unless explicitly stated. The parameters and variables in boldface indicate $3 \times 1$ dimensional vectors to represent values for the three network phases. Constraints (\ref{eq:nrt_opt_pf1})--(\ref{eq:nrt_opt_pf2}) represent the nodal power balance equations using the linearized branch flow model for unbalanced system \cite{lindistflow_extension}. Constraints (\ref{eq:nrt_opt_pf5})--(\ref{eq:nrt_opt_pf7}) impose limits on the maximum power flowing over a network line and unidirectionality of power flow over a line using the binary variable vector ${\boldsymbol{\rho}}_{ij,h}$. Constraints (\ref{eq:nrt_opt_pf8})--(\ref{eq:nrt_opt_pf10}) compute the node voltages and limit the nodal voltages within acceptable bounds. ${{\boldsymbol{\zeta }}_{ij,h}}$ is a slack variable added to avoid conflict with (\ref{eq:nrt_opt_pf9}) when any two adjoining nodes are disconnected for a given direction of power flow:
\begin{subequations}
\begin{gather}
 {\bf{P}}_{i,h}^{\text {PV}} + {\bf{P}}_{i,h}^{\text {ES}} + {\bf{P}}_{i,h}^{\text {DG}}-{\bf{P}}_{i,h}^{\text {D}}{=}\sum\limits_{j:ij \in \mathcal{E}_{i}}{{{\bf{P}}_{ij,h}}{-}\sum\limits_{i:ji \in \mathcal{E}_{i}}{{{\bf{P}}_{ji,h}}}}, \label{eq:nrt_opt_pf1} \\
{\bf{Q}}_{i,h}^{\text {PV}} + {\bf{Q}}_{i,h}^{\text {ES}} + {\bf{Q}}_{i,h}^{\text {DG}} - {\bf{Q}}_{i,h}^{\text {D}} {=} \sum\limits_{j:ij \in \mathcal{E}_{i}} {{{\bf{Q}}_{ij,h}} {-} \sum\limits_{i:ji \in \mathcal{E}_{i}} {{{\bf{Q}}_{ji,h}}} }, \label{eq:nrt_opt_pf2}\\
0 \le {{\bf{P}}_{ij,h}} \le {{\boldsymbol{\rho}} _{ij,h}}\overline {\bf{P}}, \label{eq:nrt_opt_pf5}\\
0 \le {{\bf{Q}}_{ij,h}} \le {{\boldsymbol{\rho}} _{ij,h}}\overline {\bf{Q}}, \label{eq:nrt_opt_pf6}\\
{{\boldsymbol{\rho}}_{ij,h}} + {{\boldsymbol{\rho}}_{ji,h}} \le 1, \label{eq:nrt_opt_pf7}\\
{{\bf{V}}_{i,h}} \approx {{\bf{V}}_{j,h}} - {{\bf{A}}_{ij}}{{\bf{P}}_{ij,h}} - {{\bf{B}}_{ij}}{{\bf{Q}}_{ij,h}} + {{\boldsymbol{\zeta }}_{ij,h}} ,\label{eq:nrt_opt_pf8} \\
\underline {\bf{V}}^2  \le {{\bf{V}}_{i,h}} \le \overline {\bf{V}}^2 ,\label{eq:nrt_opt_pf9} \\
- (1 - {{\boldsymbol{\rho}} _{ij,h}})\overline {\bf{V}}^2 \; \le {{\boldsymbol{\zeta }}_{ij,h}} \le (1 - {{\boldsymbol{\rho}} _{ij,h}})\overline {\bf{V}}^2,   \label{eq:nrt_opt_pf10}
\end{gather}
\end{subequations}
where
\begin{align}
\nonumber
\small {{\bf{A}}_{ij}} = \left[ {\begin{array}{*{20}{c}}
{ - 2r_{_{ij}}^{\text {aa}}}&{r_{_{ij}}^{\text {ab}} - \sqrt 3 x_{_{ij}}^{\text {ab}}}&{r_{_{ij}}^{\text {ac}} + \sqrt 3 x_{_{ij}}^{\text {ac}}}\\
{r_{_{ij}}^{\text {ba}} + \sqrt 3 x_{_{ij}}^{\text {ba}}}&{ - 2r_{_{ij}}^{\text {bb}}}&{r_{_{ij}}^{\text {bc}} - \sqrt 3 x_{_{ij}}^{\text {bc}}}\\
{r_{_{ij}}^{\text {ca}} - \sqrt 3 x_{_{ij}}^{\text {ca}}}&{r_{_{ij}}^{\text {cb}} + \sqrt 3 x_{_{ij}}^{\text {cb}}}&{ - 2r_{_{ij}}^{\text {cc}}}
\end{array}} \right]
\end{align}
and
\begin{align}
\nonumber
\small {{\bf{B}}_{ij}} = \left[ {\begin{array}{*{20}{c}}
{ - 2x_{_{ij}}^{\text {aa}}}&{x_{_{ij}}^{\text {ab}} + \sqrt 3 r_{_{ij}}^{\text {ab}}}&{x_{_{ij}}^{\text {ac}} - \sqrt 3 r_{_{ij}}^{\text {ac}}}\\
{x_{_{ij}}^{\text {ba}} - \sqrt 3 r_{_{ij}}^{\text {ba}}}&{ - 2x_{_{ij}}^{\text {bb}}}&{x_{_{ij}}^{\text {bc}} + \sqrt 3 r_{_{ij}}^{\text {bc}}}\\
{x_{_{ij}}^{\text {ca}} + \sqrt 3 r_{_{ij}}^{\text {ca}}}&{x_{_{ij}}^{\text {cb}} - \sqrt 3 r_{_{ij}}^{\text {cb}}}&{ - 2x_{_{ij}}^{\text {cc}}}
\end{array}} \right].
\end{align}
Constraints (\ref{eq:da_opt_pvds1})--(\ref{eq:da_opt_pvds5}) for PV, (\ref{eq:da_opt_esds1})--(\ref{eq:da_opt_esds2}) and (\ref{eq:da_opt_esds5})  for ES, and (\ref{eq:da_opt_dgds1})--(\ref{eq:da_opt_dgds2}) and (\ref{eq:da_opt_dgds6}) for DG can be incorporated from the EDS stage by making the following changes: replace $t$, $\mathcal{T}$, and $\Delta t$ by $h$, $\mathcal{H}_{t}$, and $\Delta h$; remove the scenario index $s$; remove NG connectivity indicator $\theta_{n,t}$; replace the single-phase variables and parameters with their equivalent three-phase vectors. Additional constraints (\ref{eq:nrt_opt_esds1})--(\ref{eq:nrt_opt_esds4}), (\ref{eq:nrt_opt_dgds3})--(\ref{eq:nrt_opt_dgds6}), and (\ref{eq:nrt_opt_dgds7}) pertaining to the ES apparent power limit, ES intertemporal SOC change, DG apparent power limit, DG phase imbalance, and fuel consumption, respectively, are added to this stage:
\begin{subequations}
\begin{gather}
%
%\underline {\bf{P}} _{i,h}^{\text {DG,NRT}} \le \gamma {\bf{P}}_{i,h}^{\text {DG}} \le \overline {\bf{P}} _{i,h}^{\text {DG,NRT}}  \label{eq:nrt_opt_dgds1}\\
%
%\underline {\bf{Q}} _{i,h}^{\text {DG,NRT}} \le \gamma {\bf{Q}}_{i,h}^{\text {DG}} \le \overline {\bf{Q}} _{i,h}^{\text {DG,NRT}}  \label{eq:nrt_opt_dgds2}\\
%
%{({\bf{P}}_{i,h}^{\text {DG}})^2} + {({\bf{Q}}_{i,h}^{\text {DG}})^2} \le {(\overline{\bf{S}}_{i,h}^{\text {DG,NRT}})^2}  \label{eq:nrt_opt_dgds3}\\
%
%{\bf{P}}_i^{\text {DG,RD,NRT}} \le {\bf{P}}_{i,h}^{\text {DG}} - {\bf{P}}_{i,h - 1}^{\text {DG}} \le {\bf{P}}_i^{\text {DG,RU,NRT}}   \label{eq:nrt_opt_dgds4}\\
%
%{\bf{P}}_{i,h}^{\text {ES}} - {\bf{\tilde{P}}}_{i,h}^{\text {ES}} = {{\boldsymbol{\delta }}_{i,h}^{\text {ES}}} ,  \label{eq:nrt_opt_esds6}\\
%
%- {\overline {\boldsymbol{\delta }} _i^{\text {ES,NRT}}} \le {{\boldsymbol{\delta }}_{i,h}^{\text {ES}}} \le {\overline {\boldsymbol{\delta }} _i^{\text {ES,NRT}}}, \label{eq:nrt_opt_esds7}\\
%
%
[{\langle {\bf{1}},{\bf{P}}_{i,h}^{\text {ES}}\rangle^2} + {\langle {\bf{1}},{\bf{Q}}_{i,h}^{\text {ES}}\rangle^2}] \le {(\overline S _i^{\text {ES}})^2} , \label{eq:nrt_opt_esds1}\\
SO{C_{i,h}^{\text {ES}}} = SO{C_{i,h - 1}^{\text {ES}}} - \frac{\langle {\bf{1}},{\bf{P}}_{i,h}^{\text {ES}}\rangle}{{E_i^{\text {ES}}}}\Delta t\ , \label{eq:nrt_opt_esds4}\\
[{\langle {\bf{1}},{\bf{P}}_{i,h}^{\text {DG}}\rangle^2} + {\langle {\bf{1}},{\bf{Q}}_{i,h}^{\text {FG}}\rangle^2}] \le {(\overline S _i^{\text {DG}})^2} , \label{eq:nrt_opt_dgds3}\\
{\bf{P}}_{i,h}^{\text {DG}} - {\bf{\tilde{P}}}_{i,h}^{\text {DG}} = {{\boldsymbol{\delta }}_{i,h}^{\text {DG}}} ,  \label{eq:nrt_opt_dgds5}\\
- {\overline {\boldsymbol{\delta }} _i^{\text {DG}}} \le {{\boldsymbol{\delta }}_{i,h}^{\text {DG}}} \le {\overline {\boldsymbol{\delta }} _i^{\text {DG}}}  , \label{eq:nrt_opt_dgds6}\\
F_{i,h}^{\text {DG}} = F_{i,h - 1}^{\text {DG}} - ({\alpha_i}\langle {\bf{1}},{\bf{P}}_{i,h}^{\text {DG}}\rangle  + {\beta_i}\overline P _i^{\text {DG}})\Delta h.  \label{eq:nrt_opt_dgds7}
%
%\underline F _{i}^{\text { DG,EDS}} \le F_{i,h}^{\text {DG}} \le \overline F _{i}^{\text { DG,EDS}} \quad \forall i \in {\mathcal{N}^{\text {DG}}}  \label{eq:nrt_opt_dgds8}
\end{gather}
\end{subequations}
If any generator needs to be taken offline, ceases to operate, or is not part of the new CMG boundary, then it can be removed from the set $\mathcal{N}^{\text{PV/ES/DG}}_n$, thus eliminating it from the decision making process. Constraints (\ref{eq:nrt_opt_ld1})--(\ref{eq:nrt_opt_ld2}) compute the node load by incorporating the phase-wise load connectivity status decision variable $\textbf{x}_{i,t}$ and cold load. Equation (\ref{eq:nrt_opt_ld3}) computes the overall network load allocation phase imbalance:
\begin{subequations}
\begin{gather}
{\bf{P}}_{i,h}^{\text {D}} = {\textbf{x}_{i,t}}(\overline {\bf{P}} _{i,h}^{\text {D}} + \overline {\bf{P}} _{i,h}^{\text {D,CLPU}}), \label{eq:nrt_opt_ld1}\\
{\bf{Q}}_{i,h}^{\text {D}} = {\textbf{x}_{i,t}}(\overline {\bf{Q}} _{i,h}^{\text {D}} + \overline {\bf{Q}} _{i,h}^{\text {D,CLPU}}).  \label{eq:nrt_opt_ld2} \\
-{{\boldsymbol{\delta }}_{h}^{\text {D}}} \le {\sum_{i \in \mathcal{N}}}{\bf{P}}_{i,h}^{\text {D}} - {\sum_{i \in \mathcal{N}}}{\bf{\tilde{P}}}_{i,h}^{\text {D}} \le {{\boldsymbol{\delta }}_{h}^{\text {D}}} ,  \label{eq:nrt_opt_ld3}
%
%{\bf{P}}_{i,1}^{\text {D,CLPU}} = (1 - {x_{i}^{\text o}}){\beta _i}{\bf{P}}_{i,1}^{\text {D}}  \label{eq:nrt_opt_ld3}\\
%
%{\bf{Q}}_{i,1}^{\text {D,CLPU}} = (1 - {x_{i}^{\text o}}){\beta _i}{\bf{Q}}_{i,1}^{\text {D}} \label{eq:nrt_opt_ld4}
\end{gather}
\end{subequations}
The above constraints add the functionality of DR to the proposed approach, wherein the load is controlled by remotely connecting/disconnecting it using the smart meter. Due to the phase-wise DR control and the generator phase imbalance constraints, the CMG phase imbalance can be controlled and minimized. Before beginning the NRT stage computation, the cold load value is estimated for all the disconnected load nodes using the historical information on load disconnectivity duration and cold load estimation method described in \cite{clpu_curves}. If the load node is reconnected to the grid, then the computed cold load value is added to the load forecast and will have to be supplied in full. To ease computation burden, we place an assumption that the cold load will decay within a one hour duration. To add MSD to every load node, a new set $\mathcal{N}^{\text {MSD}}$ is introduced, which contains a tuple of nodes and their corresponding phases that must remain connected until the MSD is completed. Hence, the constraint $\textbf{x}_{i,t}(p) = 1 \hspace{0.2cm} \forall {i,p} \in \mathcal{N}^{\text {MSD}}$ is added. The set $\mathcal{N}^{\text {MSD}}$ is computed for every hour by analyzing the load connectivity status of the past hours. The load connectivity status is designed to change only at the start of every hour, thus making the optimization problem, book-keeping of the $\mathcal{N}^{\text {MSD}}$ set, and dispatching DR decisions less complex.  

In the proposed approach, $\textbf{x}_{i,t}$ is assigned to an individual network bus, and all the customers connected to the particular bus will inherit the value of $\textbf{x}_{i,t}$. However, with the increase in the network size, this approach can result in an increased computation burden. To alleviate this issue, multiple nodes can be grouped and assigned to a single binary variable $\textbf{x}_{i,t}$. These load zones can mimic the loads that are aggregated and controlled by a DR aggregator, thus making the implementation of DR decisions practical. 

Lastly, we also incorporate the issue of load interruption equity, which is absent in the existing literature on resilient energy management system design. In the current setting, we have considered CMG and the distribution network to have high BTM PV generators. Having these generators would give an unfair advantage to such load nodes since any optimization-based load curtailment framework targeted at load supply maximization will prioritize keeping such nodes always connected due to the PV generation availability. There can also be some load nodes in the system that are always selected to be supplied due to their location in the system or their load values. This can lead to an unfair advantage to such nodes, resulting in supply duration inequity. To alleviate this issue, we have modified the objective function (\ref{eq:obj_nrt}) with a dynamically changing weight, $\omega_{i,t}^{\text{2}}$, which aims at maintaining the load supply duration equity. The historical load supply decisions are analyzed and then the value of $\omega_{i,t}^{\text{2}}$ is modified such that loads with higher historical supply duration have a relatively lower weight as compared to loads with a low historical supply duration. A rolling window of fixed duration is used to keep a track of the supply duration and accordingly update $\omega_{i,t}^{\text{2}}$. Note that the duration-prioritized weights for NCLs are lower then the CLs to prevent NCLs overshadowing CLs.    

\subsection{Stage 3: Real-Time Dispatch}
This stage is solved using the RT load and PV generation forecasts, and the decisions are dispatched to the system. This stage aims to fine-tune the NRT decisions using the newly obtained forecasts in a manner that does not violate network constraints. The coupled decisions between the RT and NRT stage include the DR decisions, DG setpoints, and controllable PV generator set points. The RT stage must ensure that all the loads selected in the NRT stage are entirely supplied, irrespective of the cold-load value and forecast error. In the RT stage, the load connectivity status and DG setpoints are held fixed for the hour, thus leaving scope for making changes to the grid-following ES setpoints. Hence, the objective function of the RT stage (\ref{eq:obj_rt}) is solely focused on minimizing PV curtailment, which ensures that the excess PV generation is incentivized to charge the ES units:
\begin{equation}
\begin{aligned}
\label{eq:obj_rt}
{\min _{\textbf{P}}}{\sum\limits_{i \in \mathcal{N}^{\text {PV}}}} [ \langle {\bf{1}}, \overline{\bf{P}}_{i,k}^{\text {PV}}-{\bf{P}}_{i,k}^{\text {PV}} \rangle]^2.
\end{aligned}
\end{equation}
The RT interval $k$ inherits the NRT decisions of the corresponding sub-hourly interval $h$ within the hour $t$. The load connectivity decision is obtained as a parameter from the NRT stage solution. For every constraint incorporated from the NRT stage, $h$, $\mathcal{H}_{t}$, and $\textbf{x}_{i,t}$ are replaced by $k$, $\mathcal{K}_{t,h}$, and $\hat{\textbf{x}}_{i,t}$, respectively.  
The OPF constraints (\ref{eq:nrt_opt_pf1})--(\ref{eq:nrt_opt_pf9}), load constraints (\ref{eq:nrt_opt_ld1})--(\ref{eq:nrt_opt_ld2}), and NRT equivalent PV generation constraints (\ref{eq:da_opt_pvds1})--(\ref{eq:da_opt_pvds3}), and ES constraints (\ref{eq:da_opt_esds1})--(\ref{eq:da_opt_esds5}) are replicated from the NRT stage by incorporating the above listed changes. The CMG's primary ES unit acts in the grid-forming mode, while the remaining ES units and DGs operate in grid-following mode. To prevent the output of the DG units from fluctuating intermittently, the DG output will be fixed at the values obtained from NRT stage solution as shown in (\ref{eq:rt_opt_dgds1})--(\ref{eq:rt_opt_dgds2}). The output of the controllable PV generators is also coupled with the NRT stage as shown in (\ref{eq:rt_opt_pvds1})--(\ref{eq:rt_opt_pvds2}), while that of the grid-following ES units are modified to reflect the changes in load and generation forecasts. The power output of the grid-forming ES unit cannot be pre-specified until the RT dispatch problem is solved \cite{droop_op}: 
\begin{subequations}
\begin{gather}
{\bf{P}}_{i,k}^{\text {DG}} = {\bf{\hat P}}_{i,h}^{\text {DG}},    \label{eq:rt_opt_dgds1} \\
{\bf{Q}}_{i,k}^{\text {DG}} = {\bf{\hat Q}}_{i,h}^{\text {DG}},    \label{eq:rt_opt_dgds2}\\
{\bf{P}}_{i,k}^{\text {PV}} \le \text{max}_{h \in \mathcal{H}_t}[{\bf{\hat P}}_{i,h}^{\text {PV}}],    \label{eq:rt_opt_pvds1} \\
{\bf{Q}}_{i,k}^{\text {PV}} \le \text{max}_{h \in \mathcal{H}_t}[{\bf{\hat Q}}_{i,h}^{\text {PV}}].    \label{eq:rt_opt_pvds2}
\end{gather}
\end{subequations} 
The RT model may be infeasible due to load and DG's equality constraints. At such times, these constraints are relaxed using an equivalent inequality constraint with an upper/lower bound. Deploying the relaxed solution would translate into an operating point different from the computed one for the grid-forming ES unit. This situation is significantly mitigated due to the staggered decision-making approach of the proposed HMTS algorithm. Due to consideration of a single time interval and the limited number of binary variables, the RT stage MIQP problem is computationally efficient. It can be solved within a small time duration, ensuring computation and deployment of new decisions within time intervals of $5$ minutes. 

\begin{figure}[htb]
	\centering
	\includegraphics[width = 1\linewidth ,keepaspectratio, trim={0.5cm 13.5cm 5.5cm 1cm},clip]{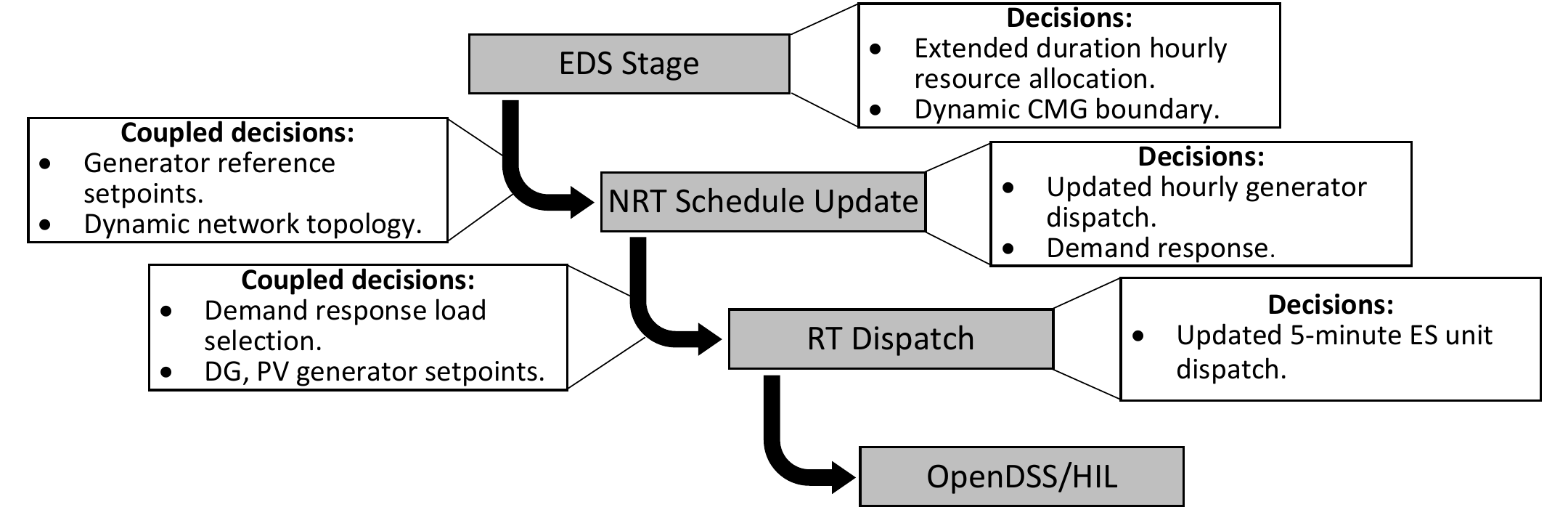}
	\caption{SA-HMTS decisions and inter-stage coupled decisions.}
	\label{fig:HMTS_decisions}
\end{figure}

Fig. \ref{fig:HMTS_decisions} shows a summary of the key decisions taken by each stage and the inter-stage coupled decisions. Segregating the complex decision making constraints across multiple stages along with inter-stage coupling help reduce the computation complexity of each stage along with ensuring decisions of hierarchically upper stage are factored in.

\section{Delayed Recourse}
The proposed HMTS algorithm is designed on the ideology that as the time-interval moves closer to the actual time of dispatch, the forecast error reduces. The reduction in forecast error coupled with staggered decision-making using updated and more accurate forecasts would make secure and proactive decisions. However, this assumption on forecast error improvement may not necessarily hold in some instances. For example, due to limited data availability and knowledge on HILF events, the forecasted data could be significantly biased \cite{biased_forecasterror1, biased_forecasterror2}. Further, due to limited generation resources at hand, there will be rolling blackouts, which will result in additional demand in the form of cold load. It is not possible to accurately estimate the cold load values making it difficult to integrate into the load forecasts. This results in a mismatch between the load forecasts and load realization, further adding to the forecast error. Under such circumstances, the HMTS framework has the following shortcomings:
\begin{itemize}
    \item Biased forecasts for multiple time intervals cause propagation of the forecast error impact with time, resulting in overconsumption or underconsumption of the generation resources with respect to their allocation. 
    \item Failure to address the forecast error impact results in the premature ceased operation of the network or under-utilization of resources.
    %\item Uniform forecast error between different stages for multiple time intervals would result in the propagation of the forecast error impact with time. This would lead to overconsumption or underconsumption of the resources concerning the hourly resource allocation. 
    %\item Failure to address the forecast error impact can result in the ceased operation of the network before the end of the projected outage duration, switching off the CMG to conserve the stored charge in grid-forming ES, or a sub-optimal operation leading to under-consumption of resources.
\end{itemize}
Under normal operating circumstances, the EDS receding horizon-based control will ensure that the decisions being taken address the impact of forecast errors and other unforeseen conditions. However, while operating under emergency conditions, the receding horizon approach may not always guarantee a secure operation with limited resources at hand. Therefore, we first mathematically demonstrate its failure under the circumstances listed above, followed by concept of the delayed recourse approach. Finally, we demonstrate the validity of the delayed recourse approach in eliminating the issues listed above. 

\begin{theorem}
\label{Thm1}
Given a a total electricity demand exceeding the available generation capacity, and a finite outage time horizon, then despite using receding-horizon control, the generation resources will exhaust before the completion of the final time interval under a uniform forecast error of $\gamma \%$, where $\gamma > 0$.
%Given a a total electricity demand exceeding the available generation capacity, and a finite outage time horizon, then despite using receding-horizon control, the generation resources will exhaust before the completion of the final time interval under a uniform forecast error of $\gamma \%$, where $\gamma > 0$.
%Given a limited available generation of ${X}$ kWh, a total demand exceeding $X$ kWh, and a time horizon of $T$ time intervals, despite using receding-horizon control, the generation resources will exhaust before the completion of the final time interval under a uniform forecast error of $\gamma \%$, where $\gamma > 0$.  
\end{theorem}

The proof of Theorem \ref{Thm1} can be found in Appendix \ref{Thm1Proof}. Theorem \ref{Thm1} highlights the limitation of receding horizon control for the applications where the total resources to be allocated are limited in quantity, and the operational constraints get stricter with time. To address this problem, we propose the concept of delayed recourse. Recourse means having the ability to take corrective actions in the face of uncertainty. Due to the lack of knowledge of the forecast errors at the time of RT dispatch, recourse actions cannot be taken in the present to counteract the forecast error impact. However, the past forecast error impact can be analyzed and accordingly be used to take certain corrective actions in the present. We term this concept as delayed recourse.

\subsection{Delayed Recourse Mathematical Formulation}
%check the notes section to make modifications to this section. VVVVV Important 
The idea behind delayed recourse is to compensate for the forecast error impact of the past on the grid-forming generator. Following the nomenclature introduced in Theorem \ref{Thm1}, for a forecast error of $\gamma\%$ during the time interval $t-1$, the consumption mismatch with regards to the EDS allocation of ${\bf{\overline{x}}}^{\text{EDS}}_{t-1}(1)$ is $0.01\gamma{\bf{\overline{x}}}^{\text{EDS}}_{t-1}(1)$. Moving to the present time interval $t$, a delayed recourse will modify the EDS allocation of ${\bf{\overline{x}}}^{\text{EDS}}_{t}(1)$ by subtracting $0.01\gamma{\bf{\overline{x}}}^{\text{EDS}}_{t-1}(1)$ from it during the NRT stage. Theorem \ref{Thm2} below shows the mathematical reasoning that establishes the validity of the delayed recourse approach. The broad assumption here is that the forecast error of $\gamma\%$ is time-invariant, and to narrow it further down, $\gamma$ is strictly positive or negative for all time intervals. However, in real-life scenarios, this assumption may hold weakly. 

\begin{theorem}
\label{Thm2}
Given a limited available generation of ${X}$ kWh, a total demand exceeding $X$ kWh, and a time horizon of $T$ time intervals, using a 1-delayed recourse approach along with receding-horizon control, the generation resources will be appropriately allocated to prevent premature exhaustion of generation resources under a uniform forecast error of $\gamma \%$, where $\gamma > 0$.
%Given a limited available generation of ${X}$ kWh, a total demand exceeding $X$ kWh, and a time horizon of $T$ time intervals, using a 1-delayed recourse approach along with receding-horizon control, the generation resources will be appropriately allocated to prevent premature exhaustion of generation resources under a uniform forecast error of $\gamma \%$, where $\gamma > 0$. %The delayed recourse analyzes the over-consumption of the immediate past time interval. In the present time interval, it takes corrective action by reducing the present allocation by a magnitude equal to the immediate past over-consumption.   
\end{theorem}

The proof of Theorem \ref{Thm2} can be found in Appendix \ref{Thm2Proof}. For time-varying forecast error, depending on the magnitude of $\gamma_t\%$, the above introduced delayed recourse approach may not necessarily eliminate the problem described in Theorem \ref{Thm1}. Moreover, based on the EDS generation allocation and generation mismatch of the past, the exact mismatch compensation cannot be made in the given hour due to load MSD constraints. Hence, the recourse approach needs to be reinforced with some heuristics. %The proposed heuristic approach combines the $n$-delayed recourse approach with $n>1$ along with a time-varying weight $\omega$. 
One such proposed heuristics is to implement the recourse decisions by analyzing the forecast error trend of multiple intervals of the past and extrapolate a value for the present time interval. The $n$ time intervals of the past that are accounted during decision-making for the present determine the reach of the delayed recourse. Hereafter, we will address this as $n$-delayed recourse. Theorem \ref{Thm2} can be easily expanded for the $n$-delayed recourse approach with $n>1$ and for time-varying forecast error.

%The goal of the proposed heuristic approach is to first analyze the trend in the forecast error of the past $n$ time intervals. Then, using the past forecast error trend, extrapolate the error for the current time interval and compensate for it along with compensating for the forecast error of the immediate past interval. 
The above introduced $n$-delayed recourse methodology will help make the framework adaptive for counteracting the time-varying forecast errors. This can be viewed similar to preemptively predicting the forecast error trend for the current time interval and modifying the load to be supplied accordingly. 
To implement this approach, we introduce the concept of linear trend estimation. The historical observed forecast errors are scaled within the range $[-1, 1]$ with respect to a predetermined maximum tolerable forecast error value. Using these scaled values, a time parameterized linear trend-line is obtained as follows:
\begin{align}
    y_t = at + b
\end{align}
where $a$ denotes the slope, $b$ denotes the intercept, and $t \in \{1,2,\cdots,n\}$. %The trend is computed by assuming a $0$ intercept, i.e., $b = 0$. The trend-line would thus rotate around the origin, with its slope ($a$) acting as a measure of the trend. 
A positive slope value would represent increasing forecast error trend and vice versa. A value close to 0 would indicate no forecast error. Once the value of $a$ is computed, it is unscaled back to its actual value denoted by $A$. Then the following constraint, which can be interpreted as a resiliency cut, is added to the NRT stage:

\begin{align}
    \sum_{i \in \mathcal{N}}\langle \bm{1}, {\bf{P}}_{i,h}^{\text{D}} \rangle \le \sum\limits_{i \in \mathcal{N}} \mathbb{E}[\hat{P}_{i,t,s}^{\text{D}}] - \langle \bm{1}, {\bf{P}}_{t-1}^{\text{G,FE}} \rangle - A,  \label{eq:recourse}
\end{align}

where ${\bf{P}}_{t-1}^{\text{G,FE}}$ is obtained by comparing the difference in grid-forming ES SOC computed using NRT forecasts and realized in OpenDSS/HIL simulation. The constraint states that the NRT load allocation must be lower than the difference between expected EDS load allocation and the estimated forecast error impact to be corrected.

\begin{figure}[htb]
  \centering
  \includegraphics[width = 1\linewidth ,keepaspectratio, trim={5cm 7.5cm 6cm 2.25cm},clip]{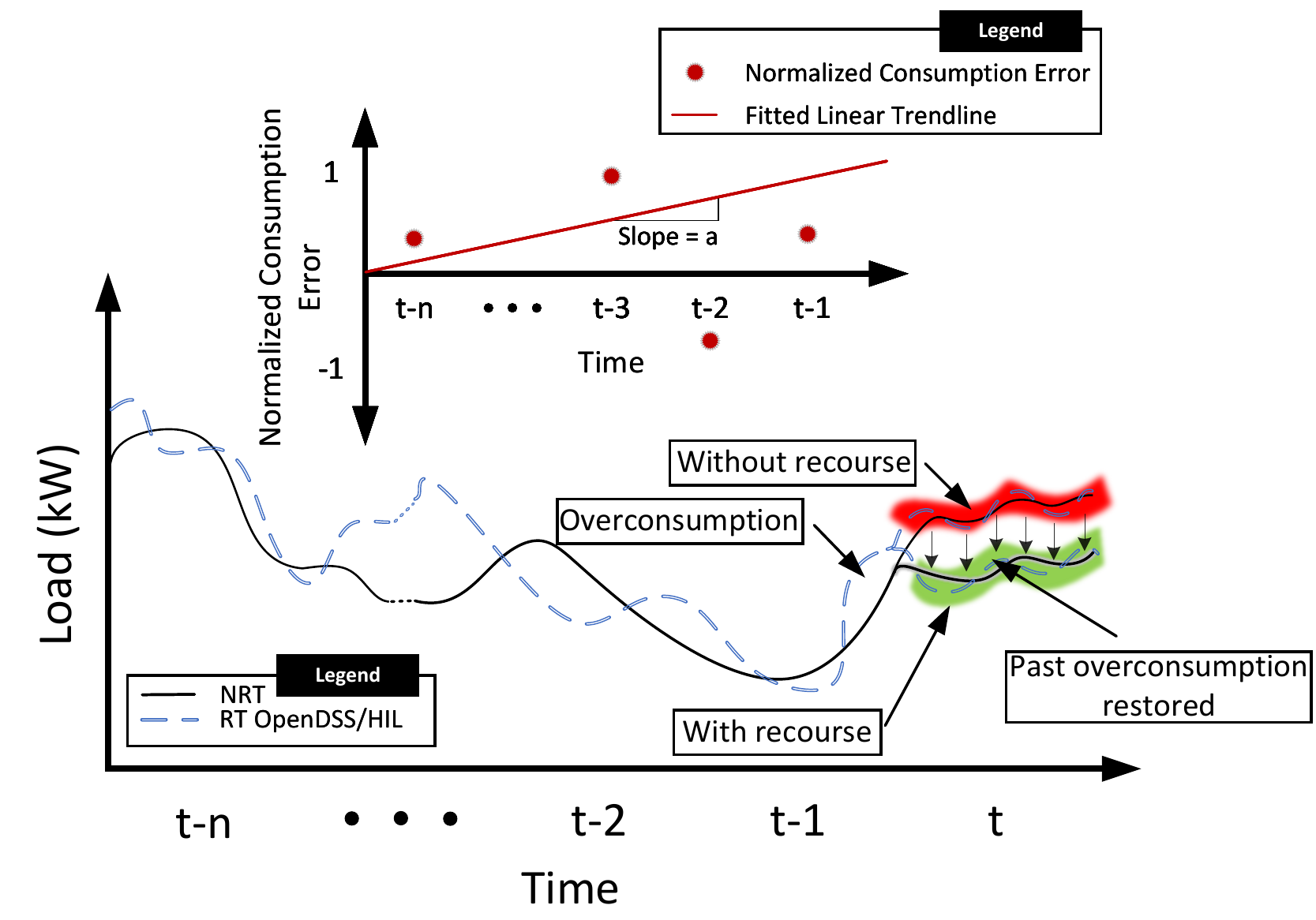}
  \caption{Delayed recourse impact from load perspective.}
  \label{fig:recourse_load}
\end{figure}

\begin{figure}[htb]
  \centering
  \includegraphics[width = 1\linewidth ,keepaspectratio, trim={1.5cm 12.25cm 10.5cm 1cm},clip]{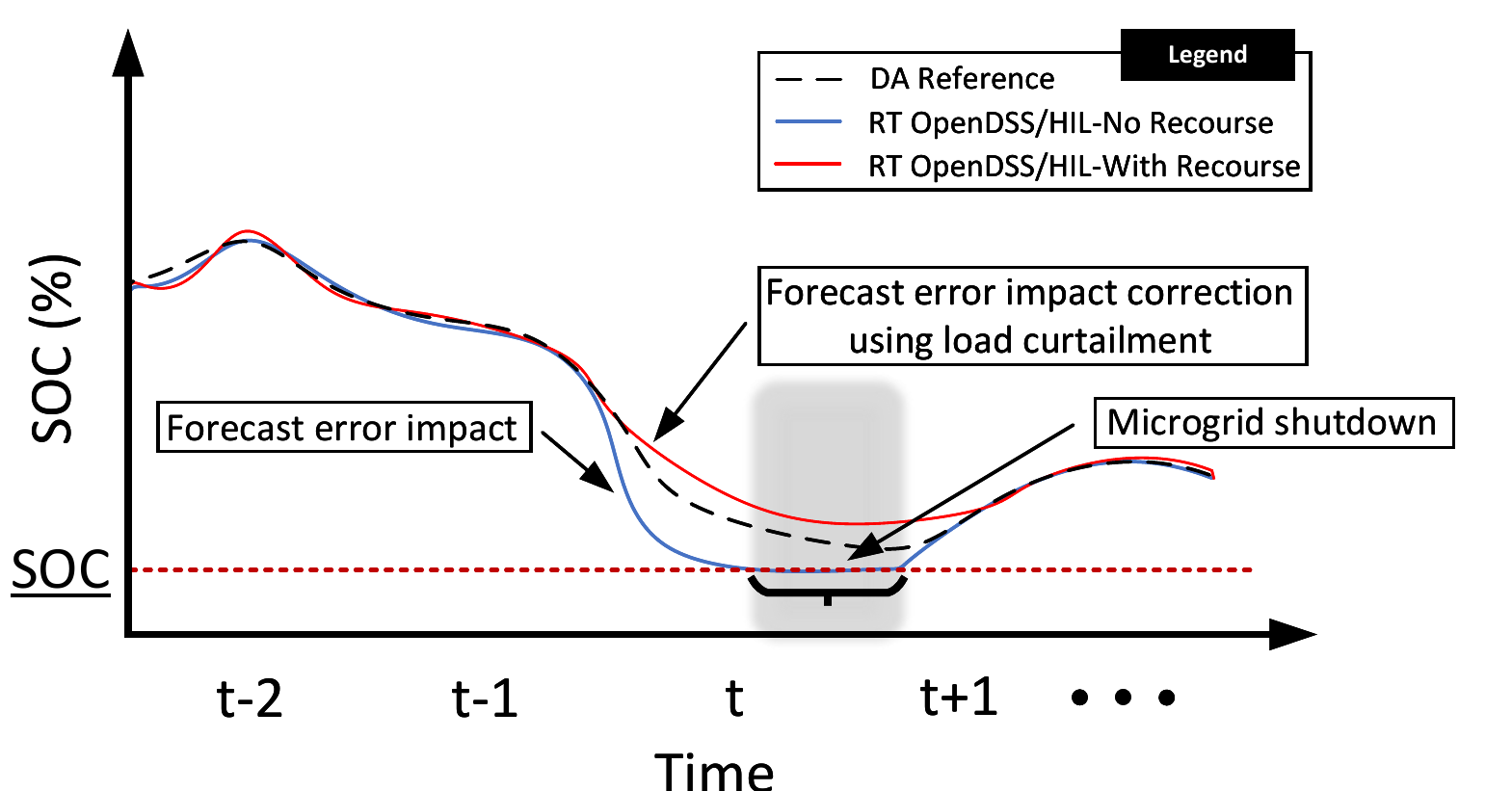}
  \caption{Delayed recourse impact from grid-forming ES SOC perspective.}
  \label{fig:recourse_soc}
\end{figure}

\begin{figure}[htb]
  \centering
  \includegraphics[width = 1\linewidth ,keepaspectratio, trim={1.25cm 3.5cm 11.25cm 1.0cm},clip]{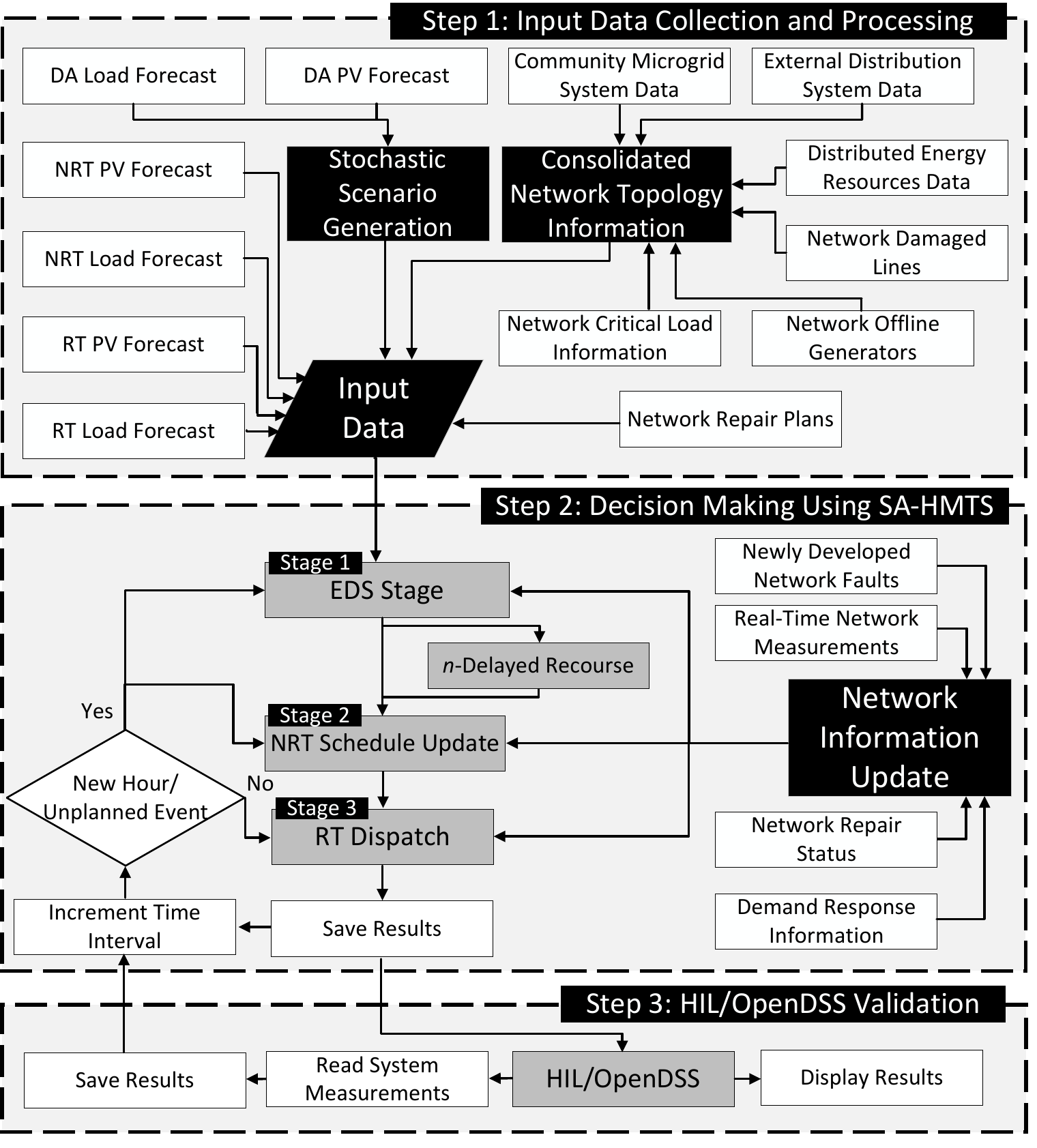}
  \caption{Overall architecture of the proposed SA-HMTS framework.}
  \label{fig:hmts_overview}
\end{figure}

Fig. \ref{fig:recourse_load} and Fig. \ref{fig:recourse_soc} show a pictorial representation of the delayed recourse idea. Implementing constraint (\ref{eq:recourse}) as is can lead to infeasible solution arising due to the MSD constraint explained in section II.B. To avoid this situation, (\ref{eq:recourse}) is converted into a soft constraint with the minimization of the penalty factor added to the objective function. Overall, this approach aims to extend the operation duration of the system, which would have been shortened under the presence of significant forecast errors. This is done by ensuring that the grid-forming ES unit SOC does not drop rapidly but follows the reference values fixed by the EDS stage that ensures extended operation of the ES unit with adequate reserve availability for all time intervals. Doing so will firstly extend the operating duration of the system under forecast errors and make the operation more secure due to the availability of an adequate amount of operating reserve. However, improving the operation security under forecast error will come at the price of compromising the total load supply amount. Furthermore, due to the interplay of numerous factors, this approach cannot guarantee sustained operation for the entire outage duration. However, it will focus on maximizing the continuous operating duration and security of the CMG. A summarized algorithm of the entire SA-HMTS framework is shown in Fig. \ref{fig:hmts_overview}.

\begin{table}[]
\caption{Interpretation of the linear trend slope}
\label{tab:trend_interpretation}
\begin{tabular}{c|c|c}
\hline
Slope value                     & Trend & Interpretation               \\ \hline \hline
$0 <  a \le 1$ & Upward        & Generation over-consumption \\ \hline
$a = 0$  & Steady          & No generation over/under-consumption \\ \hline
$ -1 \le a < 0$  & Downward          & Generation under-consumption  \\ \hline
\end{tabular}
\end{table}
%bayesian. use this terminology  

\section{Cyber Physical Layer - A New Paradigm}
The proposed SA-HMTS framework is designed to provide secure load restoration and reliable supply during extended duration outages caused by HILF events. Under normal operating conditions, the CMG will be connected to the external distribution grid with its generation resources operating in grid-following mode. However, during outages, we propose that the CMG take the role of a grid-forming resource and expand its boundaries to support the external distribution grid loads and generators. The CMG is a private entity that is connected to the distribution grid owned and operated by the utility. The CMG can operate its local resources in an islanded manner without any restrictions. However, the CMG boundary expansion and operation of external resources need to be done in conjunction with the utility. To enable this, we propose a new communication paradigm wherein the CMG will dispatch the energy management decisions routed via the utility distribution management system (DMS). 

\begin{figure}[htb]
  \centering
  \includegraphics[width = 1\linewidth ,keepaspectratio, trim={0.55cm 2cm 0.5cm 4.0cm},clip]{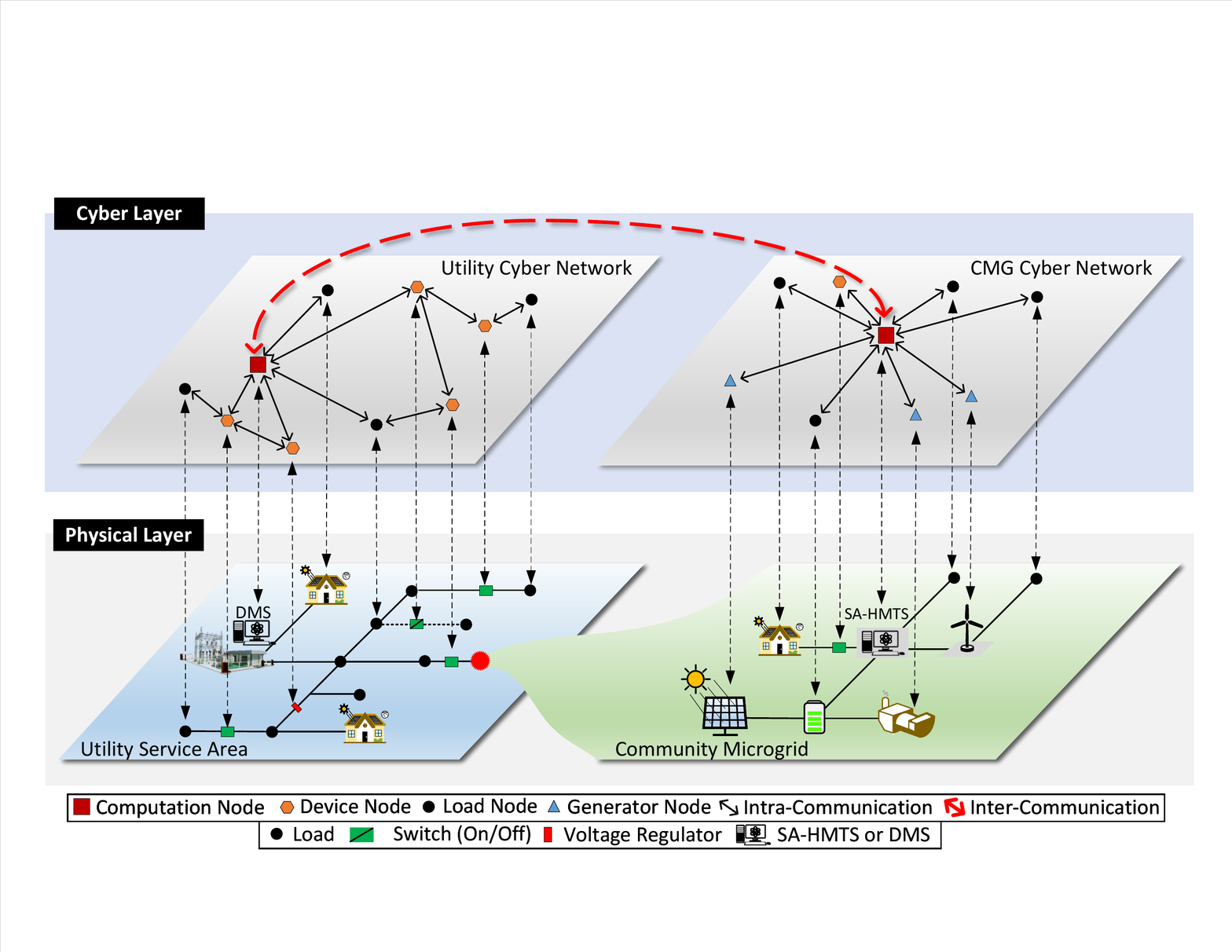}
  \caption{Cyber-physical architecture for enabling SA-HMTS framework.}
  \label{fig:hmts_cps}
\end{figure}

\begin{table}[]
\centering
\caption{Cyber layer setup of SA-HMTS framework.}
\label{tab:cps_links}
\begin{tabular}{>{\centering\arraybackslash}m{0.15\linewidth}|>{\centering\arraybackslash}m{0.45\linewidth}|>{\centering\arraybackslash}m{0.25\linewidth}}
\hline
Entities & Links                    & Channel      \\ \hline \hline
CMG & Communication between SA-HMTS framework and CMG field devices.              & Single-hop cellular and/or ethernet. \\ \hline
CMG, DMS            & Communication between SA-HMTS framework and DMS. & Single/multi-hop cellular.$^*$ \\ \hline
DMS & Communication between DMS and distribution network field devices. & Multi-hop cellular.                  \\ \hline
\end{tabular}
\begin{tablenotes}
\centering
\item[*] ${}^{*}$ = Dependent on distance between CMG and DMS. 
\end{tablenotes}
\end{table}

Fig. \ref{fig:hmts_cps} shows the representation of the proposed cyber-physical architecture. Table \ref{tab:cps_links} summarizes the three main communication links in the proposed cyber-physical architecture. Within the CMG, the proposed SA-HMTS framework acts as the central controller of the CMG communicating with field devices belonging to the CMG over a local area network. The CMG field devices include DGs, ES units, loads, switches, and voltage control devices. Next, the CMG communicates with the DMS over a wide area network (WAN). Lastly, the DMS communicates with its own field devices over a WAN. The SA-HMTS framework decisions that pertain to the local devices of the CMG are directly communicated by the CMG. However, the decisions relating to the CMG boundary expansion, operation of external generators, and DR of external loads are communicated via the DMS since the CMG does not have the authority to interact with external field devices. Also, the DMS will provide the SA-HMTS framework with the necessary information, such as anonymized network topology information and forecasts. Lastly, the DMS communicates the decisions to its own field devices over a WAN. 

The proposed setup of the cyber layer has the following advantages:
\begin{enumerate}
\item Due to the sharing of the anonymized data, the actual distribution network data will stay protected from being exploited by the CMG.
\item Since the CMG decisions about the distribution network assets are routed via the DMS, the DMS can verify the genuineness of the data before dispatching the decisions. This way, the DMS can avoid network damage if the CMG sends its decisions with malicious intent.
\item If the DMS detects any cyber-attacks, it can easily disconnect the distribution grid from the CMG, thus blocking the attack. 
\end{enumerate}
%talk about privacy
%new section for HIL setup
%%%%%%%%%%%%%%%%%%%ExperimentalResults%%%%%%%%%%%%%%%%%%
\section{Case Study and Performance Evaluation}
\begin{figure}[htb]
\vspace{-0cm}
  \centering
  \includegraphics[width = 0.96\linewidth ,keepaspectratio, trim={0cm 0cm 0cm 0cm},clip]{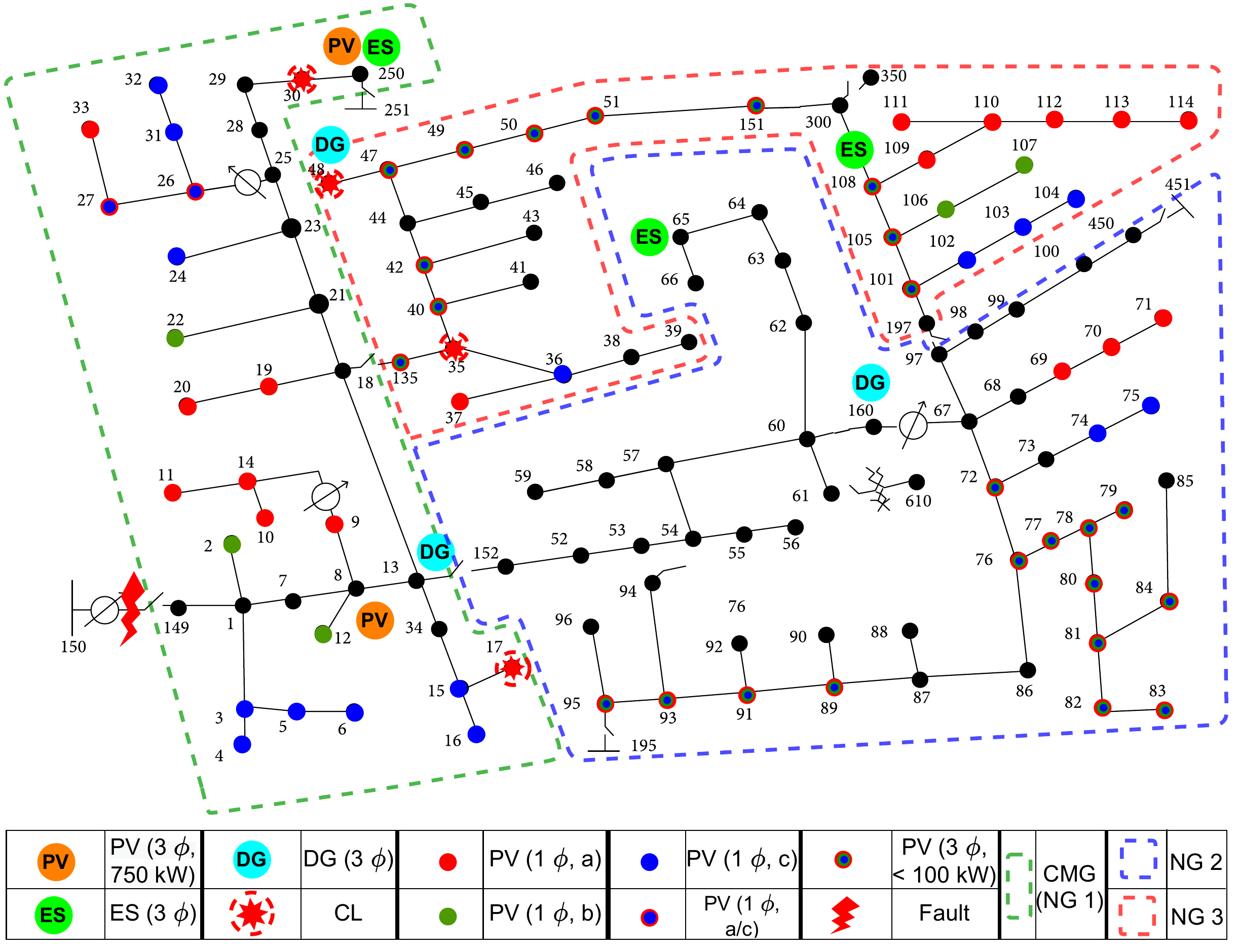}
  \vspace{-0.2cm}
  \caption{Modified IEEE 123 node system.}
  \label{fig:ieee123}
  \vspace{-0cm}
\end{figure}

\begin{figure}[htb]
\vspace{-0.0cm}
  \centering
  \includegraphics[width = 0.96\linewidth ,keepaspectratio, trim={0cm 0cm 0cm 0cm},clip]{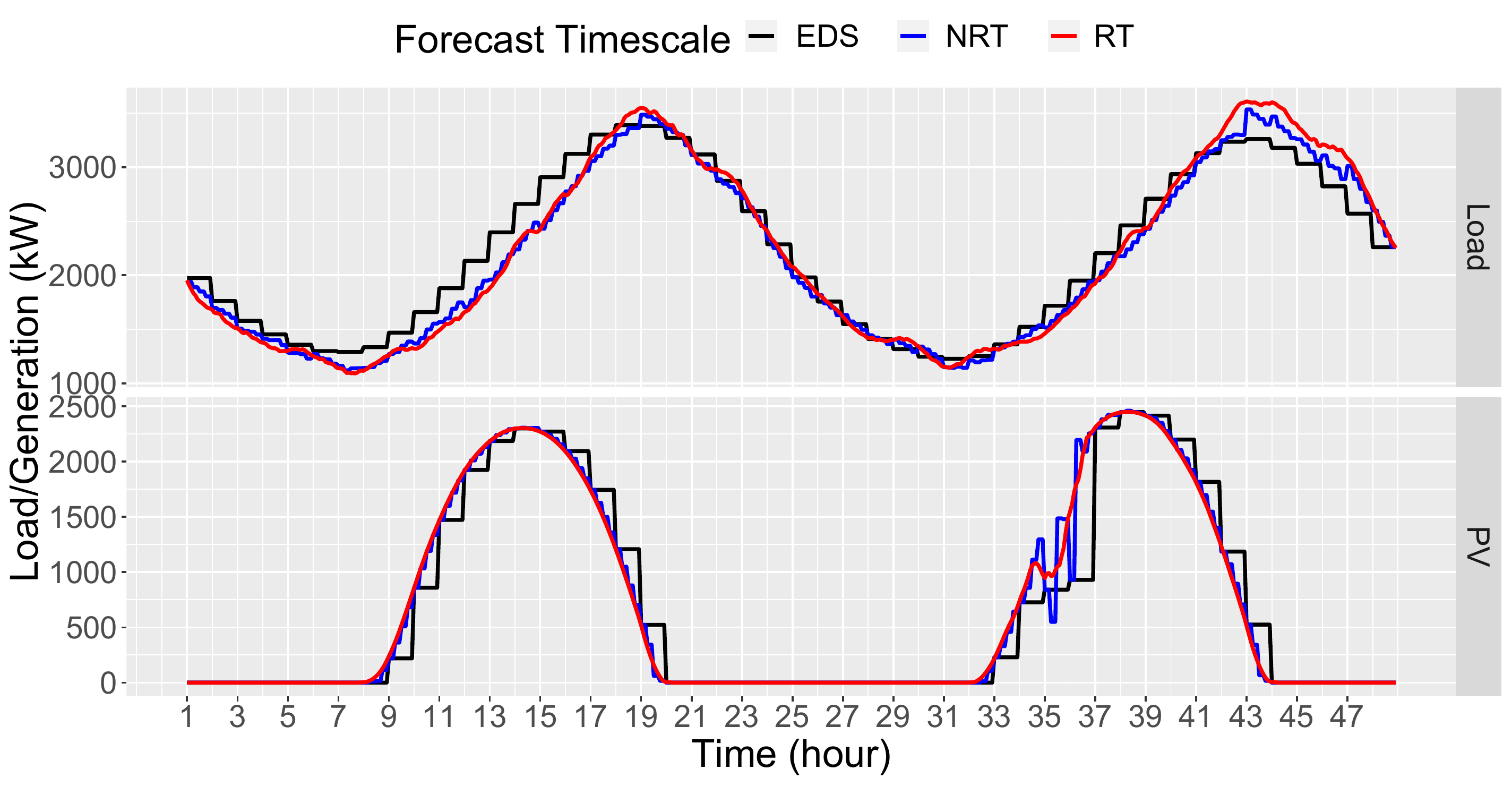}
  \vspace{-0.2cm}
  \caption{Forecast profiles for base case: (a) Total load and (b) PV generation.}
  \label{fig:profiles}
  \vspace{-0cm}
\end{figure}

In this section, the performance of the proposed SA-HMTS framework is evaluated on test distribution systems using OpenDSS/HIL co-simulations under different operating conditions. The simulations have been performed on a modified IEEE $123$ bus system, as shown in Fig. \ref{fig:ieee123}. Additional details on this test system can be found in \cite{ieee123}. The NGs are formed based on pre-existing switches in the system. The distributed generation portfolio details and the simulation parameters are listed in Table \ref{tab:dg_ratings} and Table \ref{tab:parameters}, respectively. The outage is assumed to occur at midnight and persists for $48$ hours. Fig. \ref{fig:profiles} shows the base case forecasted load and PV generation profiles for two summer days. The forecasts are obtained using the forecaster described in \cite{forecaster} which is trained on load and PV data obtained from Pecan Street dataset \cite{pecan} and a utility in North Carolina, USA, respectively. $20$ Monte-Carlo sampled scenarios are obtained for the EDS stage by sampling the forecast error distribution obtained by comparing the forecasts generated by the forecaster and the real data. The values for $\Delta t$, $\Delta h$, and $\Delta k$ are $1$ hour, $15$ minutes, and $5$ minutes, respectively. %The load priority weight for CL and NCL is fixed at $2$ and $1$, respectively.

\begin{figure} [tb]
\centering
  \includegraphics[width=1\linewidth,keepaspectratio, trim={4.25cm 9.5cm 10cm 4.0cm},clip]{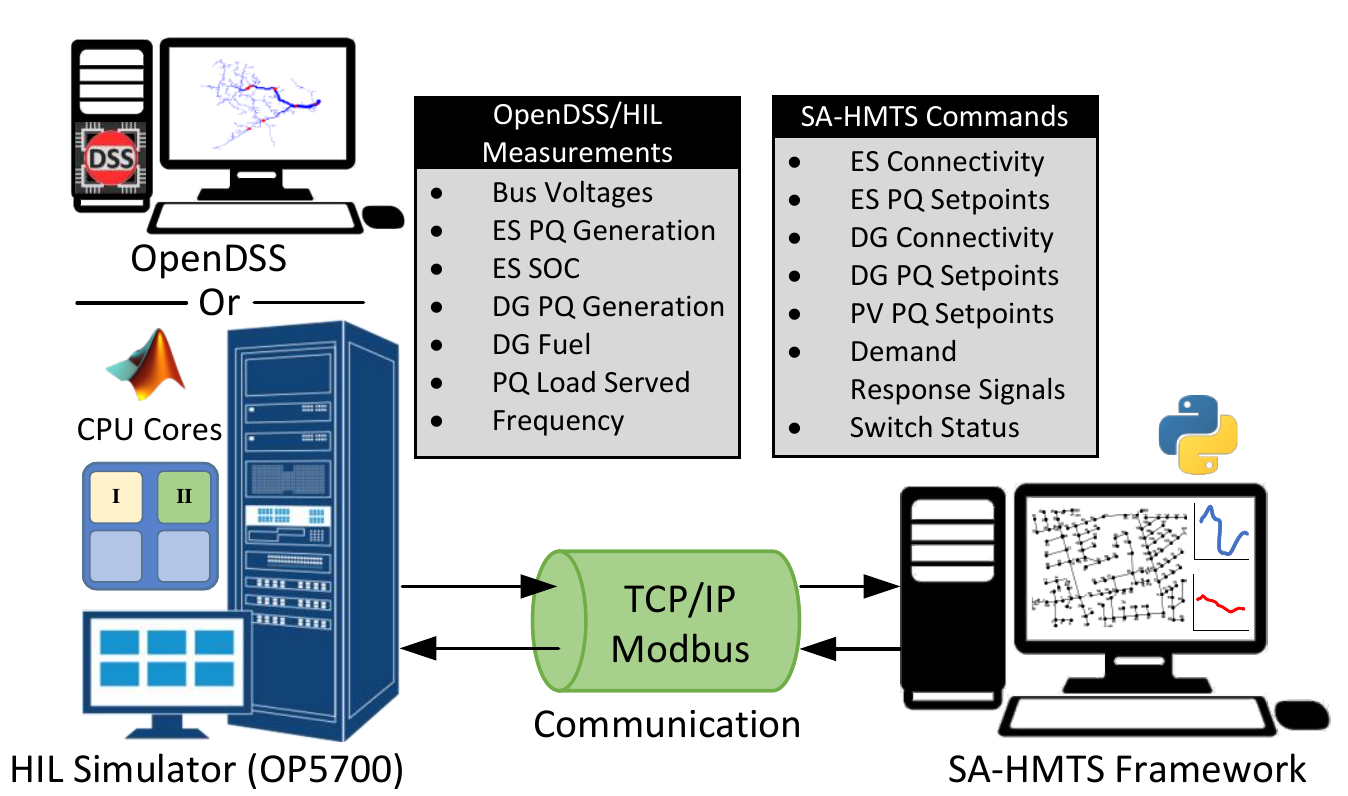} \\
  \caption{SA-HMTS and HIL simulation interface.}
  \label{fig:hil_communication} 
  \vspace{-0.5cm}
\end{figure}

The load priority weights are assigned in descending order: CL within CMG, CL external to CMG, NCL within CMG, and NCL external to CMG. The MSD for each NG and load node is $2$ hours. All ES units are assumed to be $75\%$ charged initially, and the SOC operational limits are $5\%$ and $95\%$. For scheduling purposes, the ES SOC limits are fixed at $20\%$ and $80\%$. To ensure secure operation, a $20$\% reserve availability of the grid-forming ES for all time intervals is desired and the CMG is turned-off once the SOC drops below $20\%$ and then restarted by increasing the SOC above $20\%$ using the support of the co-located PV plant. The NRT stage then takes the start-up decisions for the CMG when it deems the SOC can be raised above $20\%$ using the newly obtained NRT forecasts. A 1.04 p.u. voltage reference is set for the grid-forming ES unit. All simulations are conducted on a PC with Intel Core i9-9900K CPU @ 3.6 GHz processor and 64 GB RAM. The MILP/MIQP optimization problems were implemented and solved using Python interfaced IBM ILOG CPLEX 12.4 solver. The OpenDSS simulations are carried out using Matlab and interfaced with the SA-HMTS framework using the Matlab engine API for Python. The HIL system is modeled in RT-Lab and interfaced with the SA-HMTS framework using Modbus TCP/IP protocol. The setup of the CMG-EMS and the OpenDSS/HIL network simulator is shown in Fig. \ref{fig:hil_communication} and Table \ref{tab:data_interaction}. The analysis performed below serves the purpose of demonstrating the efficacy of the proposed approach, performance robustness against unforeseen conditions observed during emergency operations, and the generalizability of the proposed framework to different operating conditions with little to no changes. 

\begin{table}[htbp]
\vspace{-0.35cm}
\centering
\caption{Distributed generation portfolio}
\vspace{-0.25cm}
\label{tab:dg_ratings}
\scalebox{0.92}{
\begin{tabular}{c|c|c|c}
\hline  
Generator & Generator node  & Rating (kW/kWh)  & Initial Fuel/SOC             \\ \hline \hline
DG${}^{*}$       & \begin{tabular}[c]{@{}c@{}}13 \\ 48 \\ 160 \end{tabular}     & \begin{tabular}[c]{@{}c@{}}900 kW \\ 450 kW \\ 900 kW \end{tabular} & \begin{tabular}[c]{@{}c@{}}12000 liter \\ 6000 liter \\ 6000 liter \end{tabular} \\ \hline
PV${}^{*}$       & 7, 250$^{**}$          & 750 kW, 750 kW & -        \\ \hline
ES${}^{*}$ & \begin{tabular}[c]{@{}c@{}}65 \\ 108 \\ 250 \end{tabular} & \begin{tabular}[c]{@{}c@{}}500 kW/1000 kWh \\ 500 kW/1000 kWh \\ 2750 kW/5500 kWh \end{tabular} & \begin{tabular}[c]{@{}c@{}}75\% \\ 75\% \\ 75\% \end{tabular} \\ \hline
BTM PV${}^{\#}$    & See Fig. \ref{fig:ieee123} & 3 to 15 kW & -              \\ \hline
\end{tabular}}
\begin{tablenotes}
\centering
\item[*] ${}^{*}$ = Three phase, ${}^{\#}$ = Single phase, ${}^{**}$ = Grid-forming    
\end{tablenotes}
\end{table}

\begin{table}[]
\caption{Simulation parameters.}
\label{tab:parameters}
\begin{tabular}{c|c||c|c}
\hline
Parameter            & Value  & Parameter       & Value \\ \hline \hline
$\vert \mathcal{T} \vert$, $\vert \mathcal{H} \vert$, $\vert \mathcal{K} \vert$ & 48, 4, 12              & $\alpha_i, \beta_i$ & $0.244$, $0.014$ \\ \hline
$\Delta t$, $\Delta h$, $\Delta k$                                              & 1 hr., 15 min., 5 min. & $\gamma$            & 1.2              \\ \hline
$\vert \Omega \vert$ & 20     & $\eta_n$ CL     & 50\%  \\ \hline
$\pi_s$              & 0.05   & $\eta_n$ NCL    & 75\%  \\ \hline
MSD $\nu$            & 2 hrs. & $\epsilon_n$ CL & 20\%  \\ \hline
$\{\overline{SOC}, \underline{SOC}\}_{i}^{\text{ES}}$                           & 20\%, 80\%             & $\epsilon_n$ NCL    & 5\%              \\ \hline
\end{tabular}
\end{table}

\begin{table}[]
\caption{Communication details between SA-HMTS framework and HIL/OpenDSS.}
\label{tab:data_interaction}
\begin{tabular}{>{\centering\arraybackslash}m{0.25\linewidth}|>{\centering\arraybackslash}m{0.25\linewidth}|>{\centering\arraybackslash}m{0.1\linewidth} |>{\centering\arraybackslash}m{0.2\linewidth}}
\hline
Communication Direction                      & Information             & Unit     & Update Frequency \\ \hline \hline
\multirow{5}{*}{\parbox{2cm}{ \centering SA-HMTS to HIL/OpenDSS}} & ES/DG/PV connectivity   & 1/0      & 1 hour           \\ \cline{2-4} 
                                             & DG/PV PQ setpoints      & kW, kVAr & 1 hour           \\ \cline{2-4} 
                                             & ES PQ setpoints         & kW, kVAr & 5 minute         \\ \cline{2-4} 
                                             & Demand response signals & 1/0      & 1 hour           \\ \cline{2-4} 
                                             & Network switch status   & 1/0      & 1 hour           \\ \hline
\multirow{8}{*}{\parbox{2cm}{ \centering HIL/OpenDSS to SA-HMTS}} & Simulation time         & seconds   & 5 minutes        \\ \cline{2-4} 
                                             & Node voltages           & p.u.     & 5 minutes        \\ \cline{2-4} 
                                             & ES PQ generation        & kW, kVAr  & 5 minutes        \\ \cline{2-4} 
                                             & ES SOC value            & \%        & 5 minutes        \\ \cline{2-4} 
                                             & DG PQ generation        & kW, kVAr  & 5 minutes        \\ \cline{2-4} 
                                             & DG fuel                 & Liters   & 5 minutes        \\ \cline{2-4} 
                                             & PQ load served          & kW, kVAr  & 5 minutes        \\ \cline{2-4}
                                             & PCC frequency           & Hz       & 5 minutes        \\ \hline
\end{tabular}
\end{table}

%%s

The metrics that will be used to analyze and interpret the results of all the case studies are as follows. $P^{\text{NCL}}$ denotes the percent NCL real power supplied, $P^{\text{CL}}$ denotes the percent CL real power supplied, $T^{\text{ASD, NCL}}$ is the average NCL node service duration, $T^{\text{ASD, CL}}$ is the average CL node service duration, $T^{\text{AID, NCL}}$ is the average NCL node interruption duration, $T^{\text{ASD, CL}}$ is the average CL node service interruption duration, $P^{\text{PV, Total}}$ is the percent PV utilization for the purpose of network load restoration, $F^{\text{DG, Balance}}$ is the total unused fuel at the end of outage duration, $SOC^{\text{ES, Balance}}$ is the balance unused SOC of the grid-forming ES unit at the end of the outage duration, and $P^{\text{Imb.}}_k$ is the maximum network load phase imbalance across all time intervals. Lastly, $T^{\text{RL,ES}}$ is the percentage of total time intervals for which the grid-forming ES SOC violates the desired reserve limit of $20\%$ and $T^{\text{CMG,OFF}}$ is the total hours for which the CMG is turned off due to grid-forming ES SOC falling below $20\%$. The definitions of some of the metrics listed above are listed in Table \ref{tab:metrics}. Equivalent power supply metrics for reactive power are not shown since they follow a similar trend to the real power supply metrics. Case study-specific metrics are introduced within the respective case study description.  

\begin{table}[]
\footnotesize
\caption{Metrics used for analysis}
\label{tab:metrics}
\begin{tabular}{c|c}
\hline
 Metric                               & Formula            
\\ \hline \hline
 $P^{\text{NCL}}$ (\%) & $\sum\limits_{t \in \mathcal{T}, h\in\mathcal{H}_t, k \in \mathcal{K}_{t,h}, n \in \mathcal{N}^{\text{NG}}, i \in \mathcal{N}_n^{\text{NCL}}} \frac{\langle \bm{1}, {\bf{P}}_{i,k}^{\text{D}}\rangle}{\langle \bm{1}, \overline{{\bf{P}}}_{i,k}^{\text{D}}\rangle} * 100$             \\ \hline

 $P^{\text{CL}}$ (\%) & $\sum\limits_{t \in \mathcal{T}, h\in\mathcal{H}_t, k \in \mathcal{K}_{t,h}, n \in \mathcal{N}^{\text{NG}}, i \in \mathcal{N}_n^{\text{CL}}} \frac{\langle \bm{1}, {\bf{P}}_{i,k}^{\text{D}}\rangle}{\langle \bm{1}, \overline{{\bf{P}}}_{i,k}^{\text{D}}\rangle} * 100$ \\ \hline

 $T^{\text{ASD, NCL}}$ (hours)  & $\frac{\sum\limits_{t \in \mathcal{T}, n \in \mathcal{N}^{\text{NG}}, i \in \mathcal{N}_n^{\text{NCL}}} \langle \bm{1}, {\bf{x}}_{i,t} \rangle}{3\sum\limits_{i \in \mathcal{N}} \vert \mathcal{T}\vert}$                                                                                                       \\ \hline
 $T^{\text{ASD, CL}}$ (hours) & $\frac{\sum\limits_{t \in \mathcal{T}, n \in \mathcal{N}^{\text{NG}}, i \in \mathcal{N}_n^{\text{CL}}} \langle \bm{1}, {\bf{x}}_{i,t} \rangle}{3\sum\limits_{i \in \mathcal{N}^{\text{CL}}} \vert \mathcal{T}\vert}$                                                                               \\ \hline
 $T^{\text{AID, NCL}}$ (hours)  & $\frac{\sum\limits_{t \in \mathcal{T}, n \in \mathcal{N}^{\text{NG}}, i \in \mathcal{N}_n^{\text{NCL}}} \langle \bm{1}, {1-\bf{x}}_{i,t} \rangle}{3\sum\limits_{i \in \mathcal{N}} \vert \mathcal{T}\vert}$                                                                                                     \\ \hline
 $T^{\text{AID, CL}}$ (hours) & $\frac{\sum\limits_{t \in \mathcal{T}, n \in \mathcal{N}^{\text{NG}}, i \in \mathcal{N}_n^{\text{CL}}} \langle \bm{1}, {1-\bf{x}}_{i,t} \rangle}{3\sum\limits_{i \in \mathcal{N}^{\text{CL}}} \vert \mathcal{T}\vert}$                                                                             \\ \hline
 $P^{\text{PV, Total}}$ (\%) & $\sum\limits_{t \in T, h\in\mathcal{H}_t, k \in \mathcal{K}_{t,h}, i \in \mathcal{N}^{\text{PV}}} \frac{\langle \bm{1}, {\bf{P}}^{\text{PV}}_{i,k} \rangle}{\langle \bm{1}, \overline{{\bf{P}}}^{\text{PV}}_{i,k} \rangle} * 100$                          \\ \hline
 
 $F^{\text{DG, Balance}}$ (\%) & $\frac{\sum\limits_{t = |\mathcal{T}|, h=|\mathcal{H}_t|, k = \mathcal{K}_{t,h}, i \in \mathcal{N}^{\text{DG}}} {{F}}^{\text{DG}}_{i,k}} { \sum\limits_{i \in \mathcal{N}^{\text{DG}}} {{\overline{F}}^{\text{DG}}_{i}}}*100$                                                                                                                         \\ \hline
 %$SOC^{\text{ES, Balance}}$ (\%) & $\sum\limits_{t = |\mathcal{T}|, h=|\mathcal{H}_t|, k = \mathcal{K}_{t,h}, i \in \mathcal{N}^{\text{ES}}} {{SOC}}^{\text{ES}}_{i,k} * 100$                                                                                                                 \\ \hline
 $P^{\text{Imb.}}$ (\%)  & $\max_k (\frac{\vert \sum\limits_{i \in \mathcal{N}}{\bf{P}}^{\text{D}}_{i,k} - \overline{\overline{\sum\limits_{i \in \mathcal{N}}{\bf{P}}^{\text{D}}_{i,k}}} \vert}{\overline{\overline{\sum\limits_{i \in \mathcal{N}}{\bf{P}}^{\text{D}}_{i,k}}}} * 100)$                          \\ \hline

$T^{\text{RL,ES}}$ (\%) & $\frac{\sum\limits_{t \in T, h\in\mathcal{H}_t, k \in \mathcal{K}_{t,h}} \bm{1}_{SOC^{\text{ES}}_{i,k} \le 25, SOC^{\text{ES}}_{i,k} \ge 75}}{\sum\limits_{t \in T, h\in\mathcal{H}_t, k \in \mathcal{K}_{t,h}} \bm{1}}*100$ \\ \hline

%$T^{\text{CMG,OFF}}$ (hr.) & $ \sum\limits_{t \in T} \bm{1}_{SOC^{\text{ES}}_{i,k} \le 20}$ \\ \hline
\end{tabular}
\end{table}

\subsection{SA-HMTS Framework Performance Analysis}
We first begin by analyzing the performance of the SA-HMTS framework under an outage duration of $2$ days. Herein, the following assumptions are laid down: all the DER units in the system are operational; the communication networks are working as expected; all the network switches are operational; the upstream transmission network support is not available. This case will be the base case for comparing the results. The forecast profiles are shown in Fig. \ref{fig:profiles}, and the actual realization follows a similar pattern with a mean absolute percent error (MAPE) of around $5\%$. Herein, the RT validation is performed and demonstrated using OpenDSS. For the delayed recourse approach, we fix the value of $n$ at $10$, i.e., $10$ immediate hourly time intervals of the past are selected for computing the recourse decisions.    

\begin{figure}[htb]
\vspace{-0.0cm}
  \centering
  \includegraphics[width = 0.75\linewidth ,keepaspectratio, trim={2cm 1cm 0.5cm 0cm},clip]{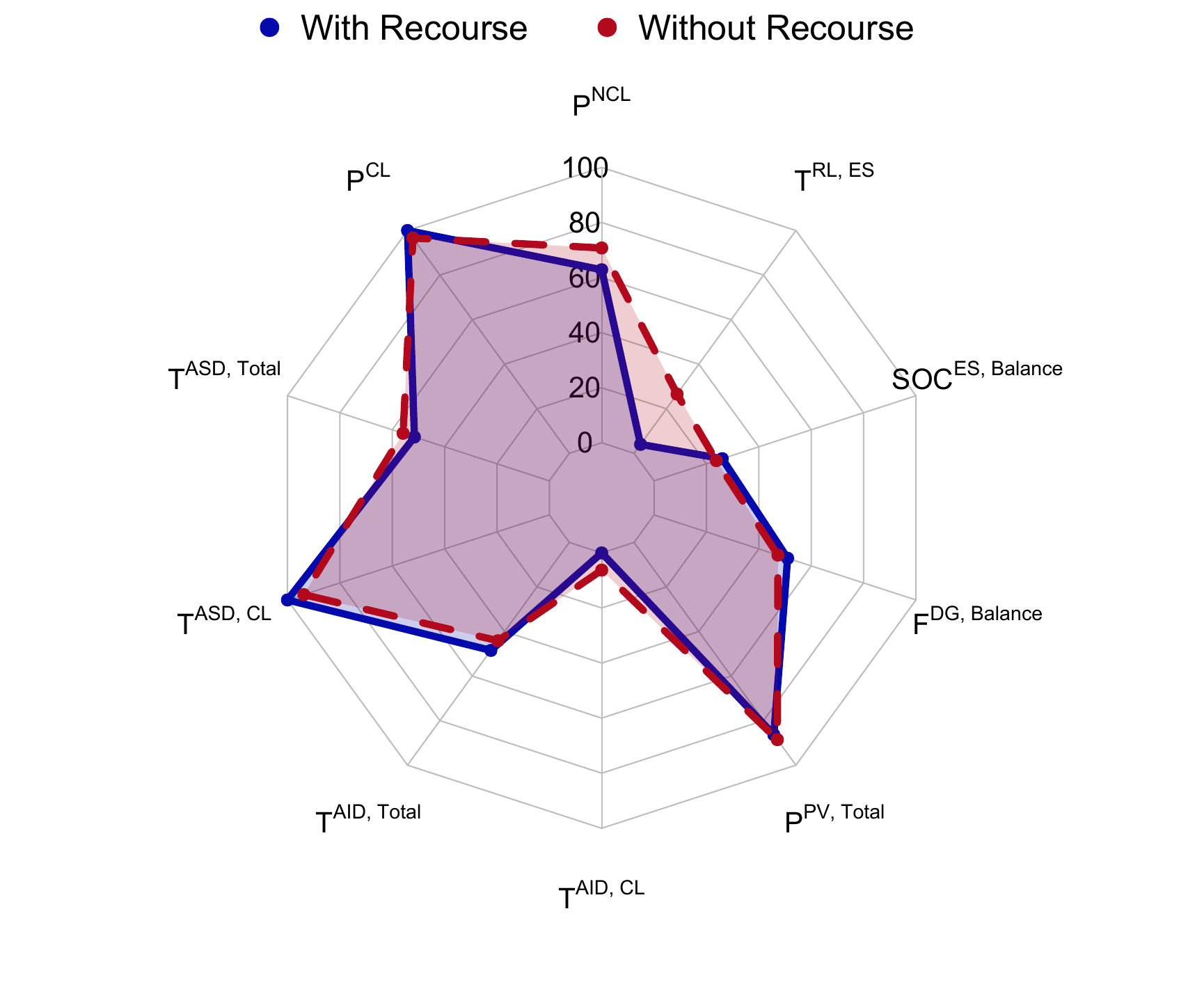}
  \vspace{-0.2cm}
  \caption{Visualizing metrics for base case with and without delayed recourse.}
  \label{fig:metrics_basecaseanalysis}
  \vspace{-0cm}
\end{figure}

\begin{figure}[htb]
\vspace{-0.0cm}
  \centering
  \includegraphics[width = 0.96\linewidth ,keepaspectratio, trim={0cm 0cm 0cm 0cm},clip]{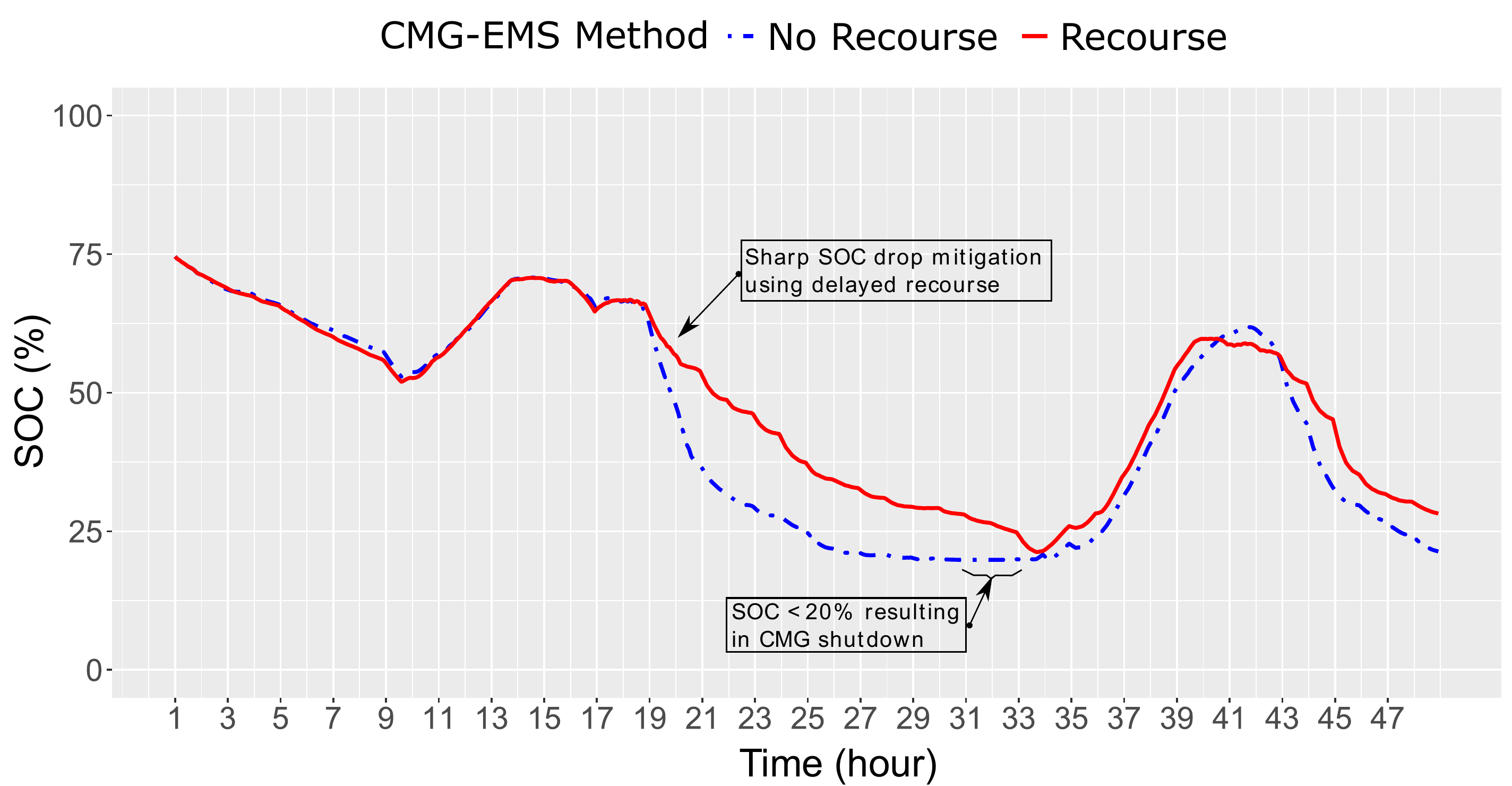}
  \vspace{-0cm}
  \caption{Grid forming ES250 SOC with and without delayed recourse.}
  \label{fig:basecase_recoursenorecourse_soc}
  \vspace{-0cm}
\end{figure}

This study aims to compare the SA-HMTS framework with and without the integration of the delayed recourse mechanism. This helps numerically demonstrate the benefits of the recourse mechanism and demonstrate the sensitivity of the forecast error on the self-learned delayed recourse constraint parameter. Table \ref{tab:metrics_basecaseanalysis} shows the metrics for this base case and a pictorial visualization is shown in Fig. \ref{fig:metrics_basecaseanalysis}. Under both the cases, the CMG expanded its support to both the NGs for the entire duration. $62.83\%$ of the total NCL was supplied using recourse actions, which is $7.91\%$ less than the load supplied without using the recourse mechanism. However, with regards to CL, the former approach supplied $100\%$ CL compared to $96.62\%$ by the latter. In the latter approach, the CL was curtailed because of the CMG shutdown caused by the grid-forming ES SOC falling below $20\%$. Another key observation is that by using the delayed recourse approach, the value of $T^{\text{RL,ES}}$ is reduced by $22.44\%$. This is a great advantage of using the delayed recourse approach and is evident from Fig. \ref{fig:basecase_recoursenorecourse_soc}. Using delayed recourse constraints, the total load supplied was impacted by a small amount at the cost of ensuring a secure operation and complete supply of the CL. With delayed recourse, the value of $T^{\text{CMG,OFF}}$ was reduced from 3 hours to 0 hours. The other metrics used for comparison do not deviate significantly. Additionally, the value of the slope of the forecast error trend line ($a$) for all the time intervals took an average of $0.085$ with a standard deviation of $0.09$. This indicates that the generation utilization for most hours was slightly less than the EDS allocated generation, depending on the overconsumption caused by the forecast errors. 

\begin{figure}[htb]
\vspace{-0.0cm}
  \centering
  \includegraphics[width = 0.96\linewidth ,keepaspectratio, trim={0cm 0cm 0cm 0cm},clip]{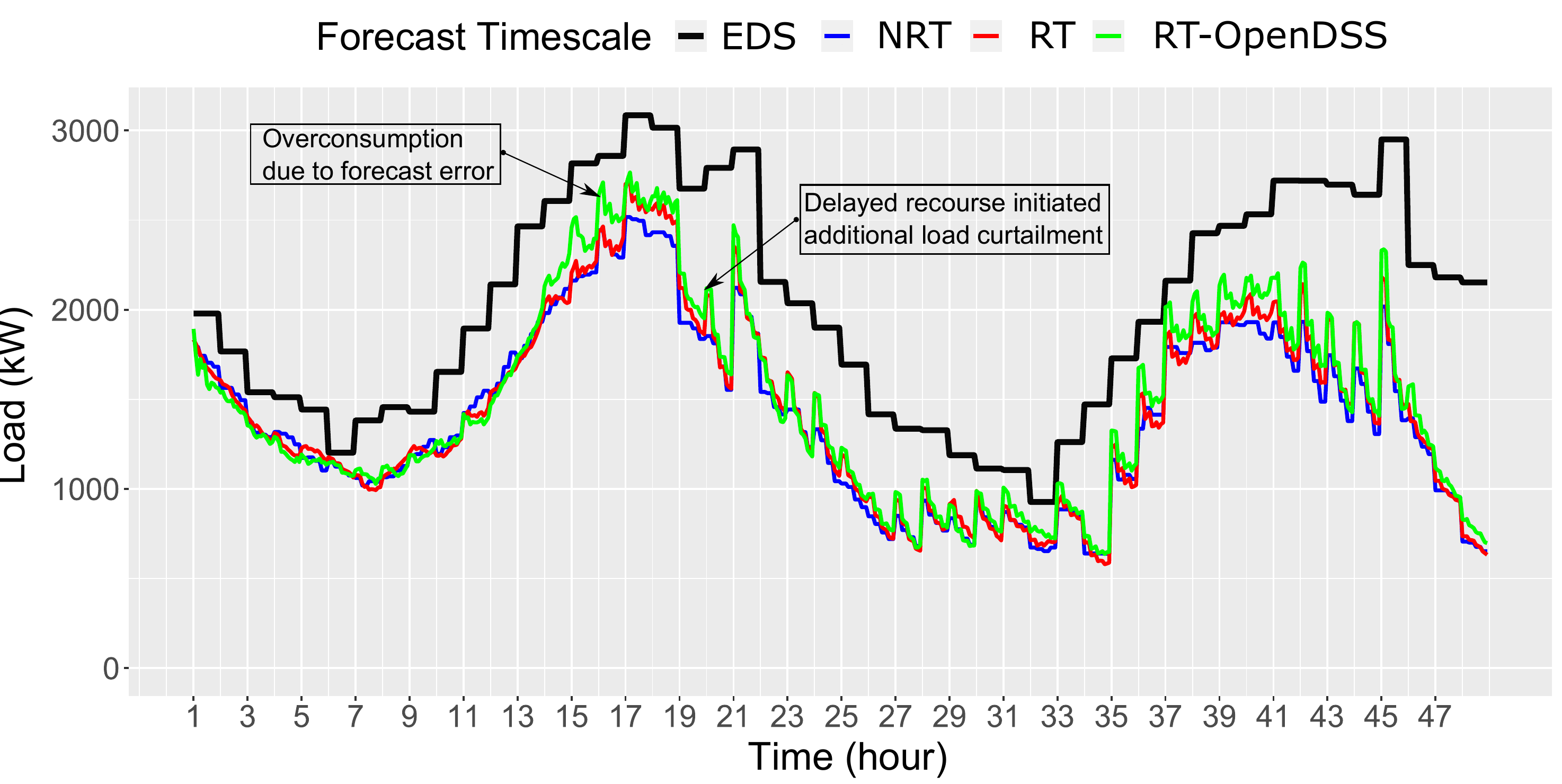}
  \vspace{-0.2cm}
  \caption{Load allocation and RT realization for base case with recourse.}
  \label{fig:basecase_recourse_loadallocation}
  \vspace{-0cm}
\end{figure}

\begin{figure}[htb]
\vspace{-0.0cm}
  \centering
  \includegraphics[width = 0.96\linewidth ,keepaspectratio, trim={0cm 0cm 0cm 0cm},clip]{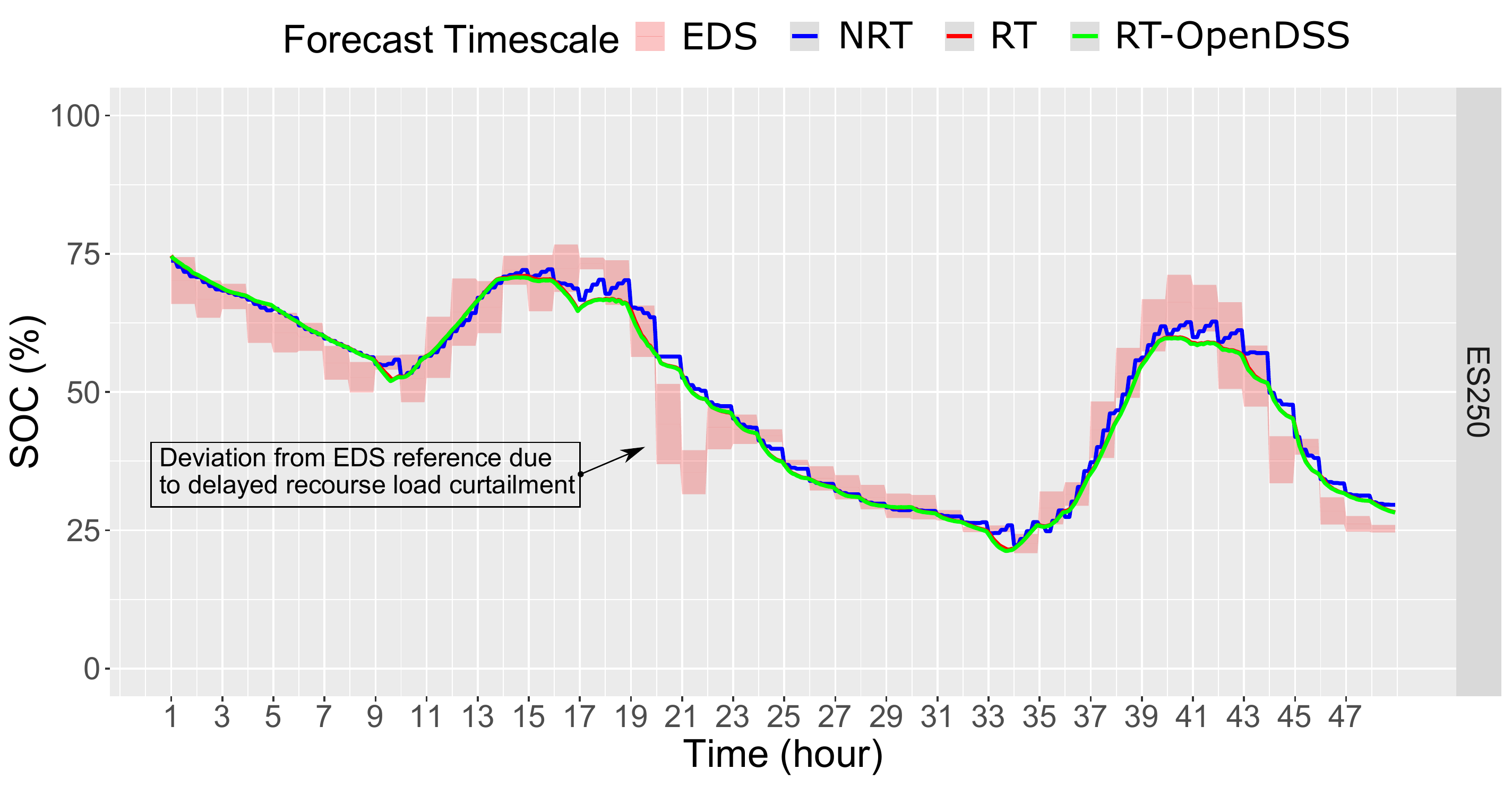}
  \vspace{-0.2cm}
  \caption{Grid forming ES250 reference SOC and RT realization for base case with recourse.}
  \label{fig:basecase_recourse_gfmbess}
  \vspace{-0cm}
\end{figure}

Next, we show the plots for the load supply and generator outputs. Fig. \ref{fig:basecase_recourse_loadallocation} shows the total EDS allocated load, followed by the total load selected to be supplied in the NRT and RT stages, and eventually, the RT realization in OpenDSS. We observe that the load is less than or equal to the EDS allocation at all times. Since the forecast error is minimum in this case, the load selection and supply go hand in hand. Also, the EDS allocated load exceeds the NRT allocation for few time intervals even when there is no additional curtailment caused by delayed recourse due to the following reasons: the EDS load forecast and the scenarios obtained forecast a higher load value than the NRT load forecasts; the EDS stage optimization problem does not contain the OPF constraints which can potentially limit the total load supplied; the NRT stage curtails some amount of load to ensure that the load selected across the three phases is balanced. Fig. \ref{fig:basecase_recourse_gfmbess} shows the SOC of the grid-forming ES unit (ES250). We observe that the SOC adheres to the reference set by the different hierarchical stages of the SA-HMTS framework except between hours $19$ and $22$ due to additional load curtailment initiated by the recourse constraints. Fig. \ref{fig:basecase_recoursenorecourse_soc} compares the SOC of ES250 with and without the delayed recourse approach. Without the recourse constraint, the SOC tends to fall sharply starting hour $19$.

\begin{figure}[htb]
\vspace{-0.0cm}
  \centering
  \includegraphics[width = 0.96\linewidth ,keepaspectratio, trim={0cm 0cm 0cm 0cm},clip]{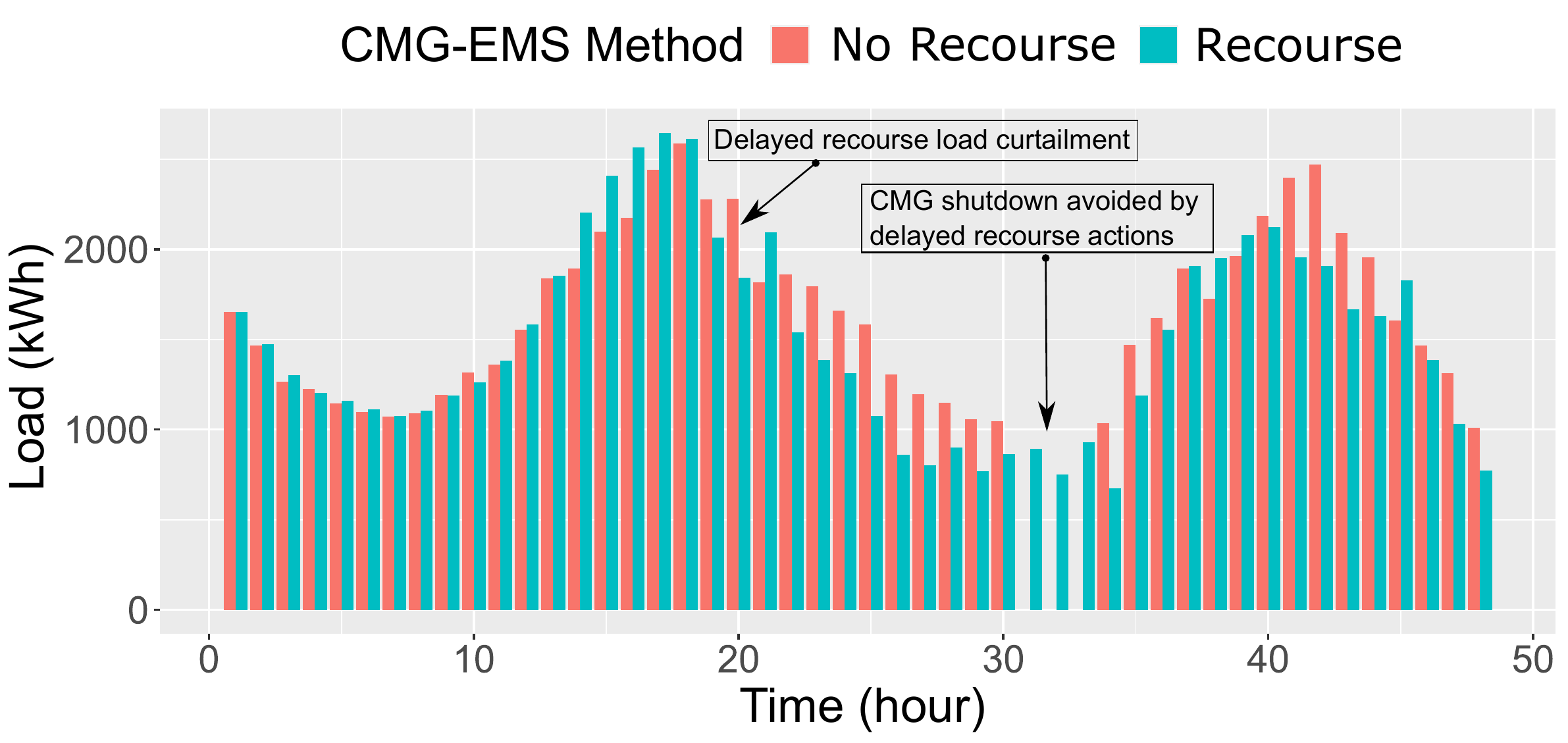}
  \vspace{-0cm}
  \caption{Comparison of hourly load supplied with and without delayed recourse.}
  \label{fig:basecase_recoursenorecourse_load}
  \vspace{-0cm}
\end{figure}

\begin{figure}[htb]
\vspace{-0.0cm}
  \centering
  \includegraphics[width = 0.96\linewidth ,keepaspectratio, trim={0cm 0cm 0cm 0cm},clip]{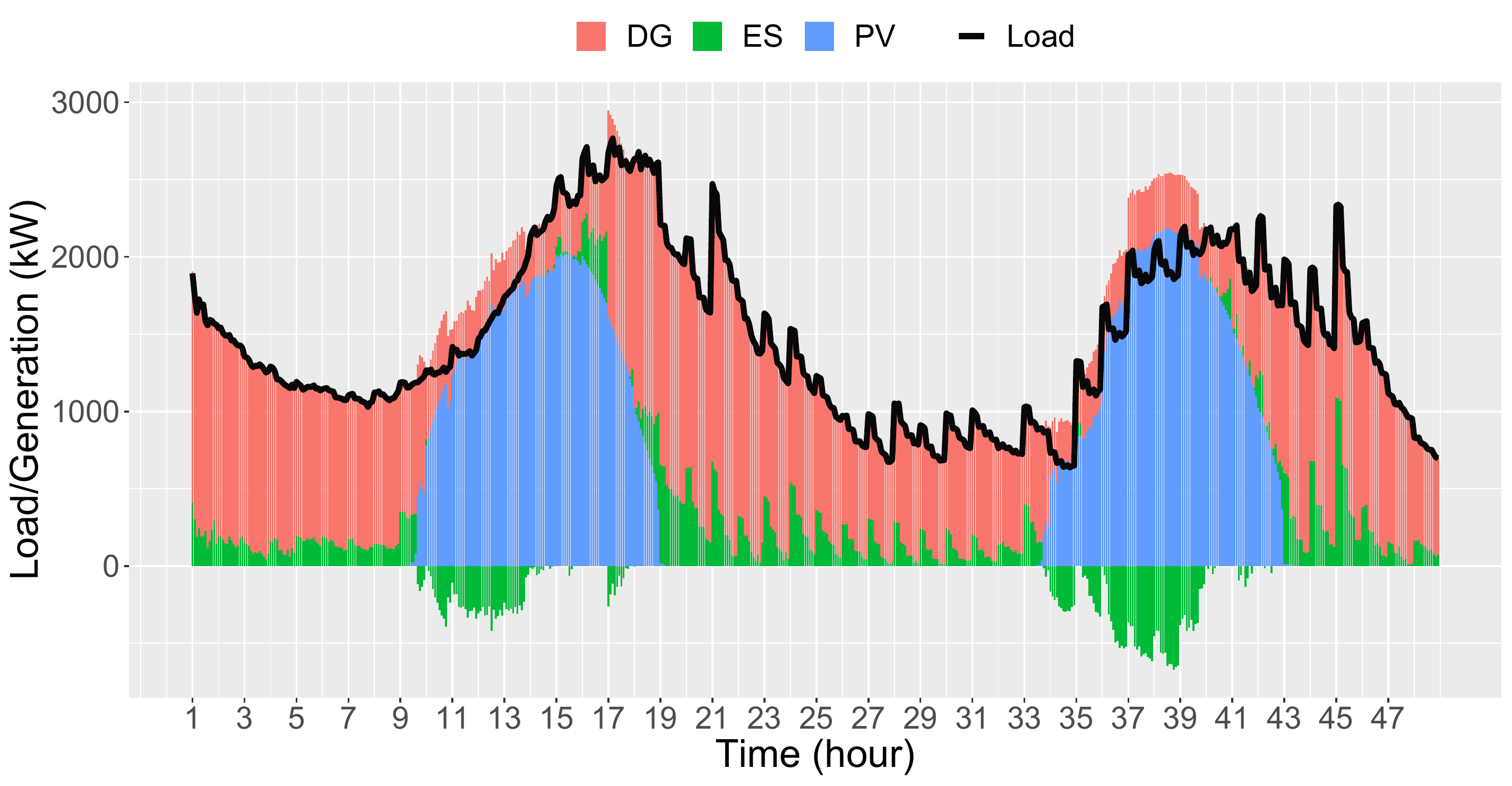}
  \vspace{-0cm}
  \caption{Grid forming ES250 reference SOC and RT realization for base case with recourse.}
  \label{fig:basecase_recourse_genportfolio}
  \vspace{-0cm}
\end{figure}

On the contrary, the recourse mechanism starts curtailing additional load starting hour $19$ to compensate for the overconsumption between hours $14$ and $19$. Due to this, the sharp drop in SOC is prevented, and the total time intervals with $20\%$ reserve violation are significantly reduced. Fig. \ref{fig:basecase_recoursenorecourse_load} compares the hourly load supplied with and without the delayed recourse mechanism. For the first $13$ hours, no significant difference between the two is observed. Starting from hour $14$, the effect of forecast error becomes evident, thus causing the delayed recourse actions to take effect. The load supplied using delayed recourse is consistently lesser than the load supplied in the latter case to counteract the forecast error-induced overconsumption. Allocating less load ensures less burden on the grid-forming ES unit to address the unforeseen forecast errors, thus preventing the CMG shut down between the hours $31$ and $33$. Fig. \ref{fig:basecase_recourse_genportfolio} shows how all the different generation resources were used to satisfy the total demand. We observe that the ES units are dispatched significantly during PV generation periods, while the DG units are dispatched during nighttime. This is because the ES dispatch can be altered faster as opposed to DG units to address the RT load and PV generation uncertainties. 

\begin{figure}[htb]
\vspace{-0.0cm}
  \centering
  \includegraphics[width = 0.96\linewidth ,keepaspectratio, trim={0cm 0cm 0cm 0cm},clip]{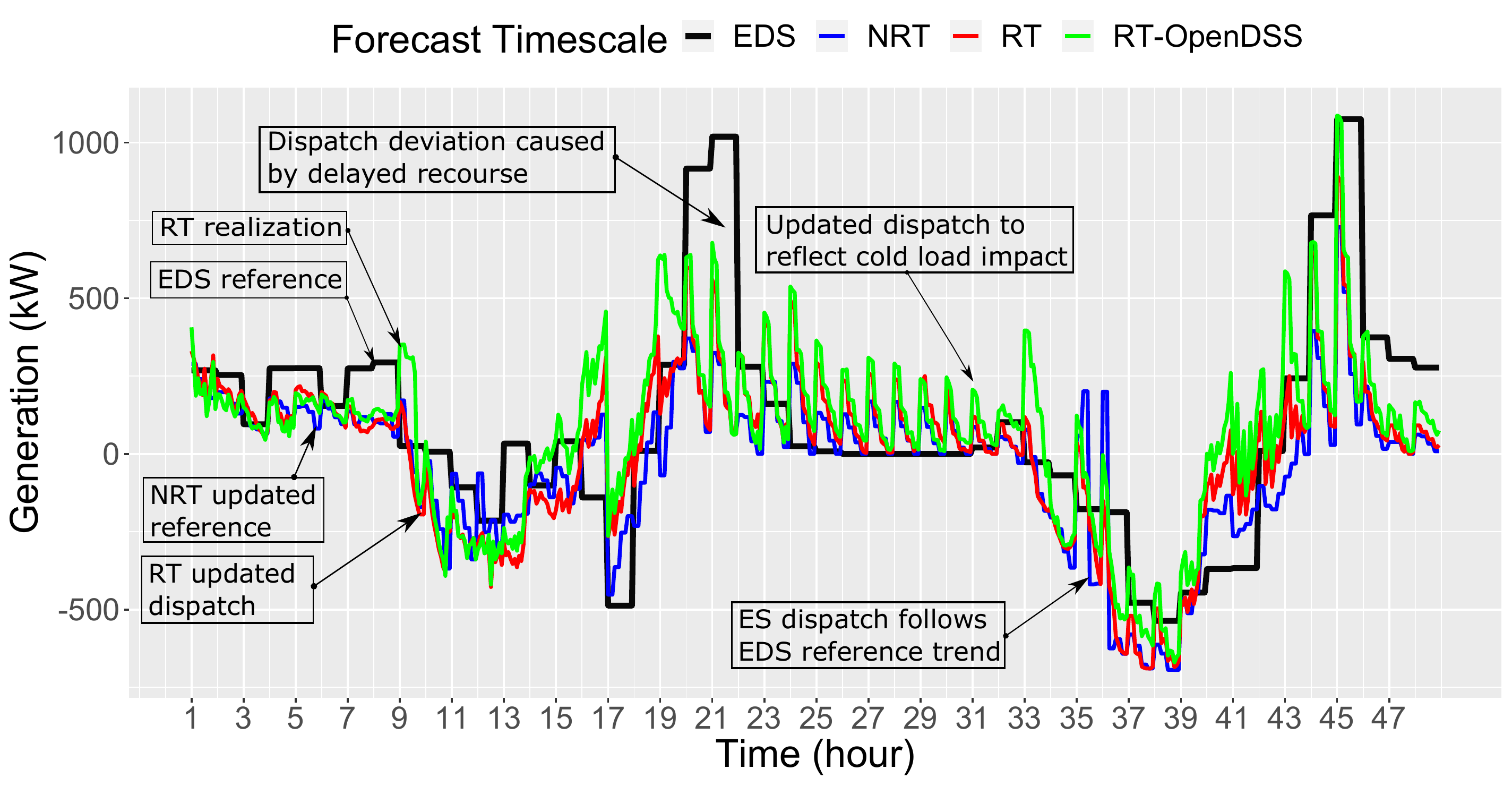}
  \vspace{-0cm}
  \caption{Total ES power schedule and its updates across the three timescales.}
  \label{fig:es_multitimescale}
  \vspace{-0cm}
\end{figure}

Fig. \ref{fig:es_multitimescale} shows the updates made to the overall ES dispatch across the three SA-HMTS timescales. For simplicity, the total ES generation, summed up over all the ES units in the network, is shown. We observe that by using an updated forecast value for each timescale, the ES dispatch is altered slightly around the EDS reference value to ensure that the demand and supply are balanced and PV curtailment is minimized. However, for some time intervals, a significant mismatch is observed between the expected EDS reference value and the NRT updated value. This can be attributed to the following: a significantly high forecast error; the effect of delayed recourse resiliency cut; and the cold load that needs to be picked up. This result highlights the necessity of three hierarchical stages, which provide multiple avenues to refine the generator dispatch under changing forecasts and operating conditions. A similar dispatch trend is observed for the DG units, wherein the EDS reference schedule refinement takes only at the NRT stage since the DG output change is limited to an hourly basis.      

Lastly, we show the results of the SA-HMTS framework with and without integrating the load supply duration equity weight $\omega_{i,t}^{\text{2}}$. Table \ref{tab:equity} shows the standard deviation of the supply duration of various NCL nodes for the three network phases. A $40\%$ improvement in terms of supply duration standard deviation reduction was observed using the dynamically changing service duration-based priority weight. Lastly, the SA-HMTS performance is also compared against the more traditional two-stage approach of DA scheduling followed by RT dispatch. Additional details on this comparison can be found in \cite{hmts_conf}. The results demonstrate a superior performance using SA-HMTS instead of a two-stage framework.

\begin{table}[]
\centering
\small
\caption{Metrics for base case analysis.}
\label{tab:metrics_basecaseanalysis}
\begin{tabular}{c|c|c}
\hline
Metric & With recourse & Without recourse  \\ \hline \hline
  $P^{\text{NCL}}$ (\%) & 62.83 & 70.74 \\ 
  $P^{\text{CL}}$ (\%) & 100 & 96.62\\
  $T^{\text{ASD, NCL}}$ (hours) &  24.73 (8.12) & 26.81 (9.80) \\
  $T^{\text{ASD, CL}}$  (hours)&  48 (0) & 45 (0) \\
  $T^{\text{AID, NCL}}$ (hours) &   23.27 (8.12) & 21.19 (9.80) \\
  $T^{\text{AID, CL}}$ (hours) &  0 (0) & 3 (0) \\
  $P^{\text{PV, Total}}$ (\%) &  86.34 & 88.53\\
  $F^{\text{DG, Balance}}$ (\%) &  51.65 & 47.33 \\
  $SOC^{\text{ES, Balance}}$ (\%) &  26.03 & 23.75 \\
  $T^{\text{RL,ES}} (\%)$  &  4.12 &  26.56 \\
  $P^{\text{Imb.}}$ (\%) &  7.13 & 7.16 \\
  %$P^{\text{DG, Imb.}}$ (\%) &  1.12  & 1.23 \\ 
  $T^{\text{CMG,OFF}}$ (hours) &  0 & 3 \\
  $a$ & 0.08 (0.09) & - \\ \hline
\end{tabular}
\end{table}

\begin{table}[]
\centering
\caption{Supply duration equity.}
\label{tab:equity}
\begin{tabular}{c|c|c}
\hline
\multirow{2}{*}{Phase} & \multicolumn{2}{c}{Standard deviation (hours)} \\ \cline{2-3} 
                       & With $\omega_{i,t}^{\text{2}}$       & Without $\omega_{i,t}^{\text{2}}$       \\ \hline \hline
A                      & 4.86       & 9.64          \\ \hline
B                      & 4.48       & 7.15          \\ \hline
C                      & 4.92       & 7.93          \\ \hline
\end{tabular}
\end{table}

\subsection{Performance Analysis of $n$-Delayed Recourse Approach}
\begin{figure}[htb]
\vspace{-0.0cm}
  \centering
  \includegraphics[width = 1\linewidth ,keepaspectratio, trim={1cm 3cm 0.5cm 0cm},clip]{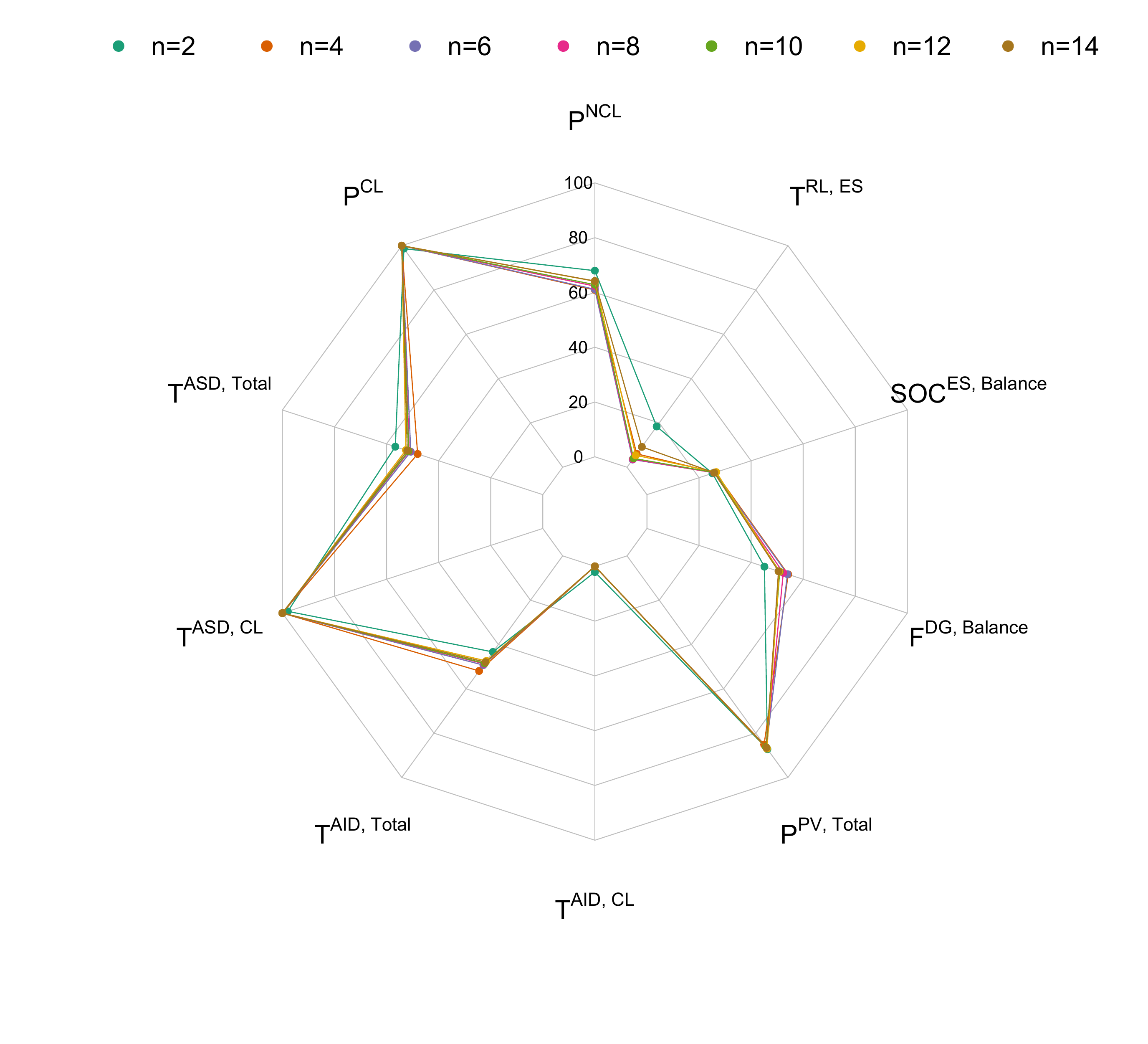}
  \vspace{-0.2cm}
  \caption{Visualizing metrics for $n$-delayed recourse with varying value of $n$.}
  \label{fig:metrics_nanalysis}
  \vspace{-0cm}
\end{figure}

In this section, we perform a detailed analysis to quantify the performance of the delayed recourse approach for selecting the best values for the hyperparameter $n$. To do so, the value for $n$, which decides the historical hours that are considered for taking the recourse action, is varied between $2$ and $14$ with an interval of $2$. Table \ref{tab:metrics_recoursecases} shows the comparison of the different metrics for the different values of $n$. For $n=2$, we observe that the value of $a$ is close to $0$ but with a high standard deviation. Due to an extremely short historical horizon, it is unable to capture the forecast error trend accurately. Starting $n = 4$, we observe that the average value of $a$ remains steady, but the standard deviation reduces. This is because, with an increase in the historical information, the linear trend estimator has more information on the past forecast error, resulting in stable trend estimation. On the contrary, for smaller values of $n$, the trend estimator will be shortsighted, resulting in fluctuating trends for each hour. Although this did not affect the performance for this case due to a small forecast error, having the value of $a$ fluctuate significantly between hours will cause frequent up/down ramps in the hourly allocated load. Hence, a higher value of $n$ is preferred. Comparing results for $n\ge 10$, no significant changes in the metrics are observed, except for $T^{\text{RL,ES}}$. Hence, out of the three, we prioritize the secure operation of the CMG and choose $n=10$. Overall, by visualizing the metrics as shown in Fig. \ref{fig:metrics_nanalysis} for all the different values of $n$, we observe that for values of $n > 2$, the metrics do not significantly vary. However, keeping in mind the drawbacks of values of $n$ ranging between $3$ and $9$, we emphasize the selection of $n$ to values greater than $9$. Hereafter, all the simulations performed have the value of $n$ fixed at $10$.

\begin{table*}[]
\centering
\small
\caption{Variation of hyperparameter $n$ within $n$-delayed recourse framework.}
\label{tab:metrics_recoursecases}
\begin{tabular}{c|c|c|c|c|c|c|c}
\hline
$n$ value & 2 & 4 & 6 & 8 & 10 & 12 & 14  \\ \hline \hline
  $P^{\text{NCL}}$ (\%) &	67.95\% &	60.91\% &	61.14\% &	62.31\% &	62.83\% &	64.09\% &	64.16\%  \\ 
  $P^{\text{CL}}$ (\%)  & 98.66\% & 100\% & 100\% & 100\% & 100\% & 100\% & 100\%\\
  $T^{\text{ASD, NCL}}$ (hours)  &  27.19 (10.12)  & 23.07 (9.31) & 24.34 (9.70) & 24.62 (9.73) & 24.73 (8.12) & 25.24 (10.03) & 24.98 (10.02)\\
  $T^{\text{ASD, CL}}$ (hours)  &  47 (0) & 48 (0) &  48 (0) &  48 (0) &  48 (0) &  48 (0) &  48 (0) \\
  $T^{\text{AID, NCL}}$ (hours)  &  20.81 (10.12) & 24.93 (9.31) & 23.66 (9.70) & 23.38 (9.73) & 23.27 (8.12) & 22.76 (10.03) & 23.02 (10.02)\\
  $T^{\text{AID, CL}}$ (hours)  &  1 (0) &  0 (0) &  0 (0) &  0 (0)  &  0 (0)  &  0 (0)  &  0 (0)    \\
  $P^{\text{PV, Total}}$ (\%)  &  87.23 & 85.16 & 86.53 & 86.55 & 86.34 & 86.95 & 86.6 \\
  $F^{\text{DG, Balance}}$ (\%)  &  45.11 & 54.24 & 53.95 & 52.35 & 51.65 & 50.79 & 50.52 \\
  $SOC^{\text{ES, Balance}}$ (\%)  &  25.15 & 26.27 & 25.99 & 25.79 & 26.03 & 26.62 & 25.86 \\
  $T^{\text{RL,ES}} (\%)$  &  18.40 & 6.02 & 3.47 & 3.64 & 4.12 & 5.38 & 9.20 \\
  $P^{\text{Imb.}}$ (\%)  &  7.99 &	7.96 & 8.01 & 7.98 & 7.13 &	8.03 &	8.11 \\
  %$P^{\text{DG, Imb.}}$ (\%)  &  1.05 &	1.10 &	1.04 &	1.14 &	1.12 &	1.03 &	1.09 \\ 
  $T^{\text{CMG,OFF}}$ (hours)  &  1 & 0 & 0 & 0 & 0 & 0 & 0\\
  $a$  &  0.04 (0.28) &	0.14 (0.18) &	0.15 (0.13) &	0.14 (0.11) &	0.08 (0.09) & 0.13 (0.09) & 0.12 (0.09) \\ \hline
\end{tabular}
\end{table*}

\subsection{Performance Under Forecast Error Scenarios}
In this section, we demonstrate case studies highlighting the performance of the SA-HMTS framework under various forecast error scenarios. This section is analogous to demonstrating the benefits of the newly proposed $n$-delayed recourse approach. We perform case studies using two types of forecast errors: 
\begin{itemize}
    \item \textbf{Case FE1:} Forecasts are biased throughout, i.e., the forecast overestimates or underestimates the demand and PV generation for all time intervals.
    
    \item \textbf{Case FE2:} The forecast overestimates and underestimates randomly.
\end{itemize}

Case FE1 has a higher probability of occurrence during outages caused by extreme events because the forecasting algorithm does not have enough information about HILF events to make meaningful predictions. Case FE2 is commonly observed under typical operation scenarios but can also occur during emergency scenarios. These two cases broadly cover the extreme forecast error scenarios observed while operating under HILF extreme events. 

\begin{figure}[htb]
\vspace{-0.0cm}
  \centering
  \includegraphics[width = 0.96\linewidth ,keepaspectratio, trim={0cm 0cm 0cm 0cm},clip]{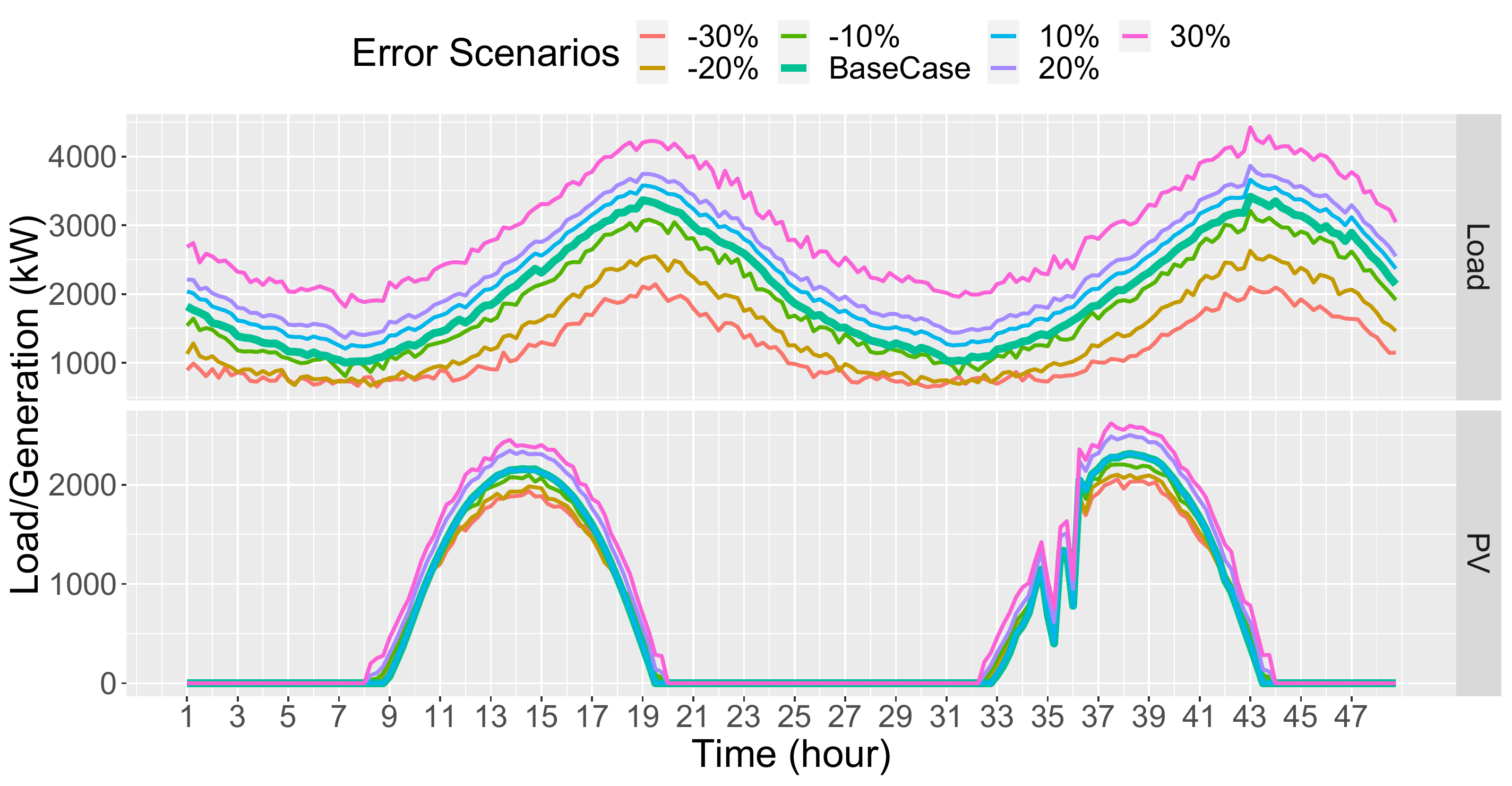}
  \vspace{-0.2cm}
  \caption{NRT forecast profiles for $6$ different scenarios in case FE1: (a) Total load and (b) PV generation.}
  \label{fig:profiles_nrt_forecasterror_uniform}
  \vspace{-0cm}
\end{figure}

To demonstrate case FE1, we introduce a uniform forecast error between the forecasts and the actual RT realization in a uniform manner. The load selection is performed in the NRT-stage using the error-introduced forecasts, and then the selected loads have to be supplied in full during RT realization. The load forecasts used for the base case are modified such that the MAPE value between the forecasts and the RT realization varies between $-30\%$ and $+30\%$. Given that MAPE score values are always positive, the negative sign associated with this score indicates that the forecasted values are lower than the actual realization. In contrast, the positive sign indicates the opposite. Similarly, the forecasts for PV generation are modified by increasing/decreasing the PV generation in proportion to the PV generator ratings. Figure \ref{fig:profiles_nrt_forecasterror_uniform} shows the NRT-stage forecasts for case study FE1. 

Table \ref{tab:metrics_forecasterroruniform_recourse} shows the computed metrics for case study FE1 with using delayed recourse. As the forecast error magnitude goes on changing from $-30\%$ to $0\%$, the service duration and the load supply metrics go on increasing, while the service interruption metrics go on decreasing. Between $10\%$ and $30\%$, we see a slight increasing trend in the load supply and the supply duration metrics. Although the recourse metric is relaxed due to the underconsumption of generation resources, the optimization problem follows the reference resource allocation computed by the stochastic EDS stage. The PV utilization is also seen increasing, with the lowest for the case with $-30\%$ error and highest for the $30\%$ case. This is because additional loads are curtailed to compensate for resource overconsumption due to forecast errors, resulting in the disconnection of BTM PV systems from the network. This results in the underutilization of available PV generators for network load restoration. Comparing the metrics, we see that the framework has tried to maximize the load supply by prioritizing the CL for the negative forecast error cases. The model performance is superior to the base case with no forecast error for the positive forecast error cases.

\begin{figure*} [tb]
\captionsetup[subfigure]{labelformat=empty}
\centering
  \subfloat[]{%
        \includegraphics[width=1\linewidth,keepaspectratio, trim={0.5cm 7.75cm 0cm 7cm},clip]{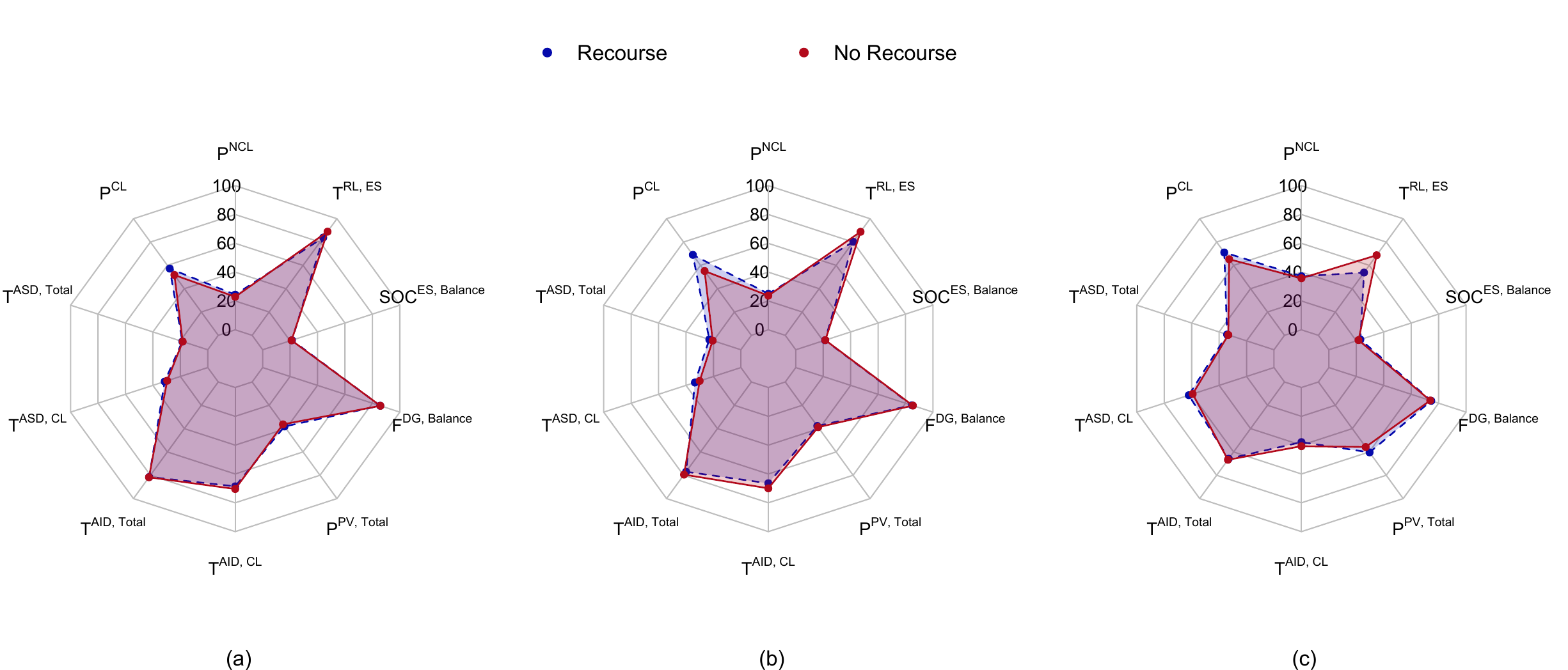}} \\
    \vfill
  \subfloat[]{%
        \includegraphics[width=1\linewidth,keepaspectratio, trim={0.5cm 7.75cm 0cm 9cm},clip]{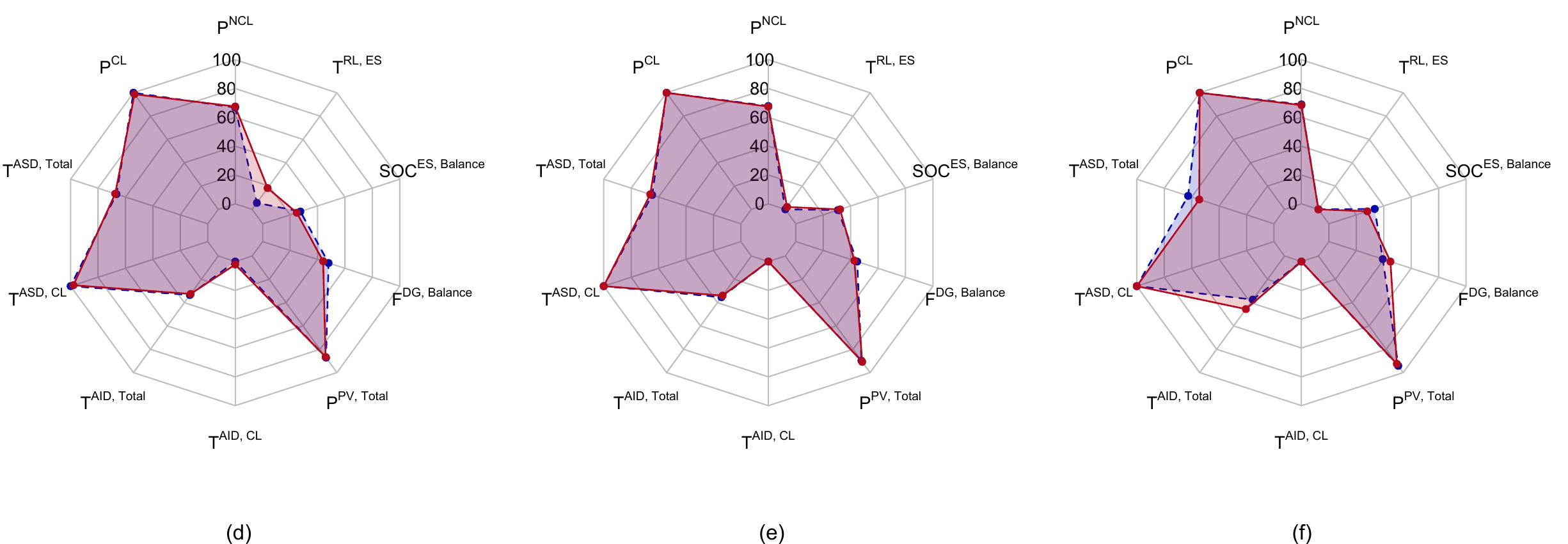}}
  \caption{Metrics for different forecast error scenarios in case FE1 with and without delayed recourse with forecast MAPE as follows: (a) -30\%, (b) -20\%, (c) -10\%, (d) 10\%, (e) 20\%, (f) 30\%}
  \label{fig:metrics_forecasterroruniform_recourse} 
  \vspace{-0.5cm}
\end{figure*}

Next, we compare the results without using the delayed recourse approach. Table \ref{tab:metrics_forecasterroruniform_norecourse} shows the metrics for the case study without using the delayed recourse approach. For the negative forecast error cases, including the delayed recourse constraints helps improve the metrics compared to without using the delayed recourse mechanism. For negative forecast error cases, we observe that the recourse constraints help maximize the load supply since its proactive load curtailment ensures a secure operation and results in reducing the hours for which the CMG needs to be turned off. For cases with positive forecast error, the performance between both approaches is similar. Although the delayed recourse constraints allow for the additional load to be supplied, it is limited by the penalty imposed on the deviation of the DG power output and ES SOC from the EDS stage reference value. To summarize, the SA-HMTS framework performance under uniform forecast error is close to or superior to its counterpart with no delayed recourse. Delayed recourse benefits are visible for negative forecast error cases, while the receding-horizon EDS stage significantly contributes to performance improvement for the positive forecast error cases. Fig. \ref{fig:metrics_forecasterroruniform_recourse} shows a comparative plot of the different metrics for case study FE1 with and without recourse. 

\begin{figure}[htb]
\vspace{-0.0cm}
  \centering
  \includegraphics[width = 0.96\linewidth ,keepaspectratio, trim={0cm 0cm 0cm 0cm},clip]{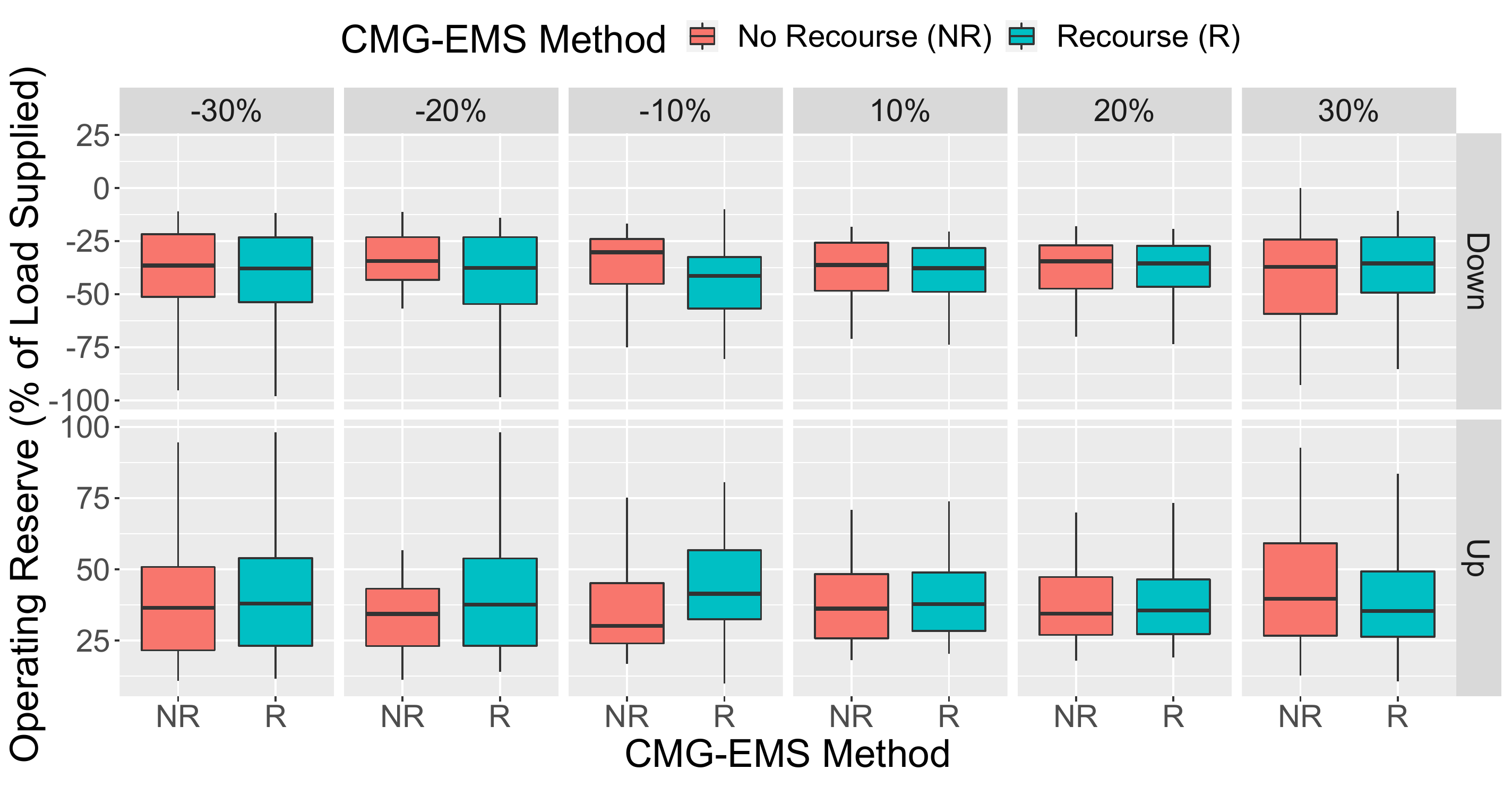}
  \vspace{-0.2cm}
  \caption{Up and down reserve analysis under case study FE1 with and without delayed recourse framework.}
  \label{fig:metrics_forecasterroruniform_reserve}
  \vspace{-0cm}
\end{figure}

Lastly, we analyze the performance from a different perspective. Since tremendous importance is placed on the secure operation of the CMG when faced with extremely high levels of uncertainty, we analyze the percent change in the supplied load that can be smoothly handled by the CMG using the $20\%$ of the grid-forming ES reserves. The time intervals for which the CMG is turned off are excluded for this analysis. The up-reserve and down-reserve provisions are analyzed. Fig.\ref{fig:metrics_forecasterroruniform_reserve} shows the results of this specific study. We observe that the percent load variation that can be handled in either direction is higher by using delayed recourse for the cases with negative forecast errors. As the forecast error in the negative direction increases, the effectiveness of the delayed recourse constraints goes on diminishing due to the long duration for which the CMG stays shut for both approaches. On the positive forecast error side, no significant difference is observed between the two proposed frameworks.  

\begin{figure}[htb]
\vspace{-0.0cm}
  \centering
  \includegraphics[width = 0.96\linewidth ,keepaspectratio, trim={0cm 0cm 0cm 0cm},clip]{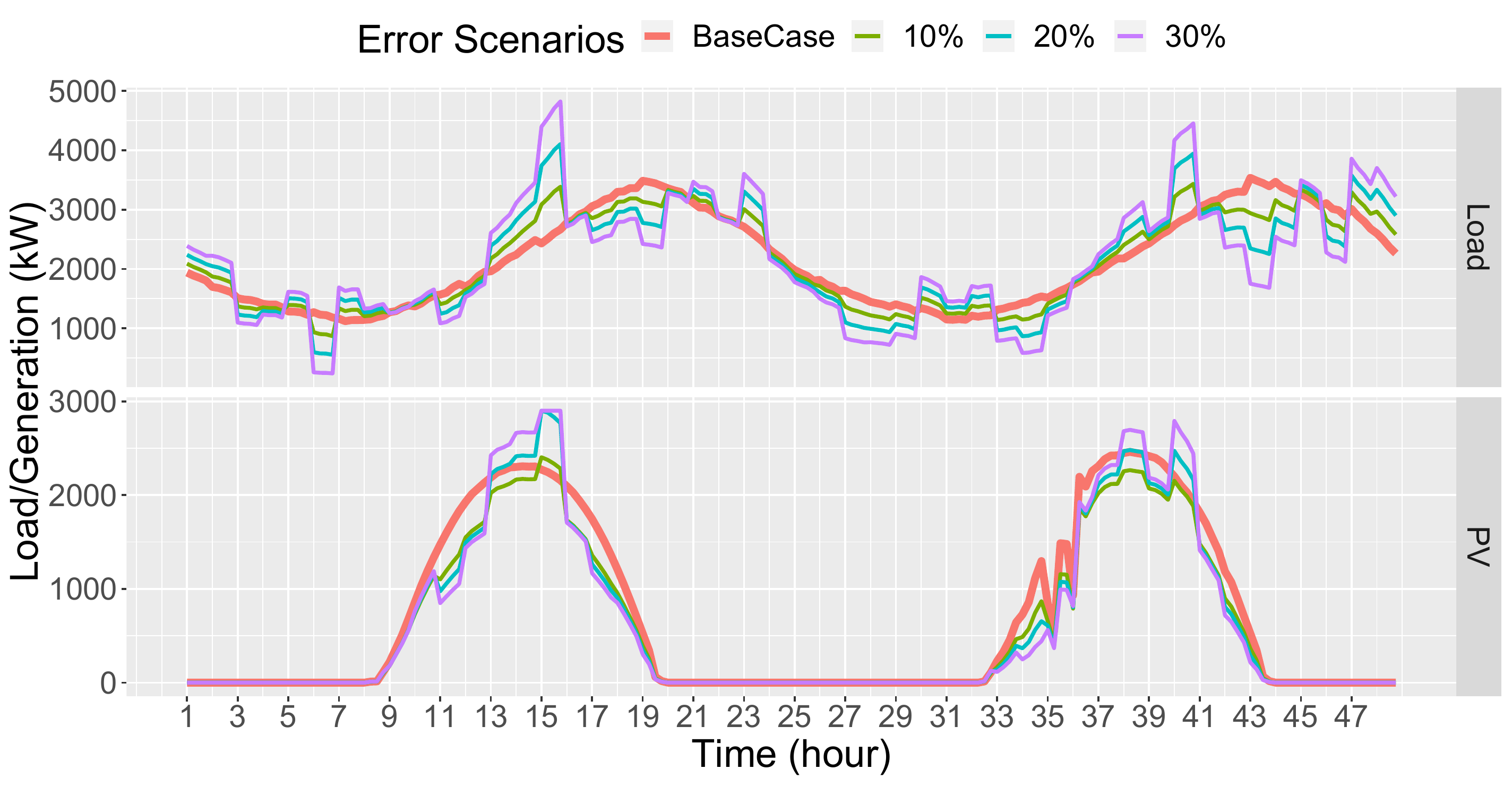}
  \vspace{-0.2cm}
  \caption{NRT forecast profiles for $3$ different scenarios in case FE2: (a) Total load and (b) PV generation.}
  \label{fig:profiles_nrt_forecasterror_random}
  \vspace{-0cm}
\end{figure}

To demonstrate case FE2, we introduce a random error between the forecasts and the actual RT realization. The forecasts used for the base case are modified such that the mean absolute percent error (MAPE) between the forecasts and the RT realization varies between $10\%$ and $30\%$. Figure \ref{fig:profiles_nrt_forecasterror_random} shows the NRT forecasts for different forecast error cases considered for case study FE2.  

\begin{figure}[htb]
\vspace{-0.0cm}
  \centering
  \includegraphics[width = 0.96\linewidth ,keepaspectratio, trim={0cm 0cm 0cm 0cm},clip]{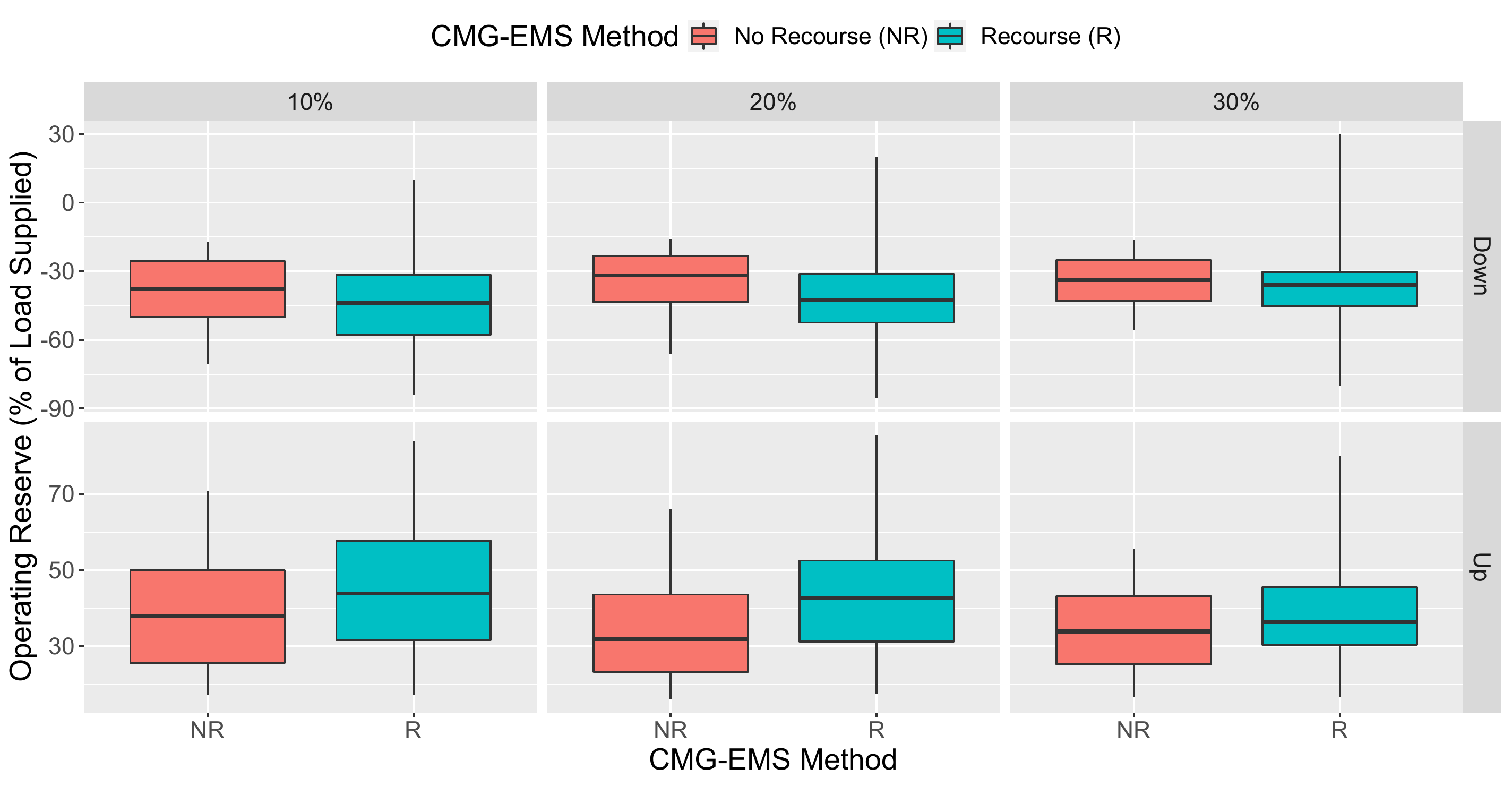}
  \vspace{-0.2cm}
  \caption{Up and down reserve analysis under case study FE2 with and without delayed recourse framework.}
  \label{fig:metrics_forecasterrorrandom_reserve}
  \vspace{-0cm}
\end{figure}

Table \ref{tab:metrics_forecasterrorrandom_recourse} shows the metrics for case study FE2. As the forecast error increases, the total load supply and the supply duration reduces while the interruption duration increases. However, we observe that the interruption of CL is very low, thus achieving the goal of CL prioritization. On comparing with Table \ref{tab:metrics_forecasterrorrandom_norecourse}, the benefit of the delayed recourse approach is visible. The recourse approach helps increase the total load supplied. It also minimizes the time intervals during which the grid-forming ES unit violates the $20\%$ reserve limit requirement, thus leading to a more secure and proactive operation. We perform a similar analysis as shown in case study FE1 to analyze the reserve availability. The results of this analysis are shown in Fig. \ref{fig:metrics_forecasterrorrandom_reserve}. The interpretations of the results are the same as those mentioned in FE1. Overall, using delayed recourse has enhanced the operating reserve availability during the extended duration operation.   

\begin{figure*} [tb]
%\captionsetup[subfigure]{labelformat=empty}
\centering
  \includegraphics[width=1\linewidth,keepaspectratio, trim={0.5cm 7cm 0.25cm 7.5cm},clip]{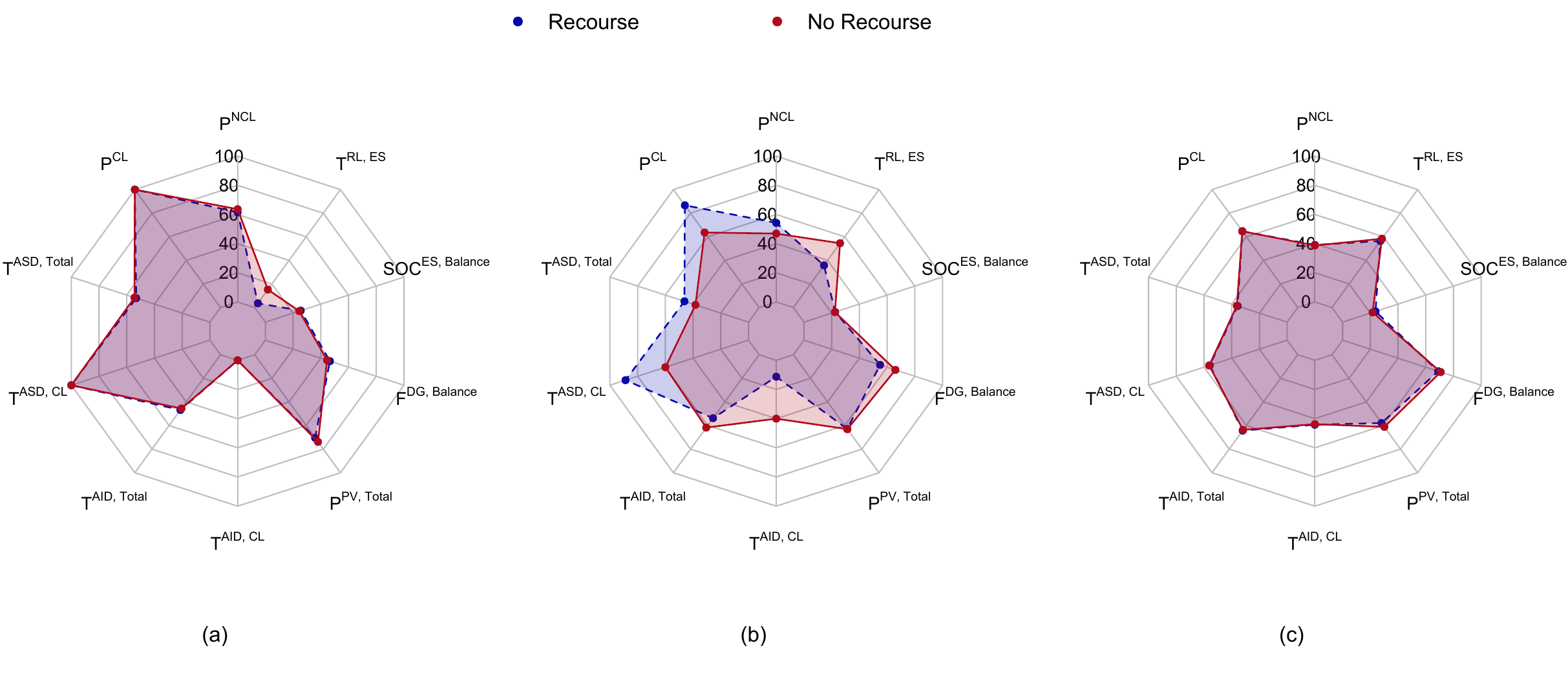} \\
  \caption{Metrics for different forecast error scenarios in case FE2 with and without delayed recourse with forecast MAPE as follows: (a) 10\%, (b) 20\%, (c) 30\%}
  \label{fig:metrics_forecasterrorrandom_recourse} 
  \vspace{-0.5cm}
\end{figure*}

Fig. \ref{fig:metrics_forecasterrorrandom_recourse} shows a comparative plot of the different metrics for case study FE2 with and without recourse. We can see that a significant benefit of the recourse mechanism is seen when the forecast error is $10\%$ and $20\%$. However, we observe a very marginal improvement for a forecast error of $30\%$. This helps conclude that the proposed SA-HMTS framework is effective against forecast errors up to a certain extent. However, beyond the threshold, the delayed recourse approach offers minimal benefit. The reason behind this is the shut down of CMG once the grid-forming ES unit SOC reduces below $20\%$. To increase the range of forecast errors for which the delayed recourse framework can provide operational improvements, changes to the linear trend estimator to make it more conservative will be required to be made. However, doing so will require a trade-off between maximizing the total load supplied and increasing the continuous CMG operation duration.

\begin{figure}[htb]
\vspace{-0.0cm}
  \centering
  \includegraphics[width = 0.96\linewidth ,keepaspectratio, trim={0cm 0cm 0cm 0cm},clip]{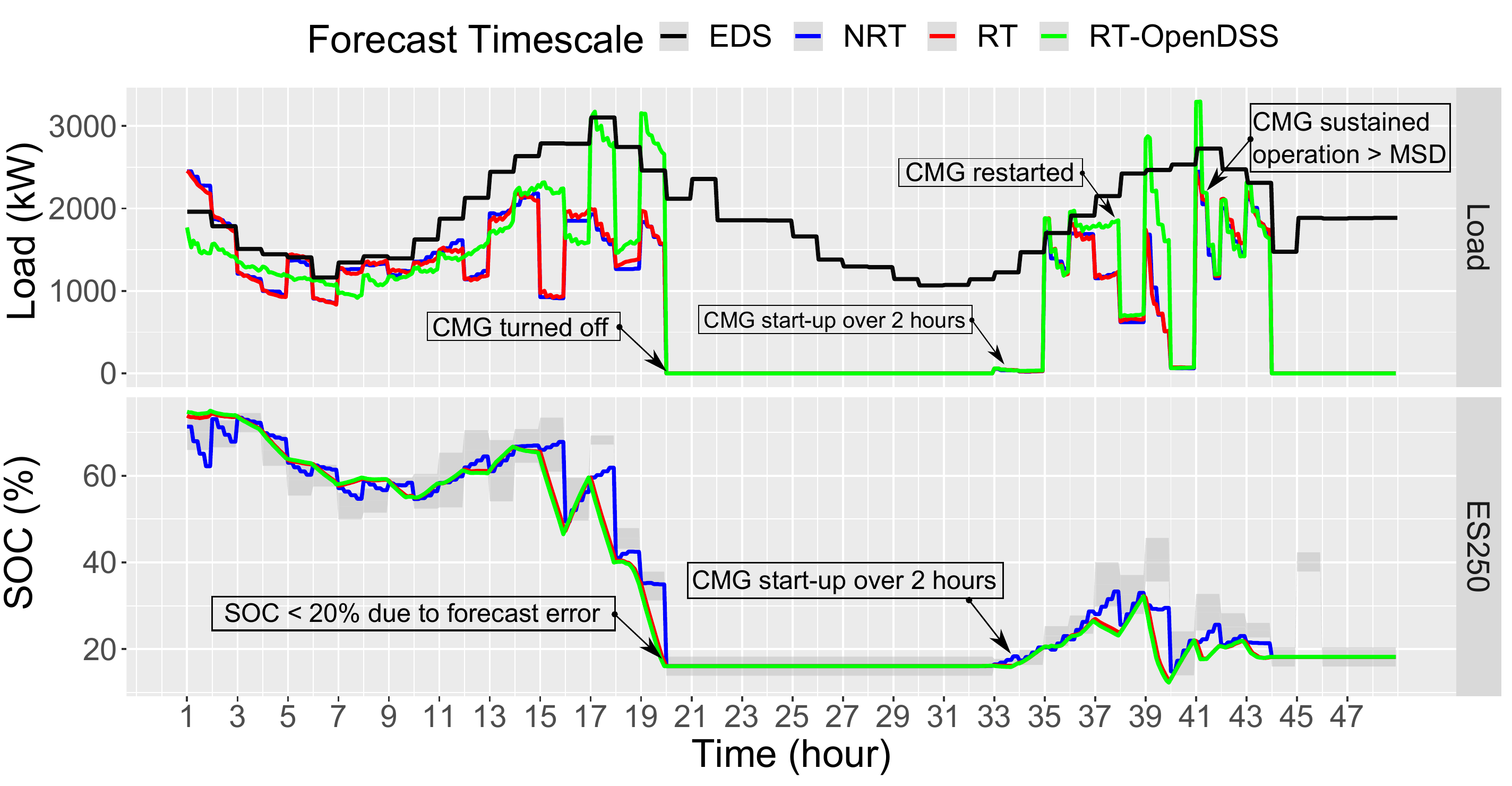}
  \vspace{-0.2cm}
  \caption{Load and ES250 SOC plots for case FE2 with forecast error of $30\%$ MAPE.}
  \label{fig:load_soc_fe2_30}
  \vspace{-0cm}
\end{figure}

Lastly, we demonstrate the system performance from the CMG shutdown and startup perspective using the $30\%$ MAPE scenario of case study FE2. Fig. \ref{fig:load_soc_fe2_30} shows the plot of the load supply and the corresponding grid-forming ES250 SOC. We observe that the moment the forecast error causes the SOC to drop below $20\%$, the CMG shuts down in the immediate next hour, ensuring that the previous hour's operation is not impacted. Due to the co-location of the ES250 and a PV generator, the ES is charged once the PV generator starts generating electricity. For the first startup instant, we observe that it takes around $2$ hours to recharge the ES unit, after which the CMG operates normally. A similar event occurs during $40$. Finally, beyond hour $44$, there are no means available to recharge the ES unit, resulting in a complete shutdown of the CMG. Also, the proactive load curtailment initiated by delayed recourse is also evident from Fig. \ref{fig:load_soc_fe2_30}.

\begin{table*}[]
\centering
\small
\caption{Metrics for different forecast error scenarios in case FE1 with delayed recourse.}
\label{tab:metrics_forecasterroruniform_recourse}
\begin{tabular}{c|c|c|c|c|c|c}
\hline
Error Scenario & -30\% & -20\% & -10\% & 10\% & 20\% & 30\%  \\ \hline \hline
  $P^{\text{NCL}}$ (\%) &	24.39\% & 24.99\% & 37.15\% & 66.82\% & 67.91\% & 68.98\%
 \\ 
  $P^{\text{CL}}$ (\%)  & 57.21\% & 69.02\% & 71.06 \% &100 \% &100 \% &100 \%
 \\
  $T^{\text{ASD, NCL}}$ (hours)  &  8.91 (4.97) & 11.05 (3.83) & 16.41 (6.66) & 31.95 (5.24) & 31.05 (4.78) & 30.05 (5.59) \\
  $T^{\text{ASD, CL}}$ (hours)  &  15.08 (1.64) & 16.13 (1.67) &  29.81 (0.44) & 48.00 (0) & 48.00 (0) & 48.00 (0) \\
  $T^{\text{AID, NCL}}$ (hours)  &  39.09 (4.97) & 36.95 (3.83) &  31.59 (6.66) & 16.05 (5.24) & 16.95 (4.78) & 17.95 (5.59) \\
  $T^{\text{AID, CL}}$ (hours)  &  32.92 (1.64) & 31.90 (1.67) & 18.19 (0.44) & 0 (0) & 0 (0) & 0 (0) \\
  $P^{\text{PV, Total}}$ (\%)  &  38.15 & 37.69 &  60.37 & 87.16 & 89.83 & 94.20 \\
  $F^{\text{DG, Balance}}$ (\%)  &  85.73 & 84.73 & 74.62 & 48.06 & 44.78 & 39.28 \\
  $SOC^{\text{ES, Balance}}$ (\%)  &  21.39 & 21.35 & 22.96 & 27.51 & 30.74 & 33.41 \\
  $T^{\text{RL,ES}} (\%)$  & 83.85 &  80.21 & 53.81 & 5.55 & 0 & 0\\
  $P^{\text{Imb.}}$ (\%)  &  7.45 & 8.12 & 9.21 & 8.77 & 10.33 & 9.23\\
  %$P^{\text{DG, Imb.}}$ (\%)  &  1.15 &	1.08 &	1.02 & 1.08 & 0.99 &	1.05\\
  $T^{\text{CMG,OFF}}$ (hours)  &  30 & 29 & 15 & 0 & 0 & 0  \\
  $a$ &  0.21 (0.24) &	0.32 (0.27) &	0.25 (0.12) & -0.11 (0.20) & -0.126 (0.25) &	-0.28 (0.16) \\ \hline
\end{tabular}
\end{table*}

\begin{table*}[]
\centering
\small
\caption{Metrics for different forecast error scenarios in case FE1 without delayed recourse.}
\label{tab:metrics_forecasterroruniform_norecourse}
\begin{tabular}{c|c|c|c|c|c|c}
\hline
Error Scenario & -30\% & -20\% & -10\% & 10\% & 20\% & 30\%  \\ \hline \hline
  $P^{\text{NCL}}$ (\%) &	22.91\% & 23.71\% & 35.79\% & 67.56\% & 67.44\% & 68.48\% \\ 
  $P^{\text{CL}}$ (\%)  & 51.72\% &55.14 \% & 65.25\% &98.73 \% & 100\% &100 \% \\
  $T^{\text{ASD, NCL}}$ (hours)  &  8.77 (4.44) & 9.87 (3.89) &15.98 (5.11) & 32.39 (5.78) & 31.71 (5.38) & 30.15 (6.60) \\
  $T^{\text{ASD, CL}}$ (hours)  &  14.20 (1.48) & 14.42 (1.14) &28.45 (0.44) & 47 (0) & 48 (0) & 48 (0) \\
  $T^{\text{AID, NCL}}$ (hours)  &  39.23 (4.44) & 38.13 (3.89) & 32.02 (5.11) & 15.61 (5.78) & 16.28 (5.38) & 17.85 (6.60) \\
  $T^{\text{AID, CL}}$ (hours)  &  33.80 (1.48) & 33.58 (1.14) & 19.55 (0.44) & 1 (0) & 0 (0) & 0 (0) \\
  $P^{\text{PV, Total}}$ (\%)  &  36.33 & 38.76 & 55.75 & 86.62 & 90.73 & 94.46 \\
  $F^{\text{DG, Balance}}$ (\%)  &  85.83 & 85.39 & 73.89 & 44.08 & 42.72 & 38.87 \\
  $SOC^{\text{ES, Balance}}$ (\%)  &  21.05 &  21.55 & 21.59  & 24.89 & 33.27 & 28.02\\
  $T^{\text{RL,ES}} (\%)$  & 88.88 &  88.88 & 68.57 & 18.23 & 2.08 & 0\\
  $P^{\text{Imb.}}$ (\%)  &  8.12 & 11.21 & 10.22 & 7.99 & 12.11 & 11.12\\
  %$P^{\text{DG, Imb.}}$ (\%)  &  1.18 &	1.10 &	1.05 & 1.09 &	1.04 &	1.09\\ 
  $T^{\text{CMG,OFF}}$ (hours)  &  33 & 32 & 18 & 1 & 0 & 0  \\ \hline
\end{tabular}
\end{table*}

\begin{table*}[]
\centering
\small
\caption{Metrics for different forecast error scenarios in case FE2 with delayed recourse.}
\label{tab:metrics_forecasterrorrandom_recourse}
\begin{tabular}{c|c|c|c}
\hline
Error Scenario & 10\% & 20\% & 30\%  \\ \hline \hline
  $P^{\text{NCL}}$ (\%) &	61.51\% &	54.25\% &	39.12\% \\ 
  $P^{\text{CL}}$ (\%)  & 100\% & 87.54\% & 64.69\% \\
  $T^{\text{ASD, NCL}}$ (hours)  &  25.54 (7.61)  & 22.24 (8.57) & 17.15 (5.78) \\
  $T^{\text{ASD, CL}}$ (hours)  &  48.00 (0) & 42.60 (0.89) &  26.80 (0.44) \\
  $T^{\text{AID, NCL}}$ (hours)  &  22.46 (7.61) & 25.76 (8.57) & 30.85 (5.78) \\
  $T^{\text{AID, CL}}$ (hours)  &  0 (0) &  5.4 (0.89) &  21.20 (0.44) \\
  $P^{\text{PV, Total}}$ (\%)  &  70.427 &	62.86 &	57.88 \\
  $F^{\text{DG, Balance}}$ (\%)  &  46.57 &	54.92 &	68.99 \\
  $SOC^{\text{ES, Balance}}$ (\%)  &  25.50 & 22.42 & 23.77 \\
  $T^{\text{RL,ES}} (\%)$  & 3.64 &  35.76 & 56.42 \\
  $P^{\text{Imb.}}$ (\%)  &  8.98 &	9.48 &	10.11 \\
  %$P^{\text{DG, Imb.}}$ (\%)  &  1.11 &	1.08 &	1.12 \\
  $T^{\text{CMG,OFF}}$ (hours)  &  0 & 3 & 21 \\
  $a$ & 0.16 (0.19) & 0.02 (0.13) & -0.07 (0.21)\\ \hline
\end{tabular}
\end{table*}

\begin{table*}[]
\centering
\small
\caption{Metrics for different forecast error scenarios in case FE2 without delayed recourse.}
\label{tab:metrics_forecasterrorrandom_norecourse}
\begin{tabular}{c|c|c|c}
\hline
Error Scenario & 10\% & 20\% & 30\%  \\ \hline \hline
  $P^{\text{NCL}}$ (\%) &	64.64\% &	46.91\% & 38.73\% \\
  $P^{\text{CL}}$ (\%)  & 100\% & 63.62\% & 64.69\% \\
  $T^{\text{ASD, NCL}}$ (hours)  &  26.12 (9.96)  & 18.36 (6.34) & 17.42 (6.22) \\
  $T^{\text{ASD, CL}}$ (hours)  &  48.00 (0) & 28.80 (0.44) &  27.00 (0) \\
  $T^{\text{AID, NCL}}$ (hours)  &  21.87 (9.96) & 29.64 (6.34) & 30.58 (6.22) \\
  $T^{\text{AID, CL}}$ (hours)  &  0 (0) &  19.20 (0.44) &  21.00 (0) \\
  $P^{\text{PV, Total}}$ (\%)  &  73.83 &	63.07 &	61.13 \\
  $F^{\text{DG, Balance}}$ (\%)  &  44.52 & 65.97 &	70.80 \\
  $SOC^{\text{ES, Balance}}$ (\%)  &  24.47 & 22.34 & 21.65 \\
  $T^{\text{RL,ES}} (\%)$  & 15.27 &  54.51 & 58.33 \\
  $P^{\text{Imb.}}$ (\%)  &  9.12 &	9.98 &	10.29 \\
  %$P^{\text{DG, Imb.}}$ (\%)  &  1.04 &	1.07 &	1.11 \\
  $T^{\text{CMG,OFF}}$ (hours)  &  0 & 19 & 21 \\\hline
\end{tabular}
\end{table*}

\subsection{Performance Under Variable Outage Start Time and Duration}
This analysis aims to demonstrate the applicability of the proposed SA-HMTS framework to different outage scenarios characterized by outage start time and total outage duration. Due to the effect of temporal factors on the total system load and the ES charge-discharge cycle, it is essential to validate the SA-HMTS decisions under different time-based scenarios. Hence, for obtaining scenarios on outage start time variation, the start time is varied in increments of three hours by keeping the outage duration fixed at $48$ hours. To analyze the impact of outage duration, the total duration is varied with increments of six hours by keeping the outage start time fixed at midnight. 

\begin{figure}[htb]
\vspace{-0.0cm}
  \centering
  \includegraphics[width = \linewidth ,keepaspectratio, trim={1cm 3cm 0.5cm 1cm},clip]{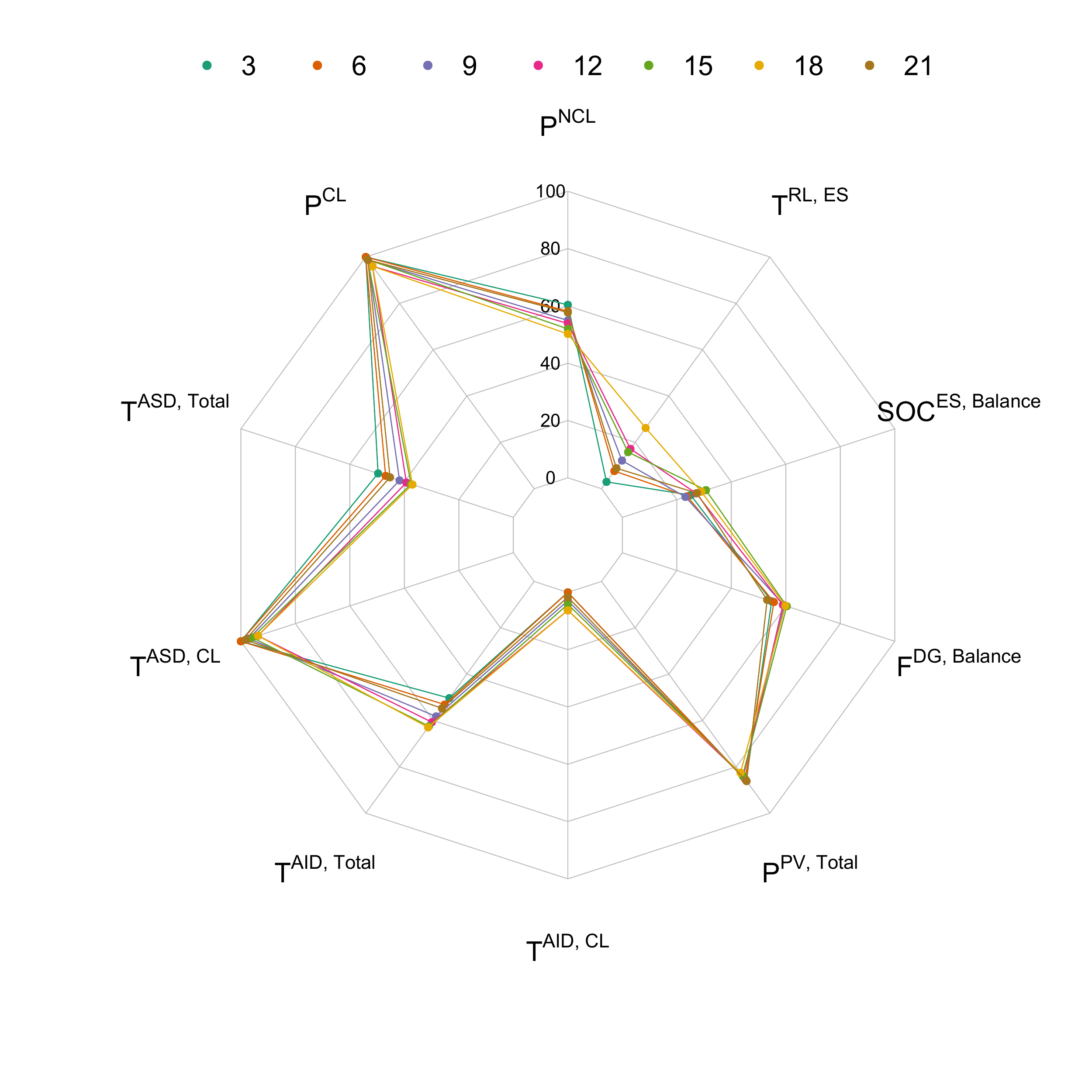}
  \vspace{-0.2cm}
  \caption{Visualizing metrics of outage start time variation analysis.}
  \label{fig:metrics_starttimeanalysis}
  \vspace{-0cm}
\end{figure}

Table \ref{tab:metrics_starttimevar} shows the metrics for outage scenarios with different start times. The similarity and dissimilarity of specific metrics over different outage start time values can be visualized using Fig. \ref{fig:metrics_starttimeanalysis}. We observe that the total load supplied follows a decreasing trend as the start time changes from hour $3$ to hour $18$, and then an increasing trend is observed. The load supply and interruption duration metrics are accordingly impacted. $95\%$ and above CL has been supplied under all cases, with the drop in supply occurring during the intervals the CMG had to be switched off. This is caused by the difference in the duration for which the system operates without any PV generation support immediately after the onset of the outage and the total system load corresponding to the outage start time. 

\begin{figure}[htb]
\vspace{-0.0cm}
  \centering
  \includegraphics[width = \linewidth ,keepaspectratio, trim={1cm 3cm 0.5cm 0.5cm},clip]{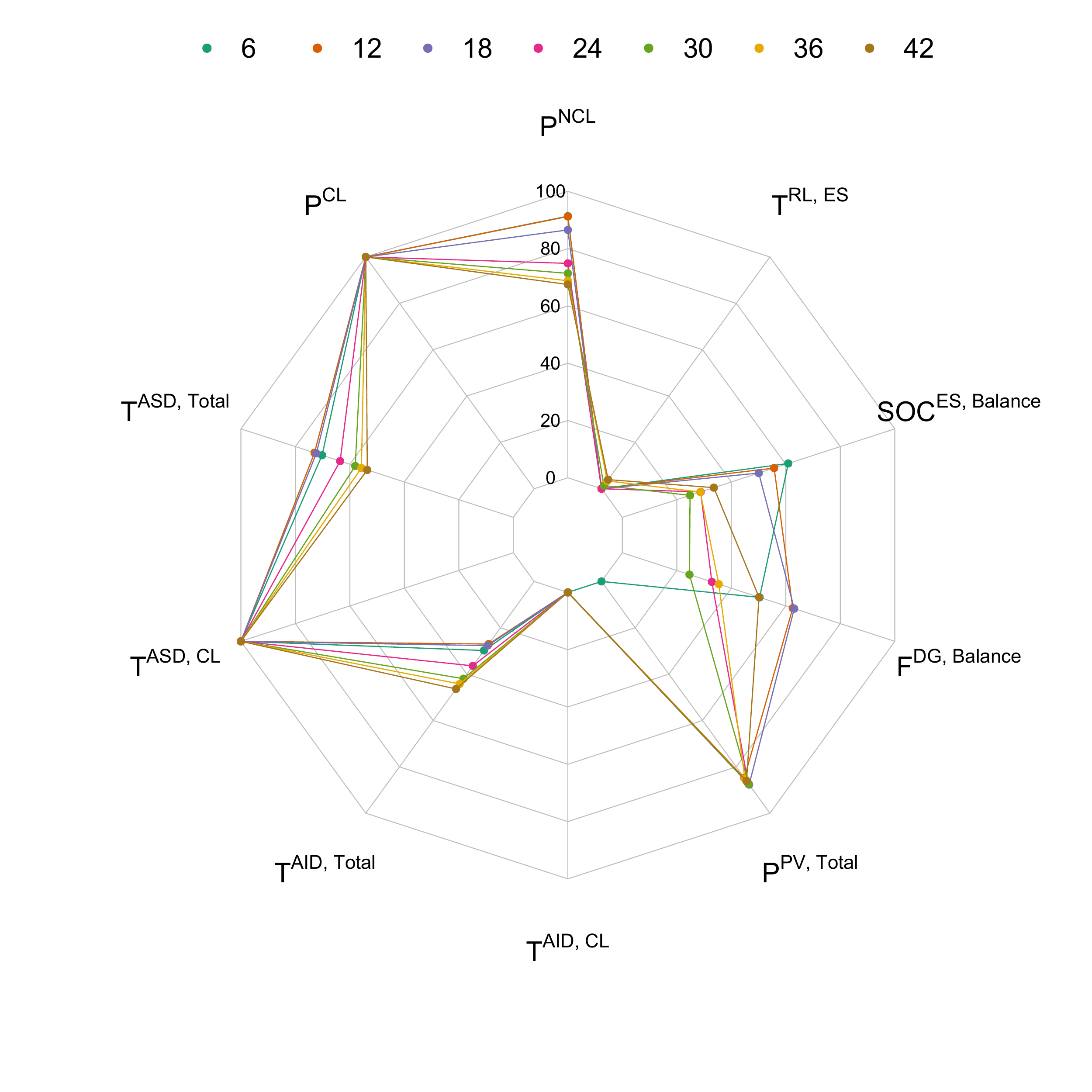}
  \vspace{-0.2cm}
  \caption{Visualizing metrics of outage duration variation analysis.}
  \label{fig:metrics_durationanalysis}
  \vspace{-0cm}
\end{figure}

Table \ref{tab:metrics_durationvar} shows the metrics for scenarios with different outage duration keeping the start time fixed at midnight. The similarity and dissimilarity of specific metrics over different outage duration values can be visualized using Fig. \ref{fig:metrics_durationanalysis}. For a shorter duration, the impact of PV generation is not very significant. However, as the duration increases, the effect of PV generation is seen from the load supply metrics. Over $90\%$ of the total load was supplied for cases with outage duration of $12$ hours and under. The entire load was not supplied in these cases due to the load imbalance mitigation constraints, which prevented supplying the total load demand due to the violation of network phase imbalance limits. Further, with an increase in the outage duration, the total load supply is observed to be decreasing, eventually reaching the base case value having an outage duration of $48$ hours. Under all cases, the CL supply is not compromised. Summarizing the above two case studies, we can conclude that the proposed SA-HMTS framework is not significantly impacted by the outage start time and duration and satisfies the goals of load supply maximization, CL prioritization, and maximization of the duration of secure operation for all the cases. Thus, this analysis validates the generalizability from the temporal features of the outage. 

\begin{table*}[]
\centering
\small
\caption{Metrics for outage start time variation analysis.}
\label{tab:metrics_starttimevar}
\begin{tabular}{c|c|c|c|c|c|c|c}
\hline
Start Hour & 3 & 6 & 9 & 12 & 15 & 18 & 21  \\ \hline \hline
  $P^{\text{NCL}}$ (\%) &	60.38\%  &58.10\%  &54.92\%  &53.85\%  &51.97\%  &50.21\%  &57.70\%  \\ 
  $P^{\text{CL}}$ (\%)  & 100\%  &100\%  &98.68\%  &96.03\%  &98.87\%  &95.994\%  &98.759\%\\
  $T^{\text{ASD, NCL}}$ (hours)  &  23.83   (9.86)  &22.50   (9.89)  &20.05   (8.71)  &18.85   (9.09)  &18.02   (9.57)  &17.77   (8.82)  &21.72   (9.78)\\
  $T^{\text{ASD, CL}}$ (hours)  &  48   (0)  &48   (0)  &46.60  (0.89)  &45   (0)  &46   (0)  &45   (0)  &47   (0) \\
  $T^{\text{AID, NCL}}$ (hours)  &  24.17   (9.86)  &25.50   (9.89)  &27.95   (8.71)  &29.15   (9.09)  &29.98   (9.57)  &30.23   (8.82)  &26.28   (9.78)\\
  $T^{\text{AID, CL}}$ (hours)  &  0 (0) &  0 (0) &  1.4 (0.89) &  3 (0)  &  2 (0)  &  3 (0)  &  1 (0)    \\
  $P^{\text{PV, Total}}$ (\%)  &  85.92  &85.30  &83.95 &84.22  &83.81  &82.54  &86.15 \\
  $F^{\text{DG, Balance}}$ (\%)  &  54.79  &55.62  &59.41  &58.96  &60.36  &59.83  &53.15 \\
  $SOC^{\text{ES, Balance}}$ (\%)  &  25.12  &23.81  &23.14  &27.13  &30.71 &29.01  &27.38 \\
  $T^{\text{RL,ES}} (\%)$  & 2.96 &  7.63 & 12.15 & 17.19 & 15.79 & 26.21 & 8.85\\
  $P^{\text{Imb.}}$ (\%)  &  9.36 &	11.27 &	14.22 &	13.69 &	9.73 &	12.60 &	11.59 \\
  %$P^{\text{DG, Imb.}}$ (\%)  &  1.21 &	1.05 &	1.01 &	1.15 &	1.02 &	1.08 &	1.14 \\ 
  $T^{\text{CMG,OFF}}$ (hours)  &  0 & 0 & 1 & 3 & 2 & 3 & 1  \\\hline
\end{tabular}
\end{table*}

\begin{table*}[]
\centering
\small
\caption{Metrics for outage duration variation analysis.}
\label{tab:metrics_durationvar}
\begin{tabular}{c|c|c|c|c|c|c|c}
\hline
Outage Duration & 6 & 12 & 18 & 24 & 30 & 36 & 42  \\ \hline \hline
  $P^{\text{NCL}}$ (\%) &	91.31\%  &91.21\%  &86.51\%  &74.84\%  &71.35\%  &68.77\%  &67.47\%  \\ 
  $P^{\text{CL}}$ (\%)  & 100\%  &100\%  &100\%  &100\%  &100\%  &100\%  &100\%\\
  $T^{\text{ASD, NCL}}$ (hours)  &  4.21 (1.38)  &8.75 (2.91)  &13.02 (4.39)  &15.25     (5.35)  &17.41  (5.95)  &20.08  (7.54)  &22.51  (8.42)\\
  $T^{\text{ASD, CL}}$ (hours)  &  6 (0) & 12 (0) &  18 (0) &  24 (0) &  30 (0) &  36 (0) &  42 (0)\\
  $T^{\text{AID, NCL}}$ (hours)  &  1.78 (1.38)  &3.24    (2.91)  &4.98     (4.39)  &8.74     (5.35)  &12.59     (5.95)  &15.92     (7.54)  &19.48     (8.42)\\
  $T^{\text{AID, CL}}$ (hours)  &  0 (0) &  0 (0) &  0 (0) &  0 (0)  &  0 (0)  &  0 (0)  &  0 (0)    \\
  $P^{\text{PV, Total}}$ (\%)  &  NA  &84.79  &87.67  &87.03  &87.49  &85.14  &86.10 \\
  $F^{\text{DG, Balance}}$ (\%)  &  50.37  &62.53  &63.09  &32.85  &24.63  &35.45  &50.17\\
  $SOC^{\text{ES, Balance}}$ (\%)  &  60.90  &55.72  &50.07  &28.75  &24.84  &28.62  &33.58 \\
  $T^{\text{RL,ES}} (\%)$  & 0 &  0 & 0 & 0 & 1.38 & 3.38 & 3.93\\
  $P^{\text{Imb.}}$ (\%)  &  8.01  &7.30  &7.07  &10.89  &13.98  &10.66  &9.29 \\
  %$P^{\text{DG, Imb.}}$ (\%)  &  1.02 &	1.15 &	1.11 &	1.09 &	1.06 &	1.08 &	1.11 \\
  $T^{\text{CMG,OFF}}$ (hours)  &  0 & 0 & 0 & 0 & 0 & 0 & 0  \\\hline
\end{tabular}
\end{table*}

\subsection{Performance Under Variable PV Hosting Capacity Factor}
\begin{figure}[htb]
\vspace{-0.0cm}
  \centering
  \includegraphics[width = \linewidth ,keepaspectratio, trim={1cm 3cm 0cm 0cm},clip]{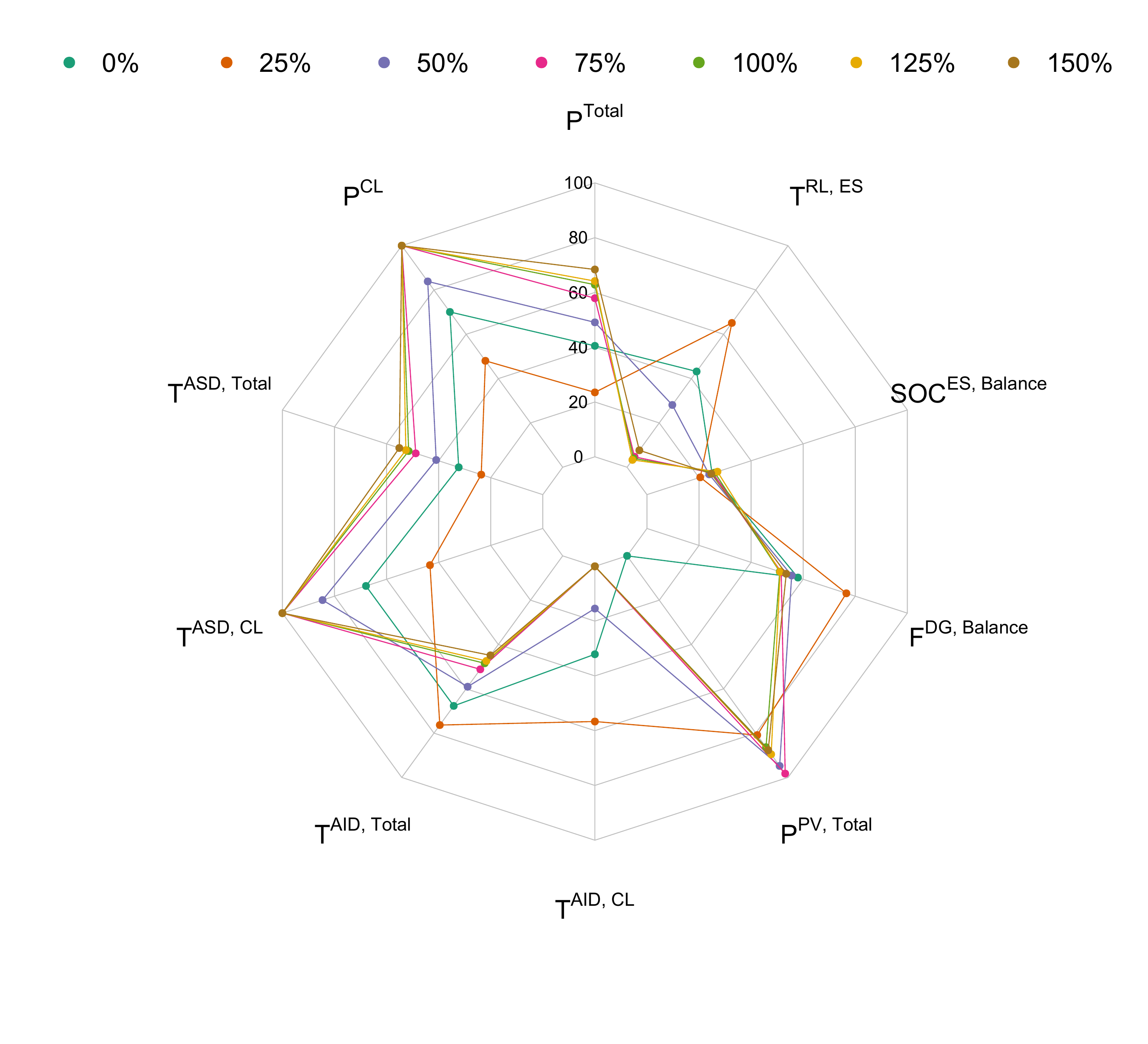}
  \caption{Visualizing metrics of PV hosting capacity variation analysis.}
  \label{fig:metrics_pvvariationanalysis}
  \vspace{-0cm}
\end{figure}

In this section, we analyze the impact of the network PV hosting capacity on the performance of the SA-HMTS framework under extended duration outages. To do so, we scale the ratings of PV units in the network between $0\%$ and $150\%$ in increments of $25\%$ with respect to the base case rating. This case study aims to analyze the impact of PV generation on the decision-making of the proposed framework. Table \ref{tab:metrics_pvhostingvar} shows the values of the metrics computed for the different network PV hosting capacities. With increasing the PV hosting capacity above $25\%$, we observe an increase in the total load supplied and load service duration. For hosting capacities $75\%$ and above, no curtailment of the CL was observed. An interesting observation can be made for the system with no PV generation ($0\%$ of the base hosting capacity). For this case, the total load supplied is greater than the case with $25\%$ PV hosting capacity. This can be addressed to the absence of the PV uncertainty in the SA-HMTS framework. The EDS stage, which runs a stochastic optimization problem, only accounts for the load uncertainty. Not having the uncertainty in PV generation causes the other grid-following generating units to operate at a different setpoint, which results in a higher load supply. This case demonstrates that integrating the uncertainty in modeling comes with a cost. However, for this case, the grid-forming ES unit discharges by the end of the first day, unlike the other cases wherein the PV generation enables the CMG to restart its operation on the second day. Fig. \ref{fig:metrics_pvvariationanalysis} shows a pictorial representation of the metrics under different PV hosting capacities. Overall, we observe that increasing the PV hosting capacity results in an enhanced load restoration when analyzed from the perspective of the different metrics listed in Table \ref{tab:metrics_pvhostingvar}.      

\begin{table*}[]
\centering
\small
\caption{Metrics for variation in network PV hosting capacities.}
\label{tab:metrics_pvhostingvar}
\begin{tabular}{c|c|c|c|c|c|c|c}
\hline
PV Hosting Capacity & 0\% & 25\% & 50\% & 75\% & 100\% & 125\% & 150\%  \\ \hline \hline
  $P^{\text{NCL}}$ (\%) & 40.50\%& 23.512\%  &49.081\%  &57.893\%  &62.83\%  &64.153\%  &68.382\% 	\\ 
  $P^{\text{CL}}$ (\%)  &  70.08\%& 48.033\%  &83.863\%  &100\%  &100\%  &100\%  &100\%  \\
  $T^{\text{ASD, NCL}}$ (hours)  &   15.50 (6.71)& 11.34  (5.17)  &19.65   (8.07)  &23.42  (9.50)  &24.73   (8.12)  &25.24  (9.86)  &26.45 (10.01) \\
  $T^{\text{ASD, CL}}$ (hours)  &  32.60 (1.41) & 20.80   (0.44)  &40.60   (0.54)  &48   (0)  &48   (0)  &48  (0)  &48 (0) \\
  $T^{\text{AID, NCL}}$ (hours)  &  32.50 (6.71) & 36.66 (5.17)  &28.35  (8.07)  &24.58 (9.50)  &23.27  (8.12)  &22.76  (9.86)  &21.55  (10.01) \\
  $T^{\text{AID, CL}}$ (hours)  & 15.40 (1.41)  & 27.2  (0.44)  &7.4  (0.54)  &0 (0)   &0 (0)   &0  (0)  &0  (0)\\
  $P^{\text{PV, Total}}$ (\%)  &  0& 80.88  &94.8  &98.26  &86.34  &89.541  &87.713 \\
  $F^{\text{DG, Balance}}$ (\%)  & 57.98 & 76.57  &55.57  &51.56  &51.65  &51.17  &53.49\\\
  $SOC^{\text{ES, Balance}}$ (\%)  &  25.01 & 20.47  &23.91  &25.38  &26.03  &27.08  &24.70\\
  $T^{\text{RL,ES}} (\%)$  &  43.22& 65.12  &28.12  &4.68  &4.12  &3.31  &7.64\\
  $P^{\text{Imb.}}$ (\%)  &   10.30& 10.71 & 11.39 & 7.97& 7.12& 8.99& 9.76\\
  %$P^{\text{DG, Imb.}}$ (\%)  & 1.06 & 1.12 &  1.08& 1.09& 1.11& 1.04& 1.05\\
  $T^{\text{CMG,OFF}}$ (hours)  &  14& 19& 5& 0& 0& 0& 0\\\hline
\end{tabular}
\end{table*}

\subsection{SA-HMTS Framework Performance Analysis in HIL Environment}
Until now, all the simulations performed were validated using OpenDSS. Although the OpenDSS results are very promising, the real-world implementability of the proposed approach is not entirely demonstrated since OpenDSS runs on a snapshot-based approach and does not capture the dynamics and transients. Hence, we perform a co-simulation study by deploying the decisions of the SA-HMTS to a HIL simulated IEEE 123 bus system. Fig. \ref{fig:hil_communication} shows the co-simulation setup and the data interactions between the CMG-EMS system and the HIL system. Additional details on the HIL modeling for the network described in Fig. \ref{fig:ieee123} can be found in \cite{hil1, hil2, hil3}. %Since the vital purpose of this setup is to analyze the dynamic performance of the system when operated using the SA-HMTS framework, we focus on the results showing the operating frequency and node voltages. 

\begin{figure}[htb]
\vspace{-0.0cm}
  \centering
  \includegraphics[width = 0.96\linewidth ,keepaspectratio, trim={0cm 0cm 0cm 0cm},clip]{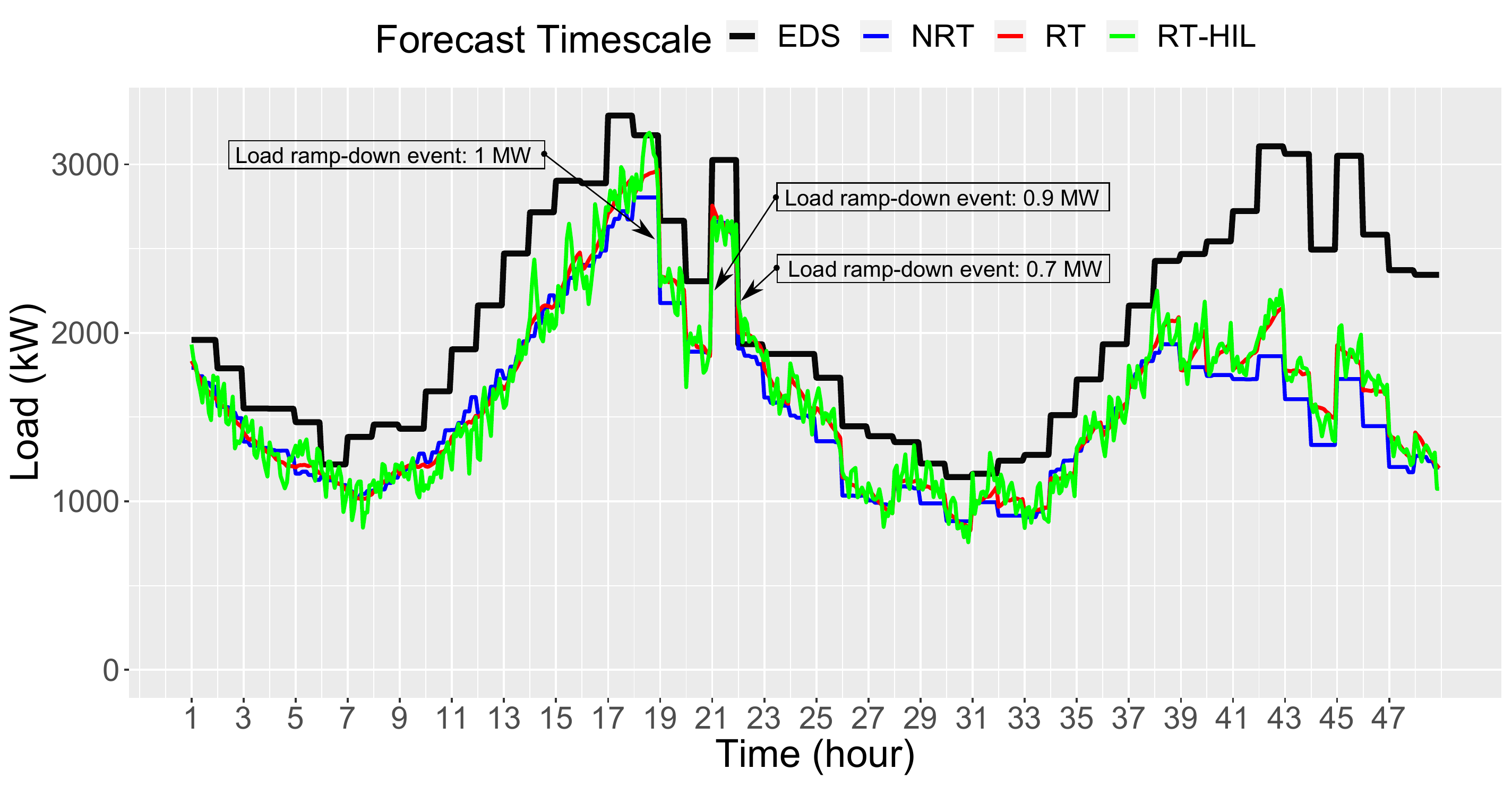}
  \vspace{-0.2cm}
  \caption{SA-HMTS load allocation and HIL RT realization.}
  \label{fig:basecase_recourse_loadallocation_hil}
  \vspace{-0cm}
\end{figure}

\begin{figure}[htb]
\vspace{-0.0cm}
  \centering
  \includegraphics[width = 0.96\linewidth ,keepaspectratio, trim={0cm 0cm 0cm 0cm},clip]{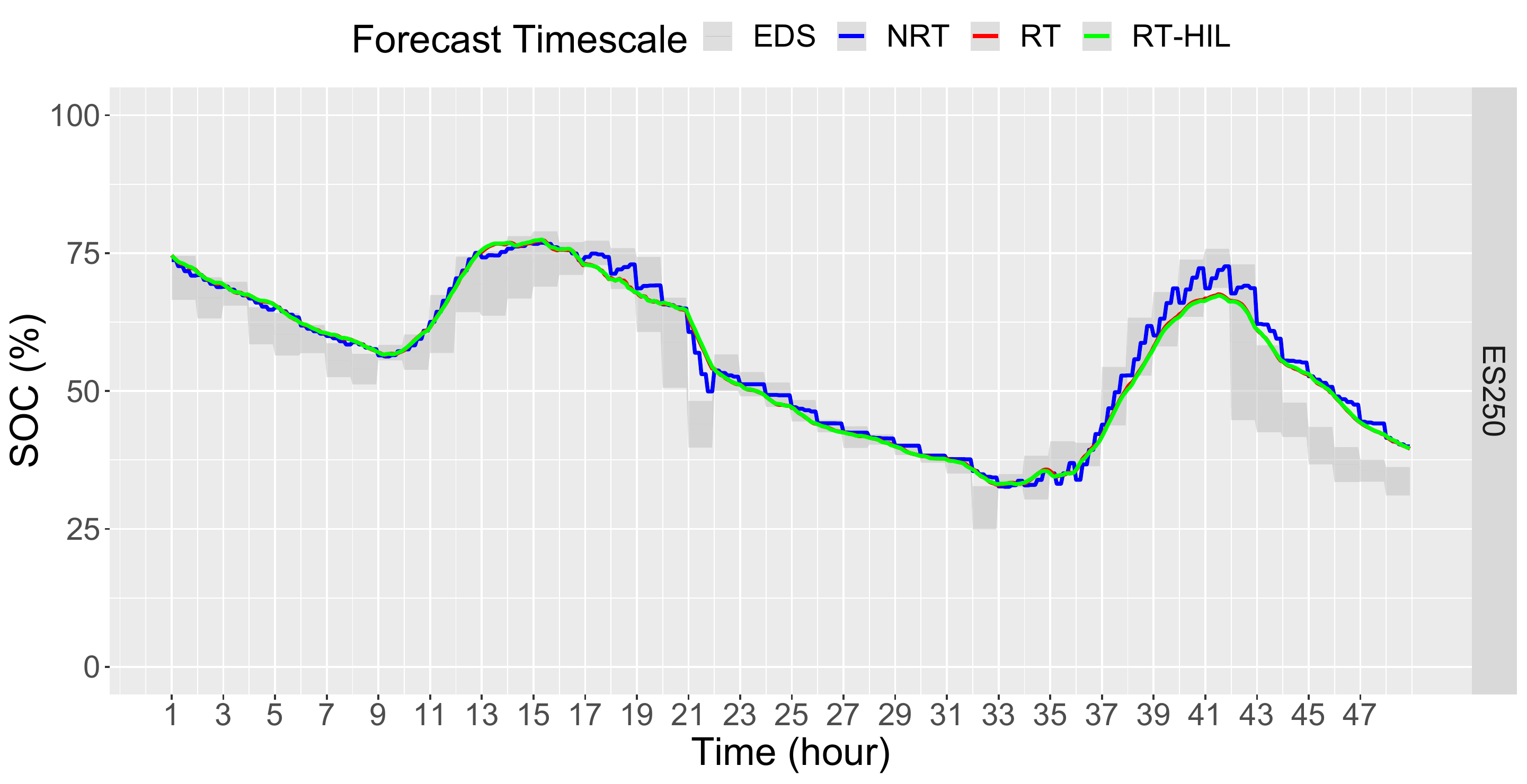}
  \vspace{-0.2cm}
  \caption{Grid forming ES250 reference SOC and HIL RT realization.}
  \label{fig:basecase_recourse_gfmbess_hil}
  \vspace{-0cm}
\end{figure}

To begin with, we run two HIL co-simulations, one with and one without the consideration of delayed recourse. Note that the cold load modeling in not accounted for in this simulation. Hence, the forecast error stemming from cold load is eliminated. The metrics are shown in Table \ref{tab:metrics_basecaseanalysis_hil}. Fig. \ref{fig:basecase_recourse_loadallocation_hil} shows the SA-HMTS load allocation and HIL RT supply throughout the outage duration using delayed recourse. Since the HIL model represents real-world operation, we observe increased load fluctuations. The impact of delayed recourse is evident from this figure, wherein the forecast error observed prior to hour $19$ has been compensated by using additional load curtailment between hours $19$ and $22$. A similar impact is observed for hours $26-28$ and $42-48$. Fig. \ref{fig:basecase_recourse_gfmbess_hil} shows the SOC of grid forming ES250. Since cold not is not incorporated here, the forecast error is lower than the one observed in the equivalent OpenDSS simulations listed above. Due to this, the metrics with and without delayed recourse differ marginally. This shows that the delayed recourse parameters have been correctly selected. We observe that the SOC follows a trend similar to the base case OpenDSS simulation shown in Fig. \ref{fig:basecase_recourse_gfmbess} with the only difference stemming from the minor forecast error and the integration of cold load. Fig. \ref{fig:basecase_recourse_gfmbess_hil} shows the comparison between the power interaction of the grid-forming ES as computed by the SA-HMTS RT stage and that observed in HIL simulation. The grid-forming ES handles the demand-supply imbalance observed in RT. We observe that the grid-forming ES interactions observed in HIL are similar to those computed by the SA-HMTS RT stage.

\begin{figure}[htb]
\vspace{-0.0cm}
  \centering
  \includegraphics[width = 0.96\linewidth ,keepaspectratio, trim={0cm 0cm 0cm 0cm},clip]{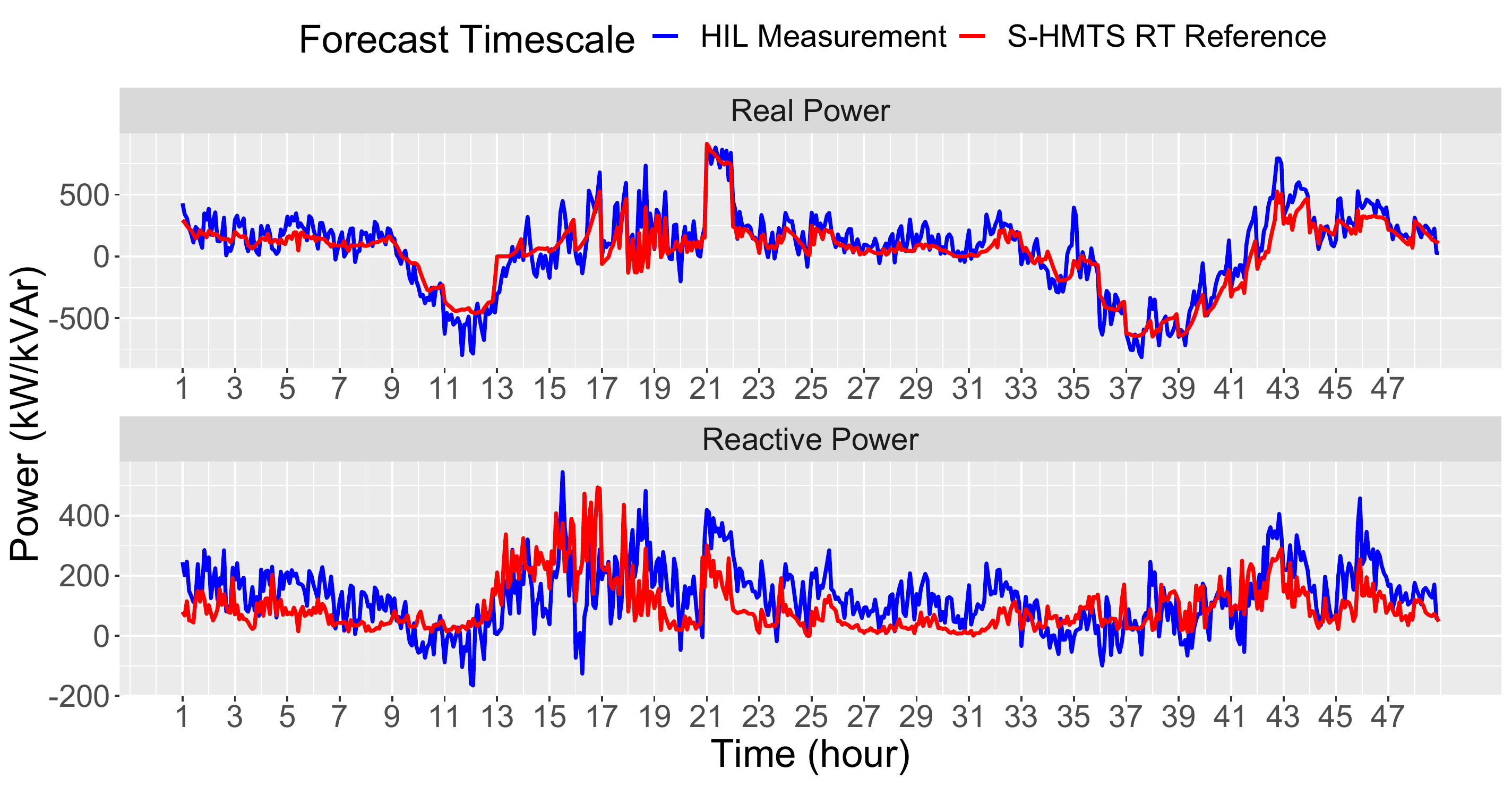}
  \vspace{-0.2cm}
  \caption{Grid forming ES250 SA-HMTS RT stage reference and HIL RT realization of real and reactive power.}
  \label{fig:basecase_hmtsrt_hilrt_voltage}
  \vspace{-0cm}
\end{figure}

\begin{figure}[htb]
\vspace{-0.0cm}
  \centering
  \includegraphics[width = 0.96\linewidth ,keepaspectratio, trim={0cm 0cm 0cm 0cm},clip]{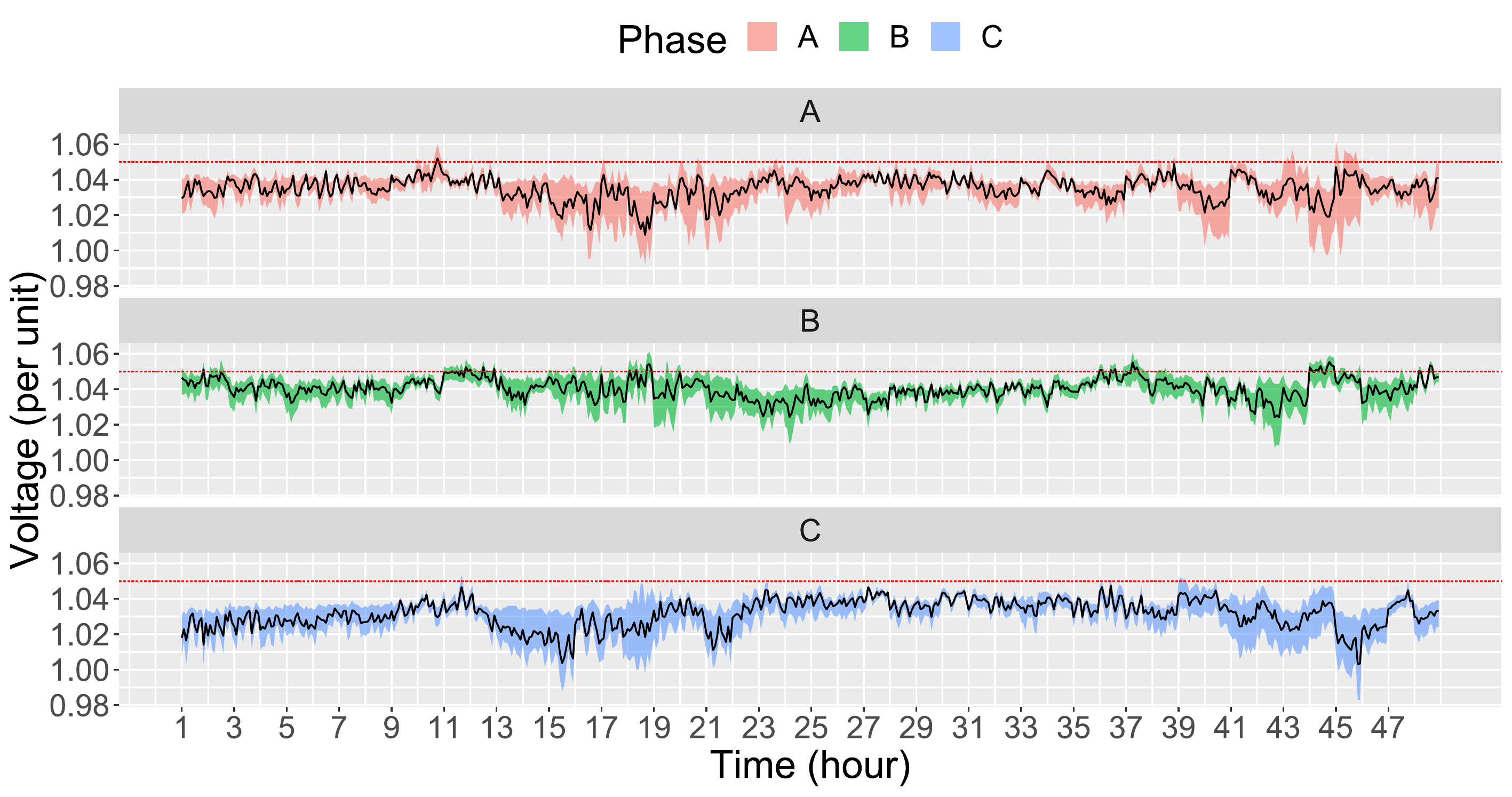}
  \vspace{-0.2cm}
  \caption{Average node voltage and the range across all nodes.}
  \label{fig:basecase_hmtsrt_hilrt_voltagetime}
  \vspace{-0cm}
\end{figure}

Given that this simulation considers the lowest forecast error, the similarity between the SA-HMTS values and the HIL measurements validates the exactness of the two models. It also demonstrates that an RT dispatch at every $5$ minute intervals is sufficient to manage the CMG operation smoothly. %A higher mismatch exists for the grid-forming ES unit reactive power, which is mainly caused by the network model mismatch between the linearized one used in the SA-HMTS model and the actual HIL model. 
Fig. \ref{fig:basecase_hmtsrt_hilrt_voltage} and Fig. \ref{fig:basecase_hmtsrt_hilrt_voltagetime} show the voltage observed throughout the outage duration at all the network nodes. We observe that the voltage observed across all the network nodes varies within small bounds, especially between the ANSI specified limits of $0.95$ and $1.05$ p.u. The voltage variations across the nodes are primarily observed during time intervals with PV generation or when the generation resources reach near exhaustion towards the end of the outage period. The necessity of any voltage regulators or capacitor banks is not observed because the voltage stays within limits across all the network nodes. Hence, these devices are not modeled in the SA-HMTS framework. Further, these devices will operate using local control in the HIL system and take decisions independently based on the local measurements. 

Lastly, the load supply will have sudden ramps due to the forecast errors, cold load, or the proactive curtailment using delayed recourse. Care needs to be taken that the system can handle these sudden changes in load without losing its stability or having impacts on the network voltage. Fig. \ref{fig:basecase_recourse_loadallocation_hil} shows some steep load ramps caused by the delayed recourse constraint. A drop in load is observed during hour $19$, and a jump in load is observed during hour $21$. These sudden load changes have a magnitude of $\pm 1$ MW. Fig. \ref{fig:basecase_hmtsrt_hilrt_voltage} shows some voltage variations corresponding to these two hours; however, the voltage magnitudes do no deviate off-limits. From Fig. \ref{fig:basecase_recourse_gfmbess}, we also observe that the grid-forming ES unit smoothly supports such load ramps without negatively affecting the system operation. Further, at system startup, we observe that the HIL system can supply a starting load of around $1.8$ MW. This ensures that any additional shutdown and startup events occurring in the middle of the system operation can be smoothly handled.

\begin{table}[]
\centering
\small
\caption{Metrics for base case analysis in HIL environment.}
\label{tab:metrics_basecaseanalysis_hil}
\begin{tabular}{c|c|c}
\hline
Metric & With recourse & Without recourse  \\ \hline \hline
 $P^{\text{NCL}} (\%)$ & 73.81 & 76.90\\ 
  $P^{\text{CL}} (\%)$  & 100  & 100\\
  $T^{\text{ASD, NCL}}$ (hours)  &  29.81 (10.21)  & 30.51 (9.85) \\
  $T^{\text{ASD, CL}}$ (hours)  &  48 (0) & 48 (0)\\
  $T^{\text{AID, NCL}}$ (hours)  &   18.19 (10.21)   & 17.49 (9.85) \\
  $T^{\text{AID, CL}}$ (hours)  &  0 (0)   & 0 (0)\\
  $P^{\text{PV, Total}} (\%)$  &  90.25 & 92.15\\
  $F^{\text{DG, Balance}} (\%)$  &  49.25 & 48.67\\
  $SOC^{\text{ES, Balance}} (\%)$  &  39.74  & 40.06\\
  $T^{\text{RL,ES}} (\%)$  &  0   &  0\\
  $P^{\text{Imb.}} (\%)$  &  8.25  & 10.11\\
  %$P^{\text{DG, Imb.}} (\%)$  &  1.12  & \\ 
  $T^{\text{CMG,OFF}}$ (hours) & 0   & 0\\
  $a$ & 0.10 (0.08)  &  - \\ \hline

\end{tabular}
\begin{tablenotes}
\centering
\item[*] Note: Cold load impact is not considered.
\end{tablenotes}
\end{table}

\begin{figure}[htb]
\vspace{-0.0cm}
  \centering
  \includegraphics[width = 0.96\linewidth ,keepaspectratio, trim={0.25cm 0cm 0.3cm 0.25cm},clip]{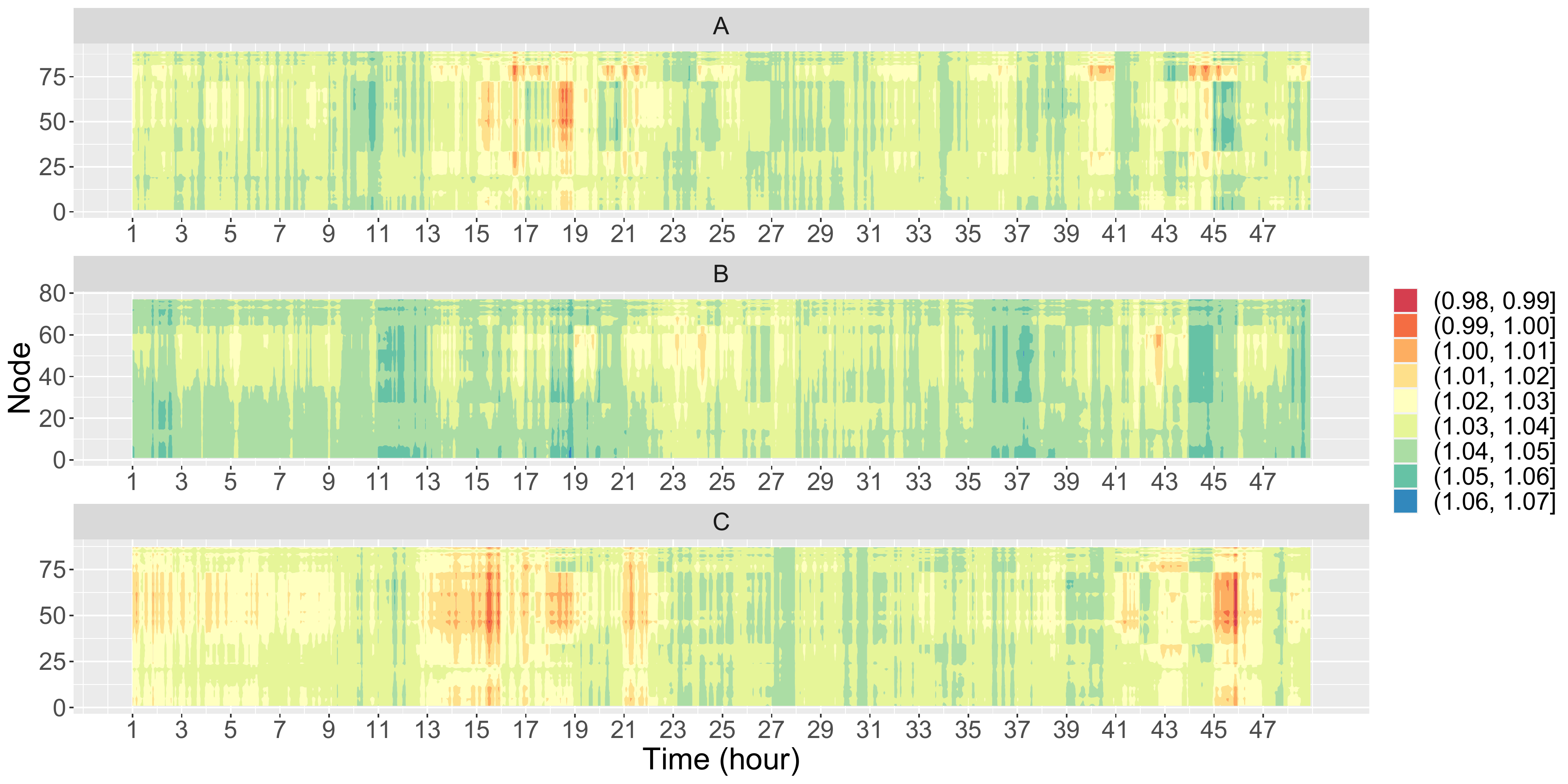}
  \vspace{-0.2cm}
  \caption{Phase-wise nodal voltage variation throughout the outage duration.}
  \label{fig:basecase_hmtsrt_hilrt_gfmpq}
  \vspace{-0cm}
\end{figure}

\subsection{SA-HMTS Framework Computation Time Performance}
\begin{figure} [tb]
%\captionsetup[subfigure]{labelformat=empty}
\centering
  \includegraphics[width=1\linewidth,keepaspectratio, trim={0cm 0cm 0cm 0.25cm},clip]{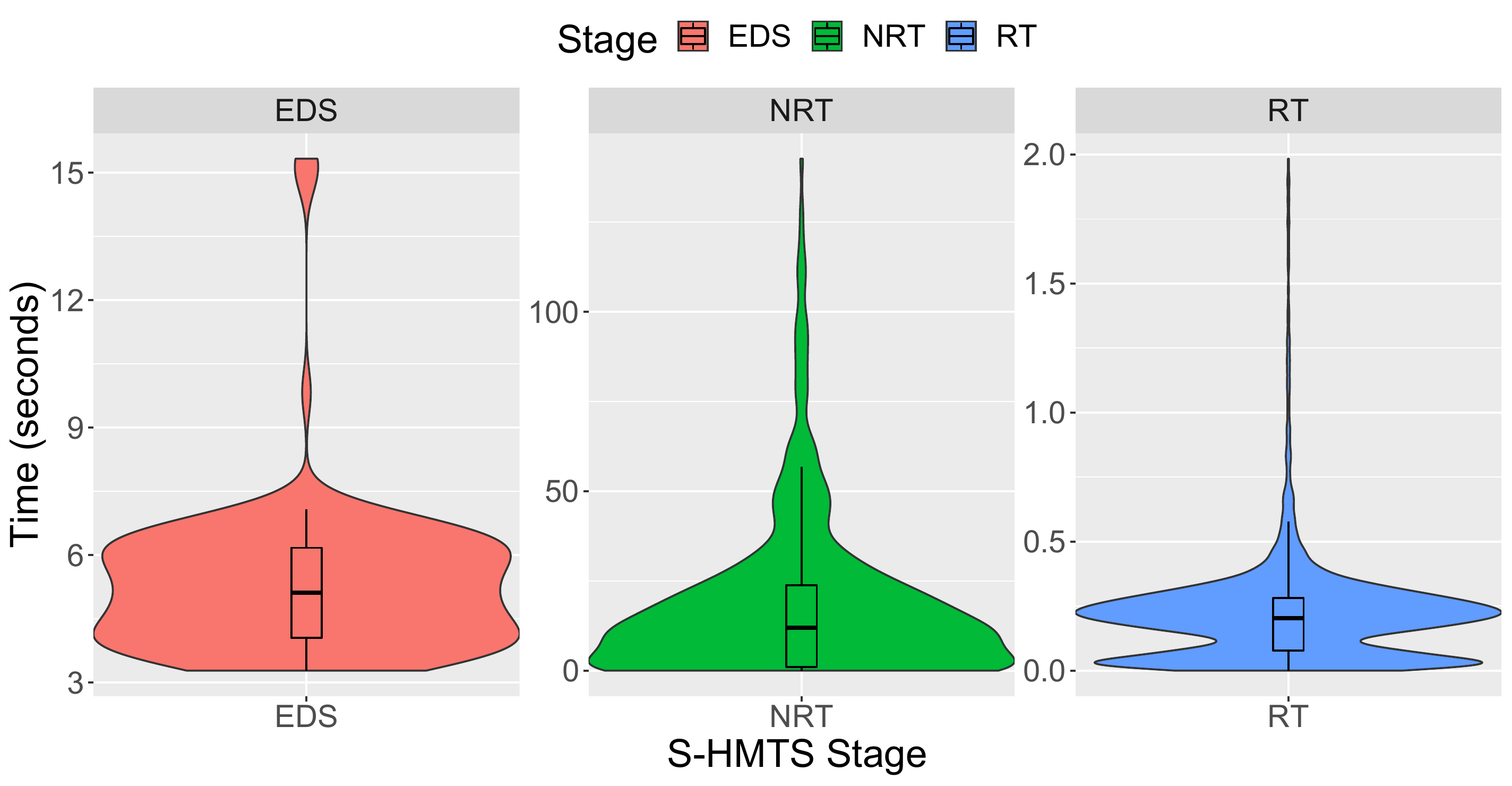} \\
  \caption{SA-HMTS stage-wise computation time.}
  \label{fig:computation_time} 
  \vspace{-0.5cm}
\end{figure}

This section analyzes the computation time requirements of each stage of the SA-HMTS framework. We gather the computation time of each stage of the SA-HMTS framework for all the different simulations performed above. The violin plot showing the distribution of the computation time is shown in Fig. \ref{fig:computation_time}. The average computation time for EDS, NRT, and RT stages is $5.33$ s, $18.40$ s, and $0.22$ s, respectively. The maximum time required to compute each stage is $15.11$ s for the EDS stage, $147.12$ s for the NRT stage, and $2$ s for the RT stage. Given that the EDS stage and NRT stages are computed at least 15 minutes in advance, the above values of computation time are reasonable to obtain the computed decisions and communicate them to the field devices in time. The average computation time of $0.22$ s for the RT stage ensures that the RT dispatch can be smoothly conducted at an interval of $5$ minutes.   

%%%%%%%%%%%%%%%%%%%Conclusion%%%%%%%%%%%%%%%%%%
\vspace{-0cm}
\section{Conclusion}
This paper proposes an SA-HMTS framework for proactive and resilient scheduling and dispatch of CMGs during emergency conditions. A buffer NRT stage between the EDS scheduling and the RT dispatch is added to update the EDS schedule using newly obtained forecasts closer to the actual RT dispatch time. The CMG scheduling and dispatch constraints are implemented at different hierarchical stages based on the relevance of a particular constraint to a specific timescale, which minimizes the computational burden of each stage and allows room to incorporate any unforeseen impacts that unfold as timescale moves closer to RT dispatch. Furthermore, a novel $n$-delayed recourse approach is proposed for proactively handling the impact of forecast error and ensure a secure and continuous operation of the system. Our results using OpenDSS and HIL simulations show that for different outage temporal variations and extreme forecast error scenarios, the SA-HMTS decisions result in maximized load supply by prioritizing CL, maximize continuous operating duration, and ensure secure operation with adequate reserve availability. Our future work will involve more comprehensive criteria for making decisions on CMG boundary expansion, analyzing performance under cyber-security attacks, test under numerous communication failure scenarios, and expand the framework for operating multiple CMGs using a distributed multi-agent control approach.   
%%%%%%%%%%%%%%%%%%%Acknowledgement%%%%%%%%%%%%%
\vspace{-0.2cm}
\section*{Acknowledgement}
The authors thank PJ Rhem with ElectriCities, Paul Darden, Steven Hamlett, and Daniel Gillen with Wilson Energy for their inputs, suggestions, and technical guidance.
\bibliographystyle{IEEEtran}
% \bibliography{IEEEabrv,references}
\vspace{-0.20cm}
\bibliography{references}

\appendix

  \subsection{Hexagonal Relaxation}
  \label{appendix:Hex}
The quadratic constraints relating to the apparent power limits of the different generators are linearized to ensure a LP formulation. The generator apparent power limit equation takes the form of a Euclidean ball. The idea behind using polygon approximation is to express the Euclidean ball using a set of linear equations. Below, we show how to linearize the entire Euclidean ball. However, in the HMTS model, the constraints that linearize the specific quadrants of the Eucldean ball representing the operating quadrants of the respective generators are used. In the equations below, $\tau=1.1$, is a coefficient that determines the exactness of the linearization.

\begin{subequations}
\begin{gather}
P^2 + Q^2 \le S^2 \\
-\sqrt 3 (P + \tau S) \le Q \le  -\sqrt 3 (P - \tau S) \\
-\frac{\sqrt 3}{2} \tau S \le Q \le \frac{\sqrt 3}{2} \tau S \\
\sqrt3 (P - \tau S) \le Q \le \sqrt 3 (P + \tau S) \\
- \frac{\sqrt 3}{2}\tau S \le P \le \frac{\sqrt 3}{2}\tau S.
\end{gather}
\end{subequations}

\subsection{Chance Constraint Approximation}
  \label{appendix:cc_approx}
We use the scenario method for for the chance constraint approximation \cite{cc_scenarioapprox2}. Below, we show the methodology for the constraint approximation. Let $\bm{x}$ be the uncertain variable having the feasible set $P$. Hence,
\begin{align}
    \text{Pr}(\bm{x} \in P) = \int_{\bm{x}\in P} f(\bm{x}) d\bm{x} \ge 1-\alpha,  \label{eq:cc_act_formulation}
\end{align}
where, $f(\bm{x})$ is the unknown probability density function of the random variable $\bm{x}$ and $\alpha$ is the maximum constraint violation probability threshold. $f(\bm{x})$ can be then approximated as:
\begin{align}
    f(\bm{x}) \simeq \frac{1}{s}\sum_{s\in \Omega} \delta(\bm{x}-\bm{d}_s),  \label{eq:probab_fcn_approx}
\end{align}
where, $s \in \Omega$ are the scenarios, $\bm{d}_j$ are the samples from $\bm{x}$, and $\delta$ is the dirac function. From (\ref{eq:cc_act_formulation}) and (\ref{eq:probab_fcn_approx}), we get:
\begin{align}
   \text{Pr}(\bm{x} \in P) = \frac{1}{s}\sum_{s\in \Omega} \bm{I}_P(\bm{d}_j),  \label{eq:cc_new_formulation}
\end{align}
 where, $\bm{I}_P(\bm{x}) = 1$ if $\bm{x} \in P$, and vice-versa. Combining the above equations, we get the following final approximation:
\begin{align}
    \text{Pr}(\bm{x} \in P) \simeq \frac{1}{s}\sum_{s\in \Omega} \bm{I}_P(\bm{d}_j) \ge 1-\alpha.  \label{eq:final_formulation}
\end{align}

\subsection{Proof of Theorem \ref{Thm1}}
\label{Thm1Proof}

\begin{proof}
Given a finite total outage duration of $T$ time intervals, let the EDS load allocation (in kW) obtained for time interval $t$ and $T-t$ future time intervals be defined by $\mathbf{\overline y^{\text{EDS}}_{t}} \in \mathbb{R}^{T-t+1 \times 1}_{+}$. Similarly, let ${\overline y^{\text{NRT}}_{t}} \in \mathbb{R}^{1 \times 1}_{+}$ and ${\overline y^{\text{RT}}_{t}} \in \mathbb{R}^{1 \times 1}_{+}$ be the NRT load allocation and RT load realization. Herein, let RT stage be interpreted as the HIL/OpenDSS simulation step. Following the structure of HMTS, let $\mathbf{\overline x^{\text{EDS}}_{t}} \in \mathbb{R}^{T-t+1 \times 1}$ be the vector of allocated generation (in kW) for the future time horizon as computed at the $t^\text{th}$ time interval computed in a receding horizon manner, ${\overline x^{\text{NRT}}_{t}} \in \mathbb{R}^{1 \times 1}$ be the NRT revised EDS generation allocation for the $t^\text{th}$ time interval, and ${\overline x^{\text{RT}}_{t}} \in \mathbb{R}^{1 \times 1}$ be the RT generation used. For simplicity, assume no PV generation and the ES units only operate in discharging mode. The generation allocation must equal the load allocation. Let $X$ kWh be the total available generation resources. For simplicity, assume no forecast error between the EDS and NRT stages and the time interval ($\Delta t$) is one hour. Hence,  ${\overline x^{\text{NRT}}_{t}} = \mathbf{\overline x^{\text{EDS}}_{t}}(1)$. Due to the positive load forecast error of $\gamma^{\text{FE}} \% $, the RT load that needs to be supplied in full equals $(1+0.01\gamma^{\text{FE}})\mathbf{\overline y^{\text{EDS}}_{t}}(1)$. Hence, the RT generation usage is given by ${\overline x^{\text{RT}}_{t}} = (1+0.01\gamma^{\text{FE}})\mathbf{\overline x^{\text{EDS}}_{t}}(1)$, i.e., ${\overline x^{\text{RT}}_{t}} > \mathbf{\overline x^{\text{EDS}}_{t}}(1)$. The excess generation consumption of $0.01\gamma^{\text{FE}}\mathbf{\overline x^{\text{EDS}}_{t}}(1)$ is observed, and the unused generation capacity, $X_t$, equals $X - (1+0.01\gamma^{\text{FE}})\mathbf{\overline x^{\text{EDS}}_{t}}(1)$.

Continuing this trend by incrementing the time intervals by 1, and assuming that the entire available generation is used up at the end of the final time interval, the available unused generation at the end of time interval $t < T$, $X_t$, and the total generation capacity, $X$, can be generalized as follows: 
\begin{subequations}
\begin{gather}
{X_t = X - \sum_{i=1}^{t} \overline x^{\text{RT}}_{i} = X - (1+0.01\gamma^{\text{FE}})\sum_{i=1}^{t} \mathbf{\overline x^{\text{EDS}}_{i}}(1)}, \label{eq:proof_availale_gen} \\
{X = (1+0.01\gamma^{\text{FE}})\sum_{i=1}^{t-1} \mathbf{\overline x^{\text{EDS}}_{i}}(1) + \sum_{j=1}^{T-(t-1)} \mathbf{\overline x^{\text{EDS}}_{t}}(j)}, \label{eq:proof_total_gen1} \\ {= \sum_{i=1}^{t} \mathbf{\overline x^{\text{EDS}}_{i}}(1) + X^{\text{err}}_{t} + X^{\text{bal}}_{t}}. \label{eq:proof_total_gen2}
\end{gather}
\end{subequations}
From (\ref{eq:proof_total_gen1}) and(\ref{eq:proof_total_gen2}), we can show that:
\begin{subequations}
\begin{gather}
{X^{\text{err}}_{t} = 0.01\gamma^{\text{FE}}\sum_{i=1}^{t-1} \mathbf{\overline x^{\text{EDS}}_{i}}(1)}, \label{eq:err_formula} \\
{X^{\text{bal}}_{t} = \sum_{j=2}^{T-(t-1)} \mathbf{\overline x^{\text{EDS}}_{t}}(j)}, \label{eq:bal_formula}
\end{gather}
\end{subequations}
where $X^{\text{err}}_{t}$ is the total generation over-consumption in proportion to the forecast error up to any time interval $t-1<T$, and $X^{\text{bal}}_{t}$ is the unused generation allocated to the future time intervals.  
%where the value of $X_t$ and $\mathbf{\overline x^{\text{EDS}}_{i}}(1)$ go on monotonically decreasing due to $\gamma^{\text{FE}} > 0$.
For continuous and secure operation of the CMG that ensures load supply for all time intervals, the following condition must be satisfied:
\begin{subequations}
\begin{gather}
{X_{t} > 0 \text{ and } \mathbf{\overline x^{\text{EDS}}_{t}} \in \mathbb{R}^{T-(t-1) \times 1}_{>0} \quad \forall t < T}, \label{eq:proof_gen_condition1} \\
{X_{t} \ge 0  \text{ and } \mathbf{\overline x^{\text{EDS}}_{t}} \in \mathbb{R}^{1 \times 1}_{>0} \quad t=T}. \label{eq:proof_gen_condition2}
\end{gather}
\end{subequations}
Equation (\ref{eq:proof_gen_condition1}) states that for any time interval $t < T$, the unused generation and the generation allocation for all the future time intervals should be greater than $0$. Equation (\ref{eq:proof_gen_condition2}) states that for the final time interval $T$, the generation allocation should be greater than $0$ and the balance unused generation can be greater than or equal to $0$.  

Next, we evaluate the conditions (\ref{eq:proof_gen_condition1}) and (\ref{eq:proof_gen_condition2}). Consider $t < T$:
\begin{subequations}
\begin{gather}
{X_t =  X - (1+0.01\gamma^{\text{FE}})\sum_{i=1}^{t} \mathbf{\overline x^{\text{EDS}}_{i}}(1)}, \label{eq:proof_availale_gen_t_less_T1} \\
{X_t = X - (1+0.01\gamma^{\text{FE}}) (X - X_t^{\text{bal}}- X_t^{\text{err}})}. \label{eq:proof_availale_gen_t_less_T2}
\end{gather}
\end{subequations}
For $X_t^{\text{bal}} < 0.01 \gamma^{\text{FE}} X/(1+0.01\gamma^{\text{FE}})-X_t^{\text{err}}$, $X_t < 0$. This violates condition (\ref{eq:proof_gen_condition1}). \\
Consider $t=T$:
\begin{subequations}
\begin{gather}
{X_T = X - (1+0.01\gamma^{\text{FE}})\sum_{i=1}^{T} \mathbf{\overline x^{\text{EDS}}_{i}}(1)}, \label{eq:proof_final_availale_gen1}\\
{X_T = \sum_{i=1}^{T} \mathbf{\overline x^{\text{EDS}}_{i}}(1) +X_T^{\text{err}} - (1+0.01\gamma^{\text{FE}})\sum_{i=1}^{T} \mathbf{\overline x^{\text{EDS}}_{i}}(1)}, \\
{X_T = 0.01\gamma^{\text{FE}}(\sum_{i=1}^{T-1} \mathbf{\overline x^{\text{EDS}}_{i}}(1)-\sum_{i=1}^{T} \mathbf{\overline x^{\text{EDS}}_{i}}(1))}.  \label{eq:proof_final_availale_gen2}  %if you expand the RHS, the RHS comes out to be negative since the sum of the forecast error will be less than the total gen X
\end{gather}
\end{subequations}
From (\ref{eq:proof_final_availale_gen2}), we observe that for a value of $\gamma^{\text{FE}} > 0$, the condition (\ref{eq:proof_gen_condition2}) will not be satisfied. This translates to the fact that the CMG can securely operate until a threshold time interval $t'<T$ such that:
\begin{subequations}
\begin{gather}
{X_t^{\text{bal}} > 0.01 \gamma^{\text{FE}} X/(1+0.01\gamma^{\text{FE}}) - X_t^{\text{err}}  \text{ and } X_{t} > 0 \quad \forall t < t'}, \\
{X_t^{\text{bal}} = 0.01 \gamma^{\text{FE}} X/(1+0.01\gamma^{\text{FE}}) - X_t^{\text{err}} \text{ and } X_{t} = 0 \quad \text{for } t = t'}, \\
{X_t^{\text{bal}} = 0  \text{ and } X_{t'} = 0 \quad  \forall t' < t \le T}.
\end{gather}
\end{subequations}
The above proof can be generalized to $\gamma^{\text{FE}} < 0$, which would result in a secure operation of the CMG, but with significant underconsumption of resources and increased availability of unused generation for the final time intervals. %Further, this proof can be generalized to time varying values of $\gamma^{\text{FE}}$.
\end{proof}

\subsection{Proof of Theorem \ref{Thm2}}
  \label{Thm2Proof}

\begin{proof}
Let the EDS load allocation (in kW) obtained for the first time interval $t$ and $T-t$ future time intervals be defined by $\mathbf{\overline y^{\text{EDS}}_{t}} \in \mathbb{R}^{T-t+1 \times 1}_{+}$. Similarly, let ${\overline y^{\text{NRT}}_{t}} \in \mathbb{R}^{1 \times 1}_{+}$ and ${\overline y^{\text{RT}}_{t}} \in \mathbb{R}^{1 \times 1}_{+}$ be the NRT load allocation and RT load realization. 
Following the structure of HMTS, let $\mathbf{\overline x^{\text{EDS}}_{t}} \in \mathbb{R}^{T-t+1 \times 1}$ be the vector of allocated generation (in kW) for the future time horizon as computed at the $t^\text{th}$ time interval computed in a receding horizon manner, ${\overline x^{\text{NRT}}_{t}} \in \mathbb{R}^{1 \times 1}$ be the NRT revised EDS generation allocation for the $t^\text{th}$ time interval, and ${\overline x^{\text{RT}}_{t}} \in \mathbb{R}^{1 \times 1}$ be the RT generation realized value. Let $X$ kWh be the total available generation resources. The generation allocation must equal the load allocation. To factor in the $1$-delayed recourse constraint, the NRT revised EDS allocation will have to be modified using the past over-consumption. For simplicity, assume no forecast error between the EDS and NRT stages and the time interval ($\Delta t$) is one hour. Hence, ${\overline x^{\text{NRT}}_{t}} = \mathbf{\overline x^{\text{EDS}}_{t}}(1)-0.01\gamma^{\text{FE}}\mathbf{\overline x^{\text{EDS}}_{t-1}}(1)$. Due to the positive load forecast error of $\gamma^{\text{FE}} \% $, the RT load that needs to be supplied in full equals $(1+0.01\gamma^{\text{FE}})(\mathbf{\overline y^{\text{EDS}}_{t}}(1) - 0.01\gamma^{\text{FE}}\mathbf{\overline y^{\text{EDS}}_{t-1}}(1))$. Hence, the RT generation usage is given by ${\overline x^{\text{RT}}_{t}} = (1+0.01\gamma^{\text{FE}})(\mathbf{\overline x^{\text{EDS}}_{t}}(1) - 0.01\gamma^{\text{FE}}\mathbf{\overline x^{\text{EDS}}_{t-1}}(1))$, i.e., ${\overline x^{\text{RT}}_{t}} > \mathbf{\overline x^{\text{EDS}}_{t}}(1)$. The excess generation consumption of $0.01\gamma^{\text{FE}}(\mathbf{\overline x^{\text{EDS}}_{t}}(1) - 0.01\gamma^{\text{FE}}\mathbf{\overline x^{\text{EDS}}_{t-1}}(1))$ is observed, and the remaining generation capacity equals $X - (1+0.01\gamma^{\text{FE}})\mathbf{\overline x^{\text{EDS}}_{t}}(1)$. Since the forecast error of the past can be computed for any time interval $t \ge 2$, the values of $\mathbf{\overline x^{\text{EDS}}_{t-1}}(1)$ and $\mathbf{\overline y^{\text{EDS}}_{t-1}}(1)$ at $t=1$ be $0$.  

Continuing this trend by incrementing the time intervals by 1, and assuming that the entire generation capacity is used up at the end of the final time interval, the available unused generation at the end of time interval $t < T$, $X_t$, and the total generation capacity, $X$, can be generalized as follows: 

\begin{subequations}
\begin{gather}
{X_t = X - (1+0.01\gamma^{\text{FE}})\sum_{i=1}^{t} \mathbf{(\overline x^{\text{EDS}}_{i}}(1)-0.01\gamma^{\text{FE}}\mathbf{\overline x^{\text{EDS}}_{i-1}}(1))}, \label{eq:proof_availale_gen_recourse} \\
{X = (1+0.01\gamma^{\text{FE}})\sum_{i=1}^{t-1} (\mathbf{\overline x^{\text{EDS}}_{i}}(1) - 0.01\gamma^{\text{FE}} \mathbf{\overline x^{\text{EDS}}_{i-1}}(1))} \\ { + \sum_{j=1}^{T-(t-1)} \mathbf{\overline x^{\text{EDS}}_{t}}(j)}, \label{eq:proof_total_gen1_recourse} \\ {= \sum_{i=1}^{t} \mathbf{\overline x^{\text{EDS}}_{i}}(1) + X^{\text{err}}_{t} + X^{\text{bal}}_{t}}. \label{eq:proof_total_gen2_recourse}
\end{gather}
\end{subequations}
%where the value of $X_t$ and $\mathbf{\overline x^{\text{EDS}}_{i}}(1)$ go on monotonically decreasing due to $\gamma^{\text{FE}} > 0$.
From (\ref{eq:proof_total_gen1_recourse}) and(\ref{eq:proof_total_gen2_recourse}), we can show that:
\begin{subequations}
\begin{gather}
{X^{\text{err}}_{t} = 0.01\gamma^{\text{FE}}\sum_{i=1}^{t-1} (\mathbf{\overline x^{\text{EDS}}_{i}}(1) - \mathbf{\overline x^{\text{EDS}}_{i-1}}(1))}, \label{eq:err_formula_recourse} \\
{X^{\text{bal}}_{t} = \sum_{j=2}^{T-(t-1)} \mathbf{\overline x^{\text{EDS}}_{t}}(j)}, \label{eq:bal_formula_recourse}
\end{gather}
\end{subequations}
where $X^{\text{err}}_{t}$ is the total generation over-consumption in proportion to the forecast error up to any time interval $t<T$, and $X^{\text{bal}}_{t}$ is the generation allocated to the future time intervals. Comparing the values of $X^{\text{err}}_{t}$ from Theorem \ref{Thm1} and Theorem \ref{Thm2}, we observe that the the generation used for addressing forecast error is lesser by using the delayed recourse approach by $0.01\gamma^{\text{FE}}\sum_{i=1}^{t-1} (\mathbf{\overline x^{\text{EDS}}_{i-1}}(1))$. For optimal and secure operation of the CMG, the following condition must be satisfied:
\begin{subequations}
\begin{gather}
{X_{t} > 0 \text{ and } \mathbf{\overline x^{\text{EDS}}_{t}} \in \mathbb{R}^{T-t+1 \times 1}_{>0} \quad \forall t < T}, \label{eq:proof_gen_condition1_recourse} \\
{X_{t} \ge 0  \text{ and } \mathbf{\overline x^{\text{EDS}}_{t}} \in \mathbb{R}^{1 \times 1}_{>0} \quad t=T}. \label{eq:proof_gen_condition2_recourse}
\end{gather}
\end{subequations}
Equation (\ref{eq:proof_gen_condition1_recourse}) states that for any time interval $t < T$, the unused generation and the generation allocation for all the future time intervals should be greater than $0$. Equation (\ref{eq:proof_gen_condition2_recourse}) states that for the final time interval $T$, the generation allocation should be greater than $0$ and the balance unused generation can be greater than or equal to $0$.  

Next, we evaluate the conditions (\ref{eq:proof_gen_condition1_recourse}) and (\ref{eq:proof_gen_condition2_recourse}). Consider $t < T$:
\begin{subequations}
\begin{gather}
{X_t =  X - (1+0.01\gamma^{\text{FE}})\sum_{i=1}^{t} (\mathbf{\overline x^{\text{EDS}}_{i}}(1)-0.01\gamma^{\text{FE}}\mathbf{\overline x^{\text{EDS}}_{i-1}}(1))}, \label{eq:proof_availale_gen_t_less_T1_recourse} \\
{X_t = X - (1+0.01\gamma^{\text{FE}}) (X - X_t^{\text{bal}}- X_t^{\text{err}})}, \label{eq:proof_availale_gen_t_less_T2_recourse}
\end{gather}
\end{subequations}
For $X_t^{\text{bal}} < 0.01 \gamma^{\text{FE}} X/(1+0.01\gamma^{\text{FE}})-X_t^{\text{err}}-0.01\gamma^{\text{FE}}\sum_{i=1}^{t} \mathbf{\overline x^{\text{EDS}}_{i-1}}(1)$, $X_t < 0$, eventually violating the condition (\ref{eq:proof_gen_condition1_recourse}). However, in comparison with the equivalent condition in Theorem \ref{Thm1}, we can see that the violating value of $X_t^{\text{bal}}$ is much smaller when using $1$-delayed recourse. This results in the ability to extend the secure operating duration closer to the final time interval $T$.\\
Consider $t=T$:
\begin{subequations}
\begin{gather}
{X_T = X - (1+0.01\gamma^{\text{FE}})\sum_{i=1}^{T}(\mathbf{\overline x^{\text{EDS}}_{i}}(1) - 0.01\gamma^{\text{FE}}\mathbf{\overline x^{\text{EDS}}_{i-1}}(1))}, \label{eq:proof_final_availale_gen1_recourse}\\
{X_T = \sum_{i=1}^{T} \mathbf{\overline x^{\text{EDS}}_{i}}(1) + X^{\text{err}}_{T} - (1+0.01\gamma^{\text{FE}}) \sum_{i=1}^{T} (\mathbf{\overline x^{\text{EDS}}_{i}}(1) -} \nonumber \\  {0.01\gamma^{\text{FE}}\mathbf{\overline x^{\text{EDS}}_{i-1}}(1))},  \label{eq:proof_final_availale_gen2_recourse}  \\
{X_T \approx 0.01\gamma^{\text{FE}}(\overline x^{\text{EDS}}_{T-1}(1)- \overline x^{\text{EDS}}_{T}(1))}.  \label{eq:proof_final_availale_gen3_recourse}
\end{gather}
\end{subequations}
Assuming a constant load profile, the load supplied will go on monotonically decreasing with time due to depletion of generation resources. Hence, from (\ref{eq:proof_final_availale_gen3_recourse}), we observe that $X_T  \ge 0$. This means that at the end of the final time interval, there is some amount of generation unused. From $X_T \ge 0$ and (\ref{eq:proof_availale_gen_t_less_T2_recourse}), we can conclude that $X_t^{\text{bal}} > 0.01 \gamma^{\text{FE}} X/(1+0.01\gamma^{\text{FE}})-X_t^{\text{err}}-0.01\gamma^{\text{FE}}\sum_{i=1}^{t} \mathbf{\overline x^{\text{EDS}}_{i-1}}(1) \quad \forall t < T$. This translates to the fact that the conditions (\ref{eq:proof_gen_condition1_recourse}) and (\ref{eq:proof_gen_condition2_recourse}) are satisfied, allowing the CMG to operate securely for all time intervals $t \le T$.

The above proof can be generalized to $\gamma^{\text{FE}} < 0$, which would result in a secure operation of the CMG, but with significant under-consumption of resources and increased availability of unused generation at the final time interval. For such a case, the recourse action will constitute of increasing the generation allocation in the present based on the under-consumption of the immediate past. %Further, this proof can be generalized to time varying values of $\gamma^{\text{FE}}$.
\end{proof}
\end{document}